\documentclass{aastex63}

\usepackage{amssymb}
\usepackage{xcolor}

\usepackage{hyperref}

\usepackage{graphicx}

\received{TBD}
\revised{TBD}
\accepted{TBD} 

\shorttitle{BURSTT}
\shortauthors{Lin et al.}

\begin{document}

\title{BURSTT: Bustling Universe Radio Survey Telescope in Taiwan}

\correspondingauthor{Hsiu-Hsien Lin}
\email{hsiuhsien@asiaa.sinica.edu.tw}

\author[0000-0001-7453-4273]{Hsiu-Hsien Lin}
  \affiliation{Institute of Astronomy and Astrophysics, Academia Sinica, 11F of AS/NTU Astronomy-Mathematics Building, No.1, Sec. 4, Roosevelt Rd, Taipei 10617, Taiwan, R.O.C.}
  \affiliation{Canadian Institute for Theoretical Astrophysics, 60 St. George Street, Toronto, ON M5S 3H8, Canada}
\author[0000-0002-8698-7277]{Kai-yang Lin}
  \affiliation{Institute of Astronomy and Astrophysics, Academia Sinica, 11F of AS/NTU Astronomy-Mathematics Building, No.1, Sec. 4, Roosevelt Rd, Taipei 10617, Taiwan, R.O.C.}
\author[0000-0002-5830-2226]{Chao-Te Li}
  \affiliation{Institute of Astronomy and Astrophysics, Academia Sinica, 11F of AS/NTU Astronomy-Mathematics Building, No.1, Sec. 4, Roosevelt Rd, Taipei 10617, Taiwan, R.O.C.}
\author[0000-0002-6495-8600]{Yao-Huan Tseng}
  \affiliation{Institute of Astronomy and Astrophysics, Academia Sinica, 11F of AS/NTU Astronomy-Mathematics Building, No.1, Sec. 4, Roosevelt Rd, Taipei 10617, Taiwan, R.O.C.}
\author[0000-0002-9893-2433]{Homin Jiang}
  \affiliation{Institute of Astronomy and Astrophysics, Academia Sinica, 11F of AS/NTU Astronomy-Mathematics Building, No.1, Sec. 4, Roosevelt Rd, Taipei 10617, Taiwan, R.O.C.}
\author[0000-0003-4708-5964]{Jen-Hung Wang}
  \affiliation{Institute of Astronomy and Astrophysics, Academia Sinica, 11F of AS/NTU Astronomy-Mathematics Building, No.1, Sec. 4, Roosevelt Rd, Taipei 10617, Taiwan, R.O.C.}
\author{Jen-Chieh Cheng}
  \affiliation{Institute of Astronomy and Astrophysics, Academia Sinica, 11F of AS/NTU Astronomy-Mathematics Building, No.1, Sec. 4, Roosevelt Rd, Taipei 10617, Taiwan, R.O.C.}
\author[0000-0003-2155-9578]{Ue-Li Pen}
  \affiliation{Institute of Astronomy and Astrophysics, Academia Sinica, 11F of AS/NTU Astronomy-Mathematics Building, No.1, Sec. 4, Roosevelt Rd, Taipei 10617, Taiwan, R.O.C.}
  \affiliation{Canadian Institute for Theoretical Astrophysics, 60 St. George Street, Toronto, ON M5S 3H8, Canada}
  \affiliation{Canadian Institute for Advanced Research, 180 Dundas St West, Toronto, ON M5G 1Z8, Canada}
  \affiliation{David D Dunlap Institute for Astronomy and Astrophysics, University of Toronto, 50 St George Street, Toronto, ON M5S 3H4, Canada}
  \affiliation{Perimeter Institute of Theoretical Physics, 31 Caroline Street North, Waterloo, ON N2L 2Y5, Canada}
\author[0000-0001-6573-3318]{Ming-Tang Chen}
  \affiliation{Institute of Astronomy and Astrophysics, Academia Sinica, 645 N Aohoku Pl, Hilo, HI 96720 USA}
  \affiliation{Institute of Astronomy and Astrophysics, Academia Sinica, 11F of AS/NTU Astronomy-Mathematics Building, No.1, Sec. 4, Roosevelt Rd, Taipei 10617, Taiwan, R.O.C.}
\author[0000-0001-5251-7210]{Pisin Chen}
  \affiliation{Department of Physics, National Taiwan University, No. 1, Sec. 4, Roosevelt Rd., Taipei 10617, Taiwan, R.O.C.}
  \affiliation{Leung Center for Cosmology and Particle Astrophysics, National Taiwan University, No. 1, Sec. 4, Roosevelt Rd., Taipei 10617, Taiwan, R.O.C.}
\author[0000-0002-8967-4911]{Yaocheng Chen}
  \affiliation{Department of Physics, National Taiwan University, No. 1, Sec. 4, Roosevelt Rd., Taipei 10617, Taiwan, R.O.C.}
  \affiliation{Leung Center for Cosmology and Particle Astrophysics, National Taiwan University, No. 1, Sec. 4, Roosevelt Rd., Taipei 10617, Taiwan, R.O.C.}
\author[0000-0002-6821-8669]{Tomotsugu Goto}
  \affiliation{Institute of Astronomy, National Tsing Hua University, 101 Section 2 Kuang-Fu Road, Hsinchu 30013, Taiwan (ROC)}
\author[0000-0001-7228-1428]{Tetsuya Hashimoto}
  \affiliation{Department of Physics, National Chung Hsing University, No. 145, Xingda Rd., South Dist., Taichung 40227, Taiwan (R.O.C.)}
\author[0000-0002-3931-6243]{Yuh-Jing Hwang}
  \affiliation{Institute of Astronomy and Astrophysics, Academia Sinica, 11F of AS/NTU Astronomy-Mathematics Building, No.1, Sec. 4, Roosevelt Rd, Taipei 10617, Taiwan, R.O.C.}
\author{Sun-Kun King}
  \affiliation{Institute of Astronomy and Astrophysics, Academia Sinica, 11F of AS/NTU Astronomy-Mathematics Building, No.1, Sec. 4, Roosevelt Rd, Taipei 10617, Taiwan, R.O.C.}
\author{Derek Kubo}
  \affiliation{Institute of Astronomy and Astrophysics, Academia Sinica, 645 N Aohoku Pl, Hilo, HI 96720 USA}
\author[0000-0002-0563-4497]{Chung-Yun Kuo}
  \affiliation{Leung Center for Cosmology and Particle Astrophysics, National Taiwan University, No. 1, Sec. 4, Roosevelt Rd., Taipei 10617, Taiwan, R.O.C.}
  \affiliation{Department of Physics, National Taiwan University, No. 1, Sec. 4, Roosevelt Rd., Taipei 10617, Taiwan, R.O.C.}
\author{Adam Mills}
  \affiliation{Institute of Astronomy and Astrophysics, Academia Sinica, 645 N Aohoku Pl, Hilo, HI 96720 USA}
\author[0000-0001-9673-3134]{Jiwoo Nam}
  \affiliation{Department of Physics, National Taiwan University, No. 1, Sec. 4, Roosevelt Rd., Taipei 10617, Taiwan, R.O.C.}
  \affiliation{Leung Center for Cosmology and Particle Astrophysics, National Taiwan University, No. 1, Sec. 4, Roosevelt Rd., Taipei 10617, Taiwan, R.O.C.}
\author{Peter Oshiro}
  \affiliation{Institute of Astronomy and Astrophysics, Academia Sinica, 645 N Aohoku Pl, Hilo, HI 96720 USA}
\author[0000-0002-1484-105X]{Chang-Shao Shen}
  \affiliation{Institute of Astronomy and Astrophysics, Academia Sinica, 11F of AS/NTU Astronomy-Mathematics Building, No.1, Sec. 4, Roosevelt Rd, Taipei 10617, Taiwan, R.O.C.}
\author{Hsien-Chun Tseng}
  \affiliation{Institute of Astronomy and Astrophysics, Academia Sinica, 11F of AS/NTU Astronomy-Mathematics Building, No.1, Sec. 4, Roosevelt Rd, Taipei 10617, Taiwan, R.O.C.}
\author[0000-0002-0060-7975]{Shih-Hao Wang}
  \affiliation{Department of Physics, National Taiwan University, No. 1, Sec. 4, Roosevelt Rd., Taipei 10617, Taiwan, R.O.C.}
  \affiliation{Leung Center for Cosmology and Particle Astrophysics, National Taiwan University, No. 1, Sec. 4, Roosevelt Rd., Taipei 10617, Taiwan, R.O.C.}
\author{Vigo Feng-Shun Wu}
  \affiliation{Institute of Astronomy and Astrophysics, Academia Sinica, 11F of AS/NTU Astronomy-Mathematics Building, No.1, Sec. 4, Roosevelt Rd, Taipei 10617, Taiwan, R.O.C.}
\author[0000-0003-4056-9982]{Geoffrey Bower}
  \affiliation{Institute of Astronomy and Astrophysics, Academia Sinica, 645 N Aohoku Pl, Hilo, HI 96720 USA}
\author{Shu-Hao Chang}
  \affiliation{Institute of Astronomy and Astrophysics, Academia Sinica, 11F of AS/NTU Astronomy-Mathematics Building, No.1, Sec. 4, Roosevelt Rd, Taipei 10617, Taiwan, R.O.C.}
\author{Pai-An Chen}
  \affiliation{Institute of Astronomy and Astrophysics, Academia Sinica, 11F of AS/NTU Astronomy-Mathematics Building, No.1, Sec. 4, Roosevelt Rd, Taipei 10617, Taiwan, R.O.C.}
\author{Ying-Chih Chen}
  \affiliation{Department of Physics, National Taiwan University, No. 1, Sec. 4, Roosevelt Rd., Taipei 10617, Taiwan, R.O.C.}
  \affiliation{Leung Center for Cosmology and Particle Astrophysics, National Taiwan University, No. 1, Sec. 4, Roosevelt Rd., Taipei 10617, Taiwan, R.O.C.}
\author[0000-0001-6320-261X]{Yi-Kuan Chiang}
  \affiliation{Institute of Astronomy and Astrophysics, Academia Sinica, 11F of AS/NTU Astronomy-Mathematics Building, No.1, Sec. 4, Roosevelt Rd, Taipei 10617, Taiwan, R.O.C.}
\author[0000-0003-2837-3477]{Anatoli Fedynitch}
  \affiliation{Institute of Physics, Academia Sinica, Taipei, 11529, Taiwan}
\author[0000-0001-6128-3735]{Nina Gusinskaia}
  \affiliation{David D Dunlap Institute for Astronomy and Astrophysics, University of Toronto, 50 St George Street, Toronto, ON M5S 3H4, Canada}
  \affiliation{Canadian Institute for Theoretical Astrophysics, 60 St. George Street, Toronto, ON M5S 3H8, Canada}
\author[0000-0002-8560-3497]{Simon C.-C.~Ho}
  \affiliation{Institute of Astronomy, National Tsing Hua University, 101 Section 2 Kuang-Fu Road, Hsinchu 30013, Taiwan (ROC)}
\author[0000-0003-4512-8705]{Tiger Y.-Y.~Hsiao}
  \affiliation{Institute of Astronomy, National Tsing Hua University, 101 Section 2 Kuang-Fu Road, Hsinchu 30013, Taiwan (ROC)}
  \affiliation{Department of Physics and Astronomy, Johns Hopkins University, Baltimore, MD 21218, USA}
\author[0000-0001-8551-2002]{Chin-Ping Hu}
  \affiliation{Department of Physics, National Changhua University of Education, Changhua, 50007, Taiwan}
\author[0000-0001-8783-6211]{Yau De Huang}
  \affiliation{Institute of Astronomy and Astrophysics, Academia Sinica, 11F of AS/NTU Astronomy-Mathematics Building, No.1, Sec. 4, Roosevelt Rd, Taipei 10617, Taiwan, R.O.C.}
\author[0000-0002-8297-732X]{José Miguel Jáuregui García}
  \affiliation{Canadian Institute for Theoretical Astrophysics, 60 St. George Street, Toronto, ON M5S 3H8, Canada}
\author[0000-0001-9970-8145]{Seong Jin Kim}
  \affiliation{Institute of Astronomy, National Tsing Hua University, 101 Section 2 Kuang-Fu Road, Hsinchu 30013, Taiwan (ROC)}
\author[0000-0002-0563-4497]{Cheng-Yu Kuo}
  \affiliation{Physics Department, National Sun Yat-Sen University, No. 70, Lien-Hai Road, Kaosiung City 80424, Taiwan, R.O.C.}
\author{Decmend Fang-Jie Ling}
  \affiliation{Department of Physics, National Tsing Hua University, 101 Section 2 Kuang-Fu Road, Hsinchu 30013, Taiwan (ROC)}
\author[0000-0003-4479-4415]{Alvina Y.~L.~On}
  \affiliation{Institute of Astronomy, National Tsing Hua University, 101 Section 2 Kuang-Fu Road, Hsinchu 30013, Taiwan (ROC)}
  \affiliation{Department of Physics, National Chung Hsing University, No. 145, Xingda Rd., South Dist., Taichung 40227, Taiwan (R.O.C.)}
  \affiliation{Mullard Space Science Laboratory, University College London, Holmbury St Mary, Surrey RH5 6NT, UK}
\author[0000-0003-1340-818X]{Jeffrey B.~Peterson}
  \affiliation{Department of Physics, Carnegie Mellon University, 5000 Forbes Ave. Pittsburgh PA 15213 USA}
\author[0000-0003-0054-6081]{Bjorn Jasper R.~Raquel}
  \affiliation{Department of Physics, National Chung Hsing University, No. 145, Xingda Rd., South Dist., Taichung 40227, Taiwan (R.O.C.)}
  \affiliation{Department of Earth and Space Sciences, Rizal Technological University, Boni Avenue, Mandaluyong, 1550 Metro Manila, Philippines}
\author{Shih-Chieh Su}
  \affiliation{Department of Physics, National Taiwan University, No. 1, Sec. 4, Roosevelt Rd., Taipei 10617, Taiwan, R.O.C.}
  \affiliation{Leung Center for Cosmology and Particle Astrophysics, National Taiwan University, No. 1, Sec. 4, Roosevelt Rd., Taipei 10617, Taiwan, R.O.C.}
\author[0000-0003-2792-4978]{Yuri Uno}
  \affiliation{Department of Physics, National Chung Hsing University, No. 145, Xingda Rd., South Dist., Taichung 40227, Taiwan (R.O.C.)}
\author{Cossas K.-W.~Wu}
  \affiliation{Institute of Astronomy, National Tsing Hua University, 101 Section 2 Kuang-Fu Road, Hsinchu 30013, Taiwan (ROC)}
\author[0000-0002-1688-8708]{Shotaro Yamasaki}
  \affiliation{Department of Physics, National Chung Hsing University, No. 145, Xingda Rd., South Dist., Taichung 40227, Taiwan (R.O.C.)}
\author[0000-0002-8202-8642]{Hong-Ming Zhu}
  \affiliation{Canadian Institute for Theoretical Astrophysics, 60 St. George Street, Toronto, ON M5S 3H8, Canada}

\begin{abstract}

Fast Radio Bursts (FRBs) are bright millisecond-duration radio transients that appear about 1,000 times per day, all-sky, for a fluence threshold 5 Jy ms at 600 MHz. The FRB radio-emission physics and the compact objects involved in these events are subjects of intense active debate. To better constrain source models, the Bustling Universe Radio Survey Telescope in Taiwan (BURSTT) is optimized to discover and localize a large sample of rare, high-fluence, nearby FRBs. This is the population most amenable to multi-messenger, multi-wavelength follow-up, allowing deeper understanding of source mechanisms.
BURSTT will provide horizon-to-horizon sky coverage with a half power field-of-view (FoV) of $\sim$10$^{4}$ deg$^{2}$, a 400 MHz effective bandwidth between 300-800 MHz, and sub-arcsecond localization, made possible using outrigger stations hundreds to thousands of km from the main array. Initially, BURSTT will employ 256 antennas. After tests of various antenna designs and optimization of system performance we plan to expand to 2048 antennas. We estimate that BURSTT-256 will detect and localize $\sim$100 bright ($\geq$100 Jy ms) FRBs per year. Another advantage of BURSTT's large FoV and continuous operation will be greatly enhanced monitoring of FRBs for repetition. The current lack of sensitive all-sky observations likely means that many repeating FRBs are currently cataloged as single-event FRBs. 
\end{abstract}


\section{INTRODUCTION}

Fast Radio Bursts (FRBs) are bright ($\sim$1 Jy), millisecond flashes of radio light of uncertain astrophysical origin  \citep{2007Sci...318..777L, 2017ApJ...834L...7T, 2020Natur.581..391M}. With the fluence threshold above 5 Jy ms at 600 MHz, the all-sky occurrence rates of FRBs is $\sim$1,000 per day (e.g., \cite{2018ApJ...863...48C, 2021ApJS..257...59C}) . The spatial distribution is independent of the galactic latitude \citep{2021ApJ...923....2J, 2018MNRAS.475.1427B}. 

The nature of FRBs, including their emission mechanism, central object type, and environment, is one of the most perplexing enigmas in astrophysics. Deepening the mystery, about 4 percent of FRB sources emit multiple bursts \citep{2021ApJS..257...59C}, the so-called “repeaters” (e.g., \cite{2016Natur.531..202S}), while for most FRBs only one burst is observed (e.g., \cite{2019A&ARv..27....4P,2021ApJS..257...59C}). A few FRBs have been reported with periodicities ranging from sub-seconds to several months \citep{2022Natur.607..256C, 2020Natur.582..351C, 2020MNRAS.495.3551R}. It is so far unclear whether repeaters and non-repeaters originate from astrophysically different populations and whether there are multiple channels for FRB formation \citep{2019PhR...821....1P, 2020MNRAS.498.3927H}. 

About two dozen FRBs have been associated with a particular host galaxy \citep{2017Natur.541...58C, 2020ApJ...895L..37B, 2020ApJ...903..152H, 2022AJ....163...69B}, and a few repeaters have been pinpointed inside the host galaxy through Very-Long Baseline Interferometry (VLBI)\citep{2017ApJ...834L...8M, 2020Natur.577..190M, 2022Natur.602..585K}. Two repeating FRBs are associated with persistent radio sources \citep{2017ApJ...834L...8M, 2022Natur.606..873N}, which show complicated polarization properties \citep{2018Natur.553..182M, 2022arXiv220211112A, 2022arXiv220308151D}. 

Because only a small number of FRB's have been associated with host galaxies the distance to these sources has so far not been available so luminosity function of FRBs is currently poorly constrained \citep{2018MNRAS.481.2320L}.

Although almost fifteen years have passed since their discovery \citep{2007Sci...318..777L}, there is no consensus about their origin despite an increasing number of recent detections \citep{2019PhR...821....1P, 2021ApJS..257...59C}.  Beyond the central questions of the source composition and emission mechanism, it has been suggested that FRBs could be used to address key issues in cosmology and physics, including dark energy (e.g., \citet{2019PhRvD..99l3517L}; \citet{2019MNRAS.488.1908H}), dark matter (e.g., \citet{2016PhRvL.117i1301M}; \citet{2022PhRvD.106d3017L}), testing of the general relativity (e.g., \citet{2015PhRvL.115z1101W}; \citet{2021PhRvD.104l4026H}), and the missing baryon problem (e.g., \citet{2018PhRvD..98j3518M, 2020Natur.581..391M}).  

There have been three major challenges in revealing the physical origins of FRBs: (i) low detection probability. Existing telescopes have limited field of view (FoV), but FRBs randomly appear on the sky, so the great majority of detectable FRBs are currently missed., (ii) poor localization to their host, due to the insufficient spatial resolution of existing radio telescopes, and (iii) The high antenna gain of current telescopes means that FRBs currently detected are often too distant to conduct multi-wavelength/multi-messenger observations to identify their progenitors, host galaxies, and simultaneous emission counterparts. All of these issues arise because existing radio telescopes are not tailored for FRBs.

The proposed Bustling Universe Radio Survey Telescope in Taiwan (BURSTT) will address all these problems by detecting high-fluence (bright) FRBs in the nearby Universe. Because of the relatively small cosmic volume in the nearby univerese these are rare, so we have maximized the FoV to increase the detection rate.  BURSTT is a unique fisheye radio software telescope, with which we will observe at least 25 times more sky than any existing radio observatory, except the Survey for Transient Astronomical Radio Emission 2 (STARE2) \citep{2020PASP..132c4202B}, which has a much lower sensitivity than BURSTT. By viewing so much more sky than other telescopes, and hence detecting a larger sample of bright and nearby sources, BURSTT will be unique in identifying multi-frequency (i.e. optical and X-ray/gamma-ray) and multi-messenger (gravitaitonal-wave and nuetrino) counterparts. In addition, BURSTT-2048 will add substantially to the the overall world-wide FRB detection rate.

In this paper, the scientific objectives of BURSTT are shown in Section \ref{section: SCIENCE OBJECTIVES}. The potential technical aspects are illustrated in Section \ref{section: TECHNICAL OBJECTIVES}. The work is summarized in Section \ref{section: SUMMARY}.

\section{SCIENCE OBJECTIVES}\label{section: SCIENCE OBJECTIVES}

\subsection{The extremely wide field of view of BURSTT}\label{subsec:Extremely wide field of view of BURSTT}

To achieve a complete census of FRBs, monitoring observations of the nearby Universe with a very wide FoV are required \citep{2020PASP..132c4202B, 2021PASP..133g5001C}.

For any array of antennas, the FoV is inversely proportional to the effective collecting area of the array, so the wide field of BURSTT means it will have a smaller collecting area than many previous telescopes. In estimating the detection rate (see below) the area and FOV factors partially compensate and the resulting detection rate is not strongly dependent on the choice of FoV.  However, for the wide field (low collecting area) telescope, the detected sources are on average closer.

The redshift of CHIME/FRB sample is in the range of 0.3-0.5 \citep{2021ApJ...922...42R}. The CHIME/FRB team is adding outriggers to CHIME, which will localize many FRB events in this redshift range \citep{2022AJ....163...65C, 2022AJ....163...48M}. For many of these, optical follow-up observations will be needed to get the redshift of the host galaxy. In contrast, we anticipate the median redshift of the BURSTT FRB sample will be in the range of 0.03-0.05. This means that many of the BURSTT host galaxies will be present in existing optical redshift survey catalogs \citep{2013AJ....145..101K, 2003tmc..book.....C, 2014ApJS..210....9B, 2016ApJS..225....5B, 2019AJ....157..168D} allowing immediate redshift determination without the need for follow-up observations. This will allow for a rapid improvement in the measured FRB luminosity function \citep{2022arXiv220714316S}. 

To estimate the system equivalent flux density (SEFD) sensitivity of the BURSTT, we assume a conservative system temperature (Tsys) of $\sim$150 K, achievable with commercially-available LNAs, and an effective area for 256 antennas of $\sim$100 m$^{2}$ at 600 MHz. The corresponding SEFD is $\sim$5000 Jy. For an FRB with a duration of 1 ms across 400 MHz bandwidth with a fluence of 100 Jy ms, the corresponding signal-to-noise ratio (S/N) is $\sim$12. In the future, we hope to achieve Tsys of 50 K, which will improve the sensitivity.

To estimate the event rate, we assume a Euclidian distribution to scale from a previously measured reference rate, 
\begin{equation}\label{equation: event_rate}
\mathrm{N(S)} =\mathrm{N_{0}(\frac{S}{S_{0}})^{-1.5}},
\end{equation}
where N(S) is the event rate above fluence S, N$_{0}$ is the reference rate above reference fluence S$_{0}$. We use the reference event rate reported by CHIME \citep{2021ApJS..257...59C}, 300 FRBs per day per sky, with  S$_{0}$ = 5 Jy ms at 600 MHz, and two additional constraints: scattering time at 600 MHz below 10 ms and DM above 100 pc cm$^{-3}$. The FoV of BURSTT is $\sim$10$^{4}$ deg$^{2}$ or 0.24 sky. Thus, we estimate that BURSTT-256 can detect $\sim$100 bright FRBs per year with the threshold of fluence higher than 100 Jy ms at 600 MHz. Note that the CHIME FRB rate is at the low end of published rate estimates, making our estimate a conservative one. 

The properties of repeating FRBs are poorly constrained (e.g., \citet{2020ApJ...891L...6F}) because only a small number have been detected. Uncertain parameters include the event rate and its dependence on flux density, the possibility of non-Poisson distribution of events \citep{2018MNRAS.475.5109O}, and the host environment of these FRBs.  With its all-sky coverage BURSTT will likely uncover a larger fraction of repeaters, sharply localize the sources and provide the detailed data for each received pulse. This should substantially improve our understanding of repeating FRBs.

\begin{figure}
\epsscale{0.75}
\plotone{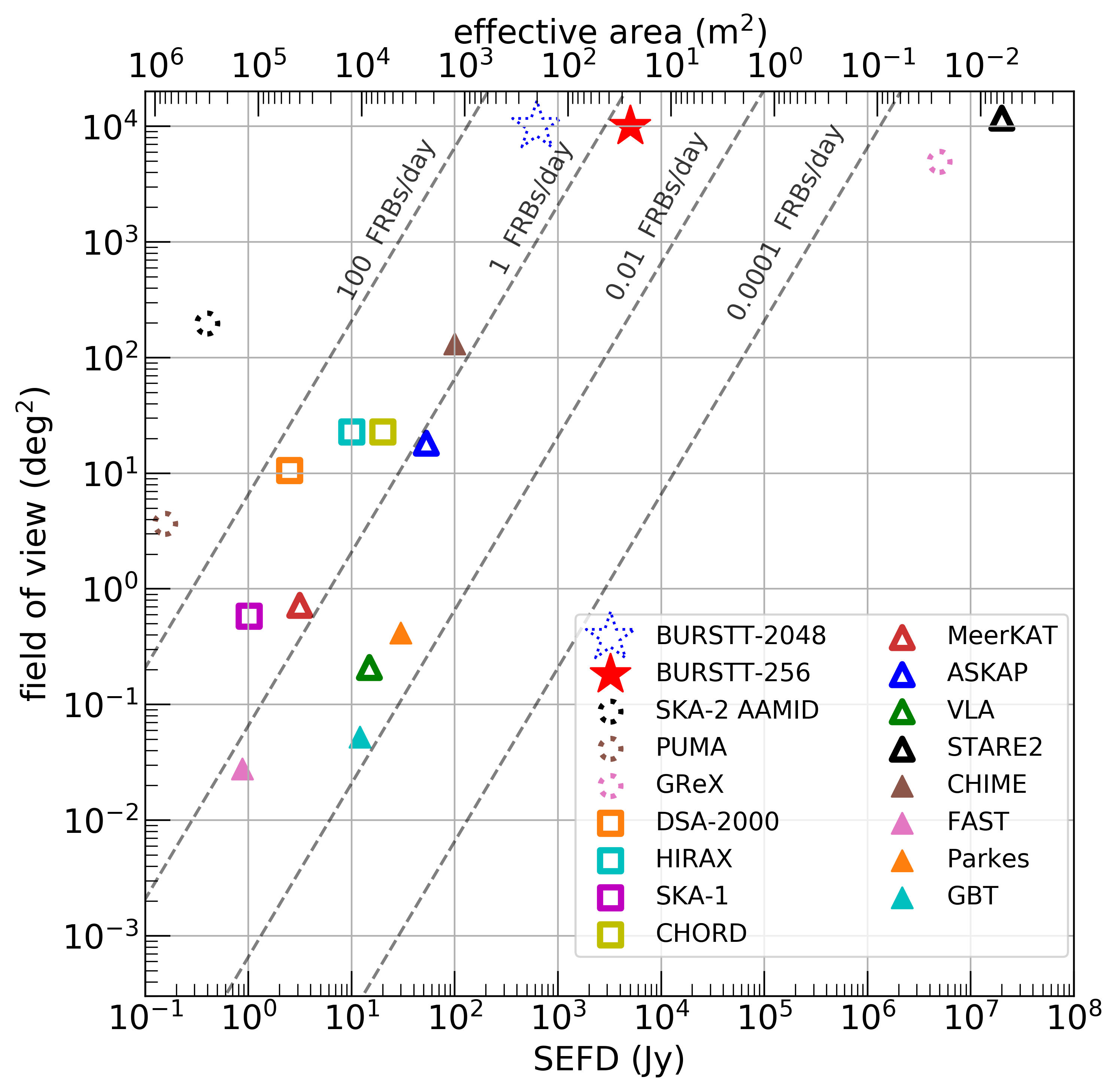}
\caption{Comparison of BURSTT’s FoV, effective collecting area, sensitivity (SEFD), and FRB detection rate (dashed lines) vs. existing (solid), planned (outline), and future-concept (dotted circle) observatories. Rates(dashed lines) were calibrated to CHIME, assuming Euclidean rates and 400 MHz bandwidth. Open triangles are sparse interferometers, which provide arcsecond localization and would require correlator upgrades to achieve these rates. The rate is a hypothetically upper limit, with the assumption of 24/7 FRB searches (which only CHIME/FRB does) as well as the optimal FRB searches with coherently beam-forming for the interferometry (ASKAP, VLA).  BURSTT is unique in the large FoV with enough sensitivity to detect a large sample of bright and nearby FRBs. 
\label{fig:telescopes_comparison}}
\end{figure}

Figure \ref{fig:telescopes_comparison} shows the FoV versus the SEFD and the effective area for the existing, planned, and future-concept FRB surveys, including CHIME \citep{2018ApJ...863...48C}, Australian Square Kilometre Array Pathfinder (ASKAP)
\citep{2010PASA...27..272M}, the Canadian Hydrogen Observatory and Radio-transient Detector (CHORD) \citep{2019clrp.2020...28V}, STARE2 \citep{2020PASP..132c4202B}, Galactic Radio Explorer (GReX) \citep{2021PASP..133g5001C}, Deep Synoptic Array 2000 (DSA-2000) \footnote{\url{ https://www.deepsynoptic.org/overview}}, Five-hundred-meter Aperture Spherical Telescope (FAST) \citep{2019SCPMA..6259502J}, Green Bank Telescope (GBT) \citep{2016MNRAS.460.1054C}, the Hydrogen Intensity and Real-time Analysis eXperiment (HIRAX) \citep{2016SPIE.9906E..5XN}, MeerKAT \citep{2021arXiv210308410R}, Parkes \citep{2014ApJ...789L..26P}, Packed Ultra-wideband Mapping Array (PUMA) \citep{2019BAAS...51g..53S}, Square Kilometre Array Phase 1 (SKA-1) \citep{2019arXiv191212699B}, SKA-2 \citep{2016arXiv161000683T}, and Very Large Array (VLA) \citep{2018ApJS..236....8L}.

\subsection{Immediate localization by BURSTT}

The critical step to reveal the origin of FRBs is to measure the accurate positions of FRBs, i.e., localization, to identify their progenitors. Out of more than 600 FRBs published to date \citep{2016PASA...33...45P, 2021ApJS..257...59C}, there has been only one successful case of localization to a stellar progenitor, which identified the FRB as a Galactic magnetar, SGR 1935+2154 \citep{2020Natur.587...54C,2020Natur.587...59B, 2021NatAs...5..414K}, and clearly indicates the magnetar as the progenitor of this particular FRB. 

Recently, three repeating FRB sources were localized in nearby star-forming regions of galaxies \citep{2017ApJ...843L...8B, 2020Natur.577..190M, 2021A&A...656L..15P}. This may indicate young stellar populations as a possible origin of repeating FRBs. In contrast, another repeating FRB source was localized in the globular cluster in M81, a nearby galaxy, suggesting FRB origins are to be found in old stellar populations \citep{2022Natur.602..585K}. This apparent contradiction could be due to the small sample size. A statistical relation between localized FRBs and their host galaxies has been studied \citep{2020ApJ...899L...6L}, while more localized samples are necessary to understand the host environments. BURSTT will detect and localize $\sim$100 bright FRB per year to further distinguish their local environment, and such statistics are important to understand the population \citep{2021AJ....161...81L, 2022AJ....163...48M, 2022AJ....163...65C}.

Very-Long Baseline Interferometry (VLBI) along with outrigger stations has been proposed to localize both non-repeating and repeating FRBs \citep{2022AJ....163...65C}. For the latter, the European VLBI Network (EVN) has localized several repeating FRBs to their hosts \citep{2020Natur.577..190M, 2022Natur.602..585K}.
Along with the VLBI outrigger stations, BURSTT can localize FRBs to their hosts. For instance, an outrigger located a few hundred km away from the main array will provide $\sim$1 arcsecond localization for the host galaxy. Such localization will lead to the identification of FRB hosts and potentially the progenitors of the nearby sources, allowing us to understand the nature of FRBs. 

\subsection{Long-term high cadence monitoring with BURSTT}

The observational classification of FRBs as repeating and non-repeating has caused a major problem over the past decade: non-repeating FRBs can be contaminated by repeating FRBs since this classification is a purely observational definition \citep{2021ApJ...906L...5A, 2022MNRAS.509.1227C}. Due to limited observational time, a significant fraction of repeating bursts may be missed and thus such FRBs could be misclassified as non-repeating FRBs. However, long-term monitoring observations are extremely expensive and difficult using current and planned radio telescopes.

BURSTT, on the other hand, will monitor a fixed large patch of the sky all the time (See Table \ref{table: BURSTT_properties}). This is essential for non-stop monitoring of repeating and non-repeating FRBs. Repeating FRBs could originate from the repeating activities of progenitors, such as pulsars and magnetars, while non-repeating FRBs may be generated by catastrophic one-off events, such as compact merger systems \citep{2019PhR...821....1P}. BURSTT will thus provide a great statistics of source repetitions as well as precise constraints on apparent non-repetition of one-off events.

BURSTT can monitor the northern hemisphere at least 7 hours per day (24 hours for North pole and $\sim$7 hours for the equator). BURSTT’s longer monitoring minimizes the chance it will miss any repeating FRBs, which will resolve the missing repeating FRB problem. \citet{2019NatAs...3..928R} compared the event rate of nearby FRB samples and those of cataclysmic progenitor events. The event rate of nearby FRBs exceeds the cataclysmic progenitor events, indicating that the FRB sample cannot be explained by only cataclysmic events. Hence, \citet{2019NatAs...3..928R} concluded that most observed cases of FRBs must originate from sources that emit several bursts over their lifetimes.

The \citet{2021ApJS..257...59C} reported 474 non-repeating FRBs and 62 repeating bursts from 18 repeaters, implying the probability that a burst originates at a repeater is equal to 62/(62+474) = 11.5\%. As exposure time increases \citep{2021ApJ...906L...5A}, we expect more repeating and bright bursts will be detected. For instance, \citet{2021ATel14556....1H} reported a high-fluence FRB (334 Jy ms) from the FRB 20201124A repeating source with observation times of 90 hours. As a result of the small number of detections, the properties of repeating FRBs are poorly constrained (e.g., \citet{2020ApJ...891L...6F}). This uncertainty includes the event rate and its dependence on flux density, the possibility of non-Poisson distribution of events \citep{2018MNRAS.475.5109O}, and the host environment of the FRB. With its all-sky, continuous sensitivity to high-fluence events, BURSTT will explore a unique parameter space for repeaters. Since BURSTT will detect $\sim$100 high-fluence FRBs from the nearby Universe, the non-detections of repeats will shed some light on the underlying population using statistical methods.

Finding repeaters requires either high sensitivity or long exposure time. In order to detect more repeaters, we take the advantage of BURSTT's long exposure time (given by the large FoV) as a compensation to the sensitivity issue. We expect that BURSTT will detect high-fluence bursts from repeaters, and other telescopes with high sensitivity can do follow-up observations to detect more fainter bursts from the sources. For instance, the bright repeaters can then be readily followed-up by larger apertures, including the Five-hundred-meter Aperture Spherical Telescope (FAST), Giant Meterwave Radio Telescope (GMRT), Very Large Array (VLA). Historically, repetition rates increase rapidly with sensitivity, and thus we expect repeaters detected by the BUTSTT to have the highest rates in follow-up campaigns.

With the high cadence, BURSTT may detect bright pulses from pulsars and rotating radio transients \citep{2021ApJ...922...43G}, and further study their connections to FRBs \citep{2021ApJ...920...38B, 2021MNRAS.508.1947T, 2021ApJ...919L...6M}. 

\subsection{An FRB telescope dedicated to the nearby Universe}\label{subsection: An FRB telescope dedicated to the nearby Universe}

Our plan to reveal the origin of FRBs is to explore the nearby Universe, because this maximizes the chance of detecting possible multi-wavelength and multi-messenger counterparts of FRBs, including FRB progenitors. For example,  follow-up was key to the association of a repeating FRB with the Galactic magnetar SGR 1935+2154 \citep{2020Natur.587...54C, 2020Natur.587...59B, 2021NatAs...5..414K}. This progenitor was identified because it is located nearby. Simultaneous X-ray emissions from SGR 1935+2154 were also detected with X-ray telescopes (e.g., \citet{2021NatAs...5..401T}). This observed association gives hope that FRBs in nearby galaxies may also have detectable  X-ray counterparts. \citep{2021NatAs...5..378L}. BURSTT will monitor a large fraction of nearby galaxies.

An important wavelength range to examine for counterparts is the optical. No optical-FRB coincidence has found \citep{2018PASJ...70..103T, 2018MNRAS.481.2479M} to date, and improved constraints would be very valuable, because the expected flux density of the optical counterpart strongly depends on the physical mechanism of the FRB, e.g., the magnetosphere model predicts optical counterparts, whereas the maser model predicts a negligible flux density in optical (e.g., \cite{2019ApJ...878...89Y, 2022MNRAS.515.5682Y}). 

BURSTT is the ideal compliment to the current generation of multi-messenger detectors such as the Laser Interferometer Gravitational-Wave Observatory (LIGO) \citep{2009RPPh...72g6901A}, IceCube \citep{2017JInst..12P3012A}, and the Rubin Observatory's Legacy Survey of Space and Time (LSST) \citep{2009arXiv0912.0201L}  which, like BURSTT, monitor the nearby Universe with near full-sky coverage. It will be instuctive to search for FRBs  associated with gravitational waves detected with the LIGO-Virgo-KAGRA Gravitational Wave Detector Network in near future \citep{2022arXiv220312038T}. Some FRB models predict an association with Gravitational-Wave  \citep{2018ApJ...860L...7W} or neutrino counterparts \citep{2020ApJ...902L..22M}. If FRBs originate from neutron star mergers, which are known to generate gravitational waves \citep{2017ApJ...848L..12A} and potentially neutrinos \citep{2017ApJ...849..153F, 2018PhRvD..98d3020K}, the expected time window of radio emission is on the millisecond time scale comparable to that of FRBs (e.g., \citet{2018PASJ...70...39Y}).  So far, no FRB has been found in association with gravitational wave sources (e.g., \cite{2016PhRvD..93l2003A, 2022arXiv220312038T}). If no significant GW-FRB associations can be detected in future despite the greatly improved detection rate of BURSTT, the neutron star merger scenario for the origin of FRBs can be strongly constrained. Scenarios involving cosmic ray acceleration (e.g., \citet{2014ApJ...797...33L, 2020ApJ...902L..22M}) can produce neutrinos that can be detected by Large Volume Neutrino detectors, such as IceCube, KM3NeT \citep{2016JPhG...43h4001A}, and Baikal-GVD \citep{2022JETP..134..399A}. Although no neutrinos from FRBs have been detected so far, the probability of finding neutrino associations significantly improves with the total number of FRB detections \citep{2018ApJ...857..117A, 2020ApJ...890..111A}. Such observations would bring clarity to the role of FRBs within the non-thermal universe and constrain the still unknown acceleration mechanisms of the highest energy cosmic rays.

In addition, the diffuse Galactic foreground is known to be pervasive, bright and dynamic, thus making it challenging to be observed properly. 
Only a few surveys have a large enough coverage to observe the low-frequency radio sky below 1 GHz 
\citep[e.g.][]{1981A&A...100..209H,1982A&AS...47....1H,2011A&A...525A.138G},including MSSS \citep{2015A&A...582A.123H}, GLEAM \citep{2015PASA...32...25W} and TGSS \citep{2017A&A...598A..78I}. It is therefore complementary to generate low-frequency high resolution sky maps using e.g, the Long Wavelength Array (LWA) \citep[e.g.][]{2018AJ....156...32E} and the Engineering Development Array 2 (EDA2) \citep[e.g.][]{2022PASA...39...17K}. With the large FoV of BURSTT below 1 GHz, it has the potential to improve the removal of the Galactic foreground contamination in many Cosmic Microwave Background radiation and Epoch of Reionization works (see e.g. \citet{2014PTEP.2014fB109I, 2021MNRAS.505.1575S}).

\section{BURSTT DESIGN}\label{section: TECHNICAL OBJECTIVES}

In this Section, we describe the initial designs of the BURSTT system with 256 antennas, which serves as a test-bed for BURSTT-2048. Our technology goal for BURSTT-256 is to optimize the system and compared component and algorithm designs.

\subsection{Overview of the BURSTT Instrument and the telescope site}

BURSTT-256 will consist of a combination one 256-antenna main station, and smaller outriggers at other sites in Taiwan and Hawaii.

We have surveyed several sites, and have permission to deploy the main station at the Fushan Botanical gardens \footnote{\url{https://fushan.tfri.gov.tw/en/index.php}} in Northern Taiwan. Our survey indicates acceptable radio frequency interference (RFI) conditions at that site. We are continuing to survey other sites, sheltered from RFI, for the outrigger stations in Taiwan and surrounding islands as well as in Hawaii. The outriggers will each employ 64 antennas, allowing VLBI source localization \citep{2021AJ....161...81L, 2022AJ....163...48M, 2022AJ....163...65C}. 

 We will process an effective bandwidth of 400 MHz selected within the 300–800 MHz analog range through direct digital polyphase filterbanks (PFB). The CHIME/FRB team has  demonstrated that processing a 400MHz bandwidth is practical \citep{2018ApJ...863...48C}. we would like to explore the FRB rate at frequencies below 400 MHz, the bottom of the CHIME band. Our design offers frequency agility. Should a shift in band improve the rate, BURSTT will be able to adapt through a simple software change. 

Due to the modular nature, BURSTT is flexible to be expanded by deploying more antennas or outrigger stations. We plan to expand the main station from 256 antennas (BURSTT-256) to 2048 antennas (BURSTT-2048), and more outrigger stations could be deployed with a longer baselines, increasing the localization precision. The main properties of the BURSTT are summarized in Table \ref{table: BURSTT_properties}.

For the BURSTT-256 project, the main station will search for bursts with fluence higher than 100 Jy ms, which is mentioned in Section \ref{subsec:Extremely wide field of view of BURSTT}. Assuming the duration of 1 ms and the bandwidth of 400 MHz, the corresponding S/N at the main station is 12. Once the main station detects a candidate burst, the data in the ring buffer in the outrigger will be copied out for the offline analysis. We expect a burst with a fluence of 100 Jy ms would yield an S/N at least of 6 in the cross-correlation analysis for the localization purpose.

We plan to use single-polarization antennas in BURSTT. These are simpler to design and build than dual-polarization antennas and optimization of the antenna-LNA match is more effective when only one polarization is involved.  We plan to start with all antennas in one polarization direction within the main station array as this enhances sensitivity for linearly polarized source like FRBs.  Later, we have the option to rotate some antennas to the orthogonal polarization if we determine that dual-polarization data is essential to test FRB source models.

Even in the initial configuration we plan to provide some polarized properties of all detected FRBs, such as rotation measure and polarization position angle swings \citep{2015Natur.528..523M} using two sub-stations with orthogonal polarization at one of the outrigger sites.

The back-end includes correlator hardware that will process the radio signals and accompanying software including fast Fourier transform (FFT) techniques and establish the capability to locate FRB events. Much of the development effort of the BURSST program lies here, in the development and testing of real-time software. BURSTT will develop an upgraded version of the real-time processing digitizer and FRB search engines employed for CHIME \citep{2022ApJS..261...29C}.

\begin{figure}[!tbp]
\epsscale{1.0}
\plotone{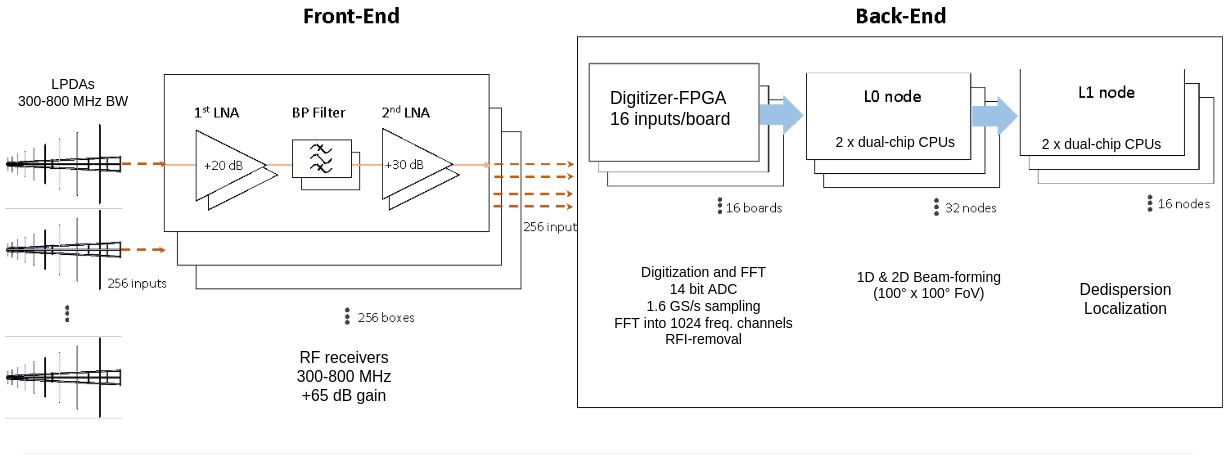}
\caption{The system diagram of the BURSTT main station. 
\label{fig:system_diagram}}
\end{figure}

An overview of the front- and back-ends of the main station is given in Figure \ref{fig:system_diagram}. The front-end receives the signal from the sky through the bandpass (BP) filter and the noise amplifiers. The back-end digitizes the signals, performs the frequency-domain channelization, forms beams on the sky, and searches for FRBs. The hardware at outrigger stations will be similar to the main station, except that no beam-forming or FRB search will be performed. Rather the outriggers just record the arriving signal streams. This is discussed further in Sections \ref{subsec:front-end} and \ref{subsec: back-end}.

\begin{deluxetable}{lcc} [!tbp]
  \tablecaption{The main properties of the BURSTT.\label{table: BURSTT_properties}}
  \tablewidth{\textwidth}
  \tablehead{
    Quantity & \multicolumn{2}{c}{Value} }
  \startdata
  Project & BURSTT-256 & BURSTT-2048\\
  SEFD & $\sim$5000 Jy &  $\sim$600 Jy\\
  Effective area & 40-200 m$^{2}$ & 320-1600 m$^{2}$ \\
  Number of antennas (main station)& 256 & 2048 \\
  $\;\;\;\;\;\;\;\;\;\;\;\;\;\;\;\;\;\;\;\;\;\;\;\;\;\;\;\;\;\;\;\;\;$(outrigger stations)      & \multicolumn{2}{c}{64}\\
  Polarization & \multicolumn{2}{c}{single} \\
  E-W FoV & \multicolumn{2}{c}{$\sim$100$^{\circ}$} \\
  N-S FoV & \multicolumn{2}{c}{$\sim$100$^{\circ}$} \\
  Daily exposure time & \multicolumn{2}{c}{24 hrs (North pole)} \\
                        & \multicolumn{2}{c}{$\sim$10 hrs (45$^\circ$)} \\
                      & \multicolumn{2}{c}{$\sim$7 hrs (Equator)} \\
  Frequency range & 300-800 MHz & TBD \\
  Bandwidth & 400 MHz & $\geq$400 MHz \\
  Number of frequency channels & 1024 & TBD \\ 
  E-W baseline & \multicolumn{2}{c}{$\sim$8000 km (Northern Taiwan to Hawaii)} \\
  N-S baseline & \multicolumn{2}{c}{$\sim$300 km (Northern to Southern Taiwan)}\\
  \enddata
\end{deluxetable}

\subsection{FRONT-END DESIGN AND DEVELOPMENT}\label{subsec:front-end}

\begin{figure}[htp]
\centering
\includegraphics[width=.3\textwidth]{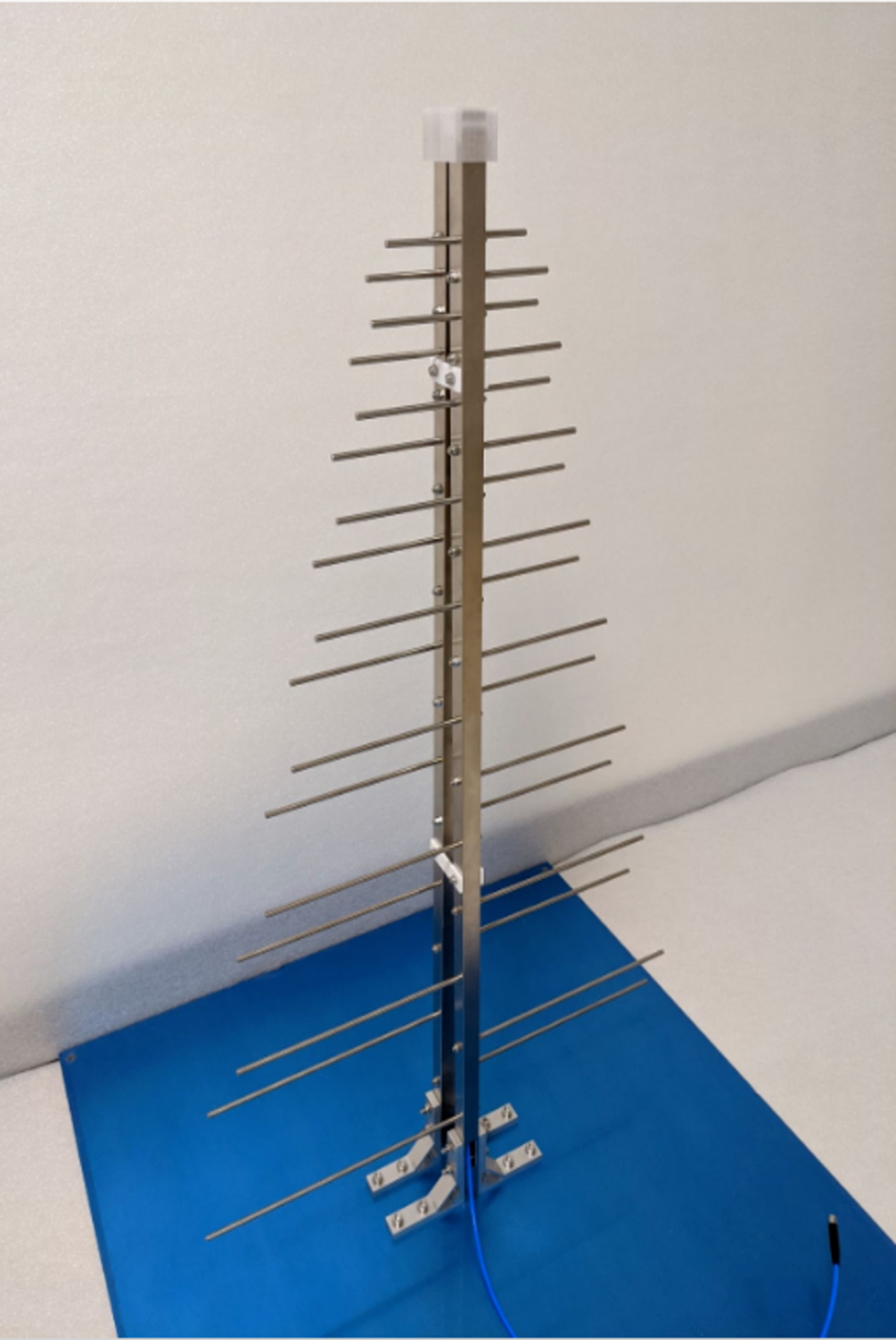}\hfill
\includegraphics[width=.3\textwidth]{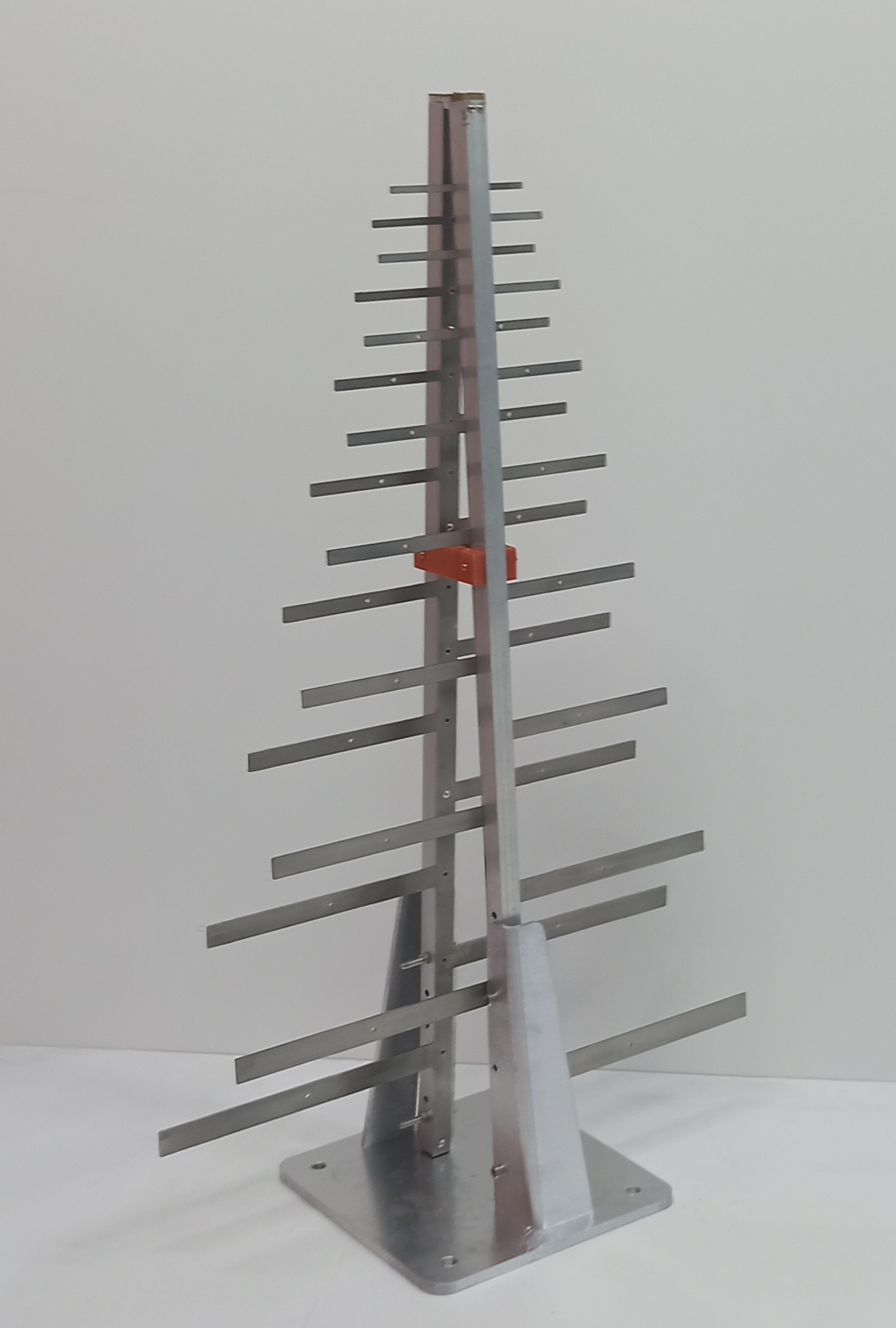}\hfill
\includegraphics[width=.3\textwidth]{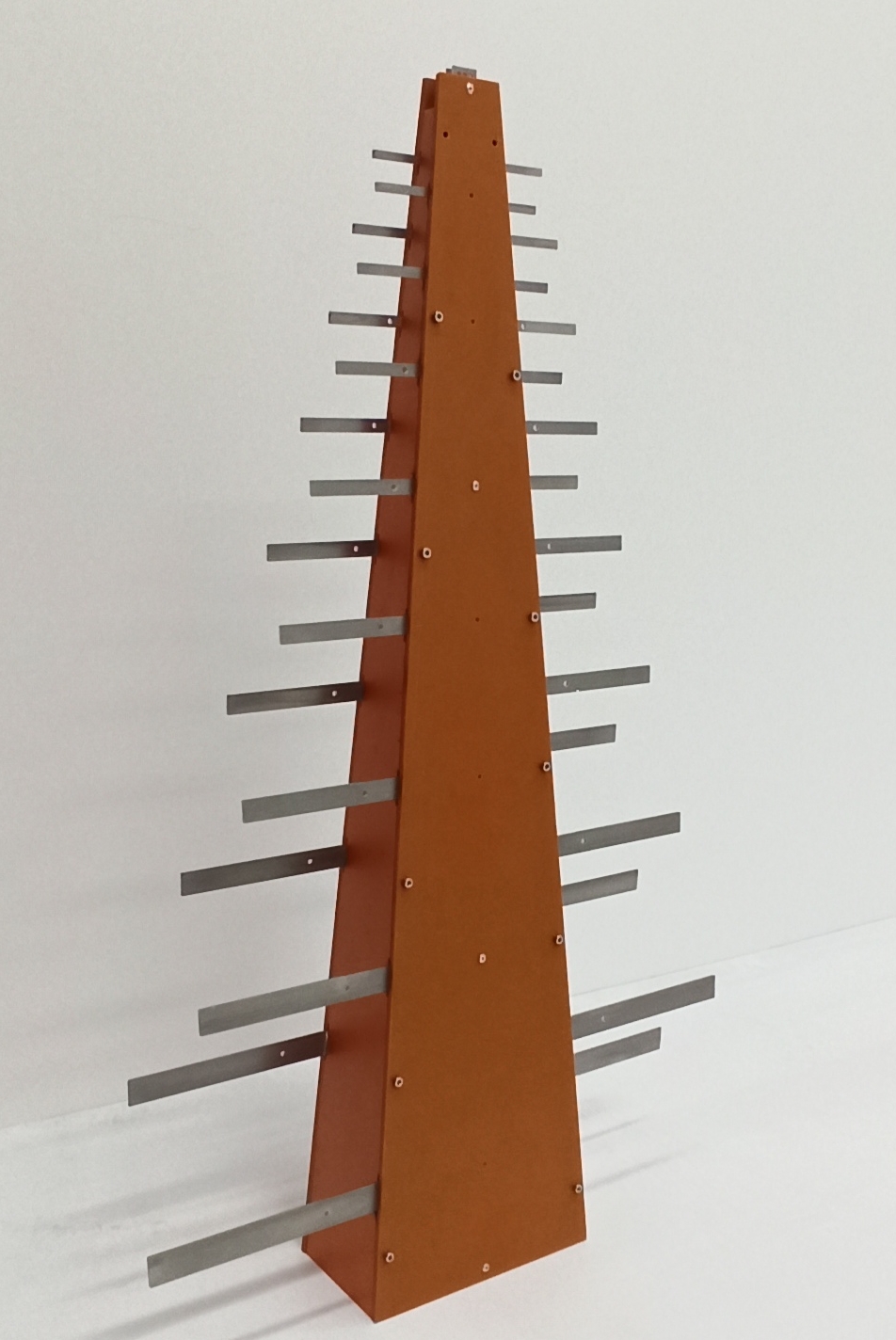}
\caption{The photo of the prototype LPDA antennas in development. Standard LPDA (left), planar LPDA (middle), and planar LPDA on a quadrangular pyramid structure (right).} 
\label{fig:LPDA}
\end{figure}

BURSTT will use Log-Periodic Dipole Array (LPDA) antennas. The LPDA antennas, which provide the required gain (7–9 dBi) with a simple structure, have been characterized and measured for use in numerous scientific and industrial applications. Millions of LPDAs have been deployed over the decades for use as home television antennas. Because LPDA antennas have a structure that can withstand strong winds, they are suitable for the environment in Taiwan, which experiences typhoons. Figure \ref{fig:LPDA} shows the photo of various prototype antennas currently in development. The first one shown on the left is a standard design, which has been characterized well, but it is expensive to manufacture, and the long elements for low frequencies would be weak against strong winds. The second shown in the middle is the planar LPDA, which is a cost-effective design that simplifies the manufacturing process. However, since the elements have a long band-like structure, they can easily vibrate in the wind. The third design shown on the right is the planar LPDA attached to a quadrangular pyramid structure composed of insulator plates, which is cost-effective and raises the structure resonance frequency of the elements so that they can withstand strong winds. However, the structure can be heated by the sun, which can degrade the amplifier noise temperature. We will continue development and will test several prototypes in the field.

\begin{figure}[!tbp]
\epsscale{0.25}
\plotone{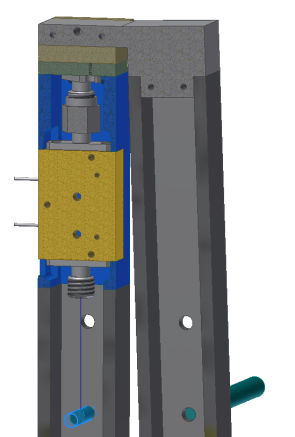}
\caption{First-stage LNA (yellow) to be integrated into the LPDA: The feed point of the LDPA is directly connected to the input port of the first-stage LNA, the amplified signal is then transmitted to the second-stage FEE module by coaxial cable inside the support rod of the LPDA.\label{fig:LNA_chassis}}
\end{figure}

\begin{figure}[htp]
\centering
\includegraphics[width=.45\textwidth]{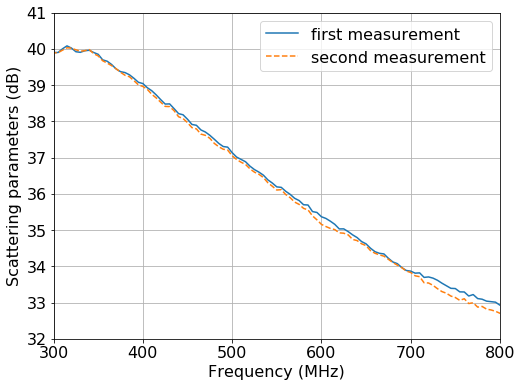}\hfill
\includegraphics[width=.45\textwidth]{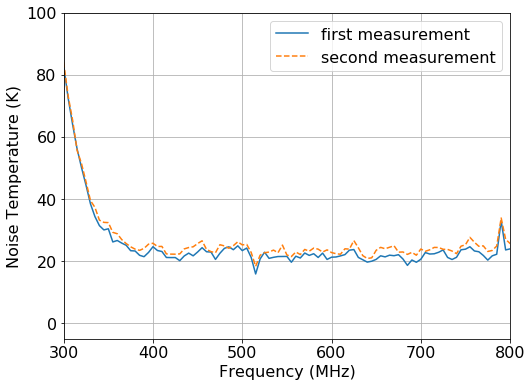}
\caption{Measured performance of the two samples of the first-stage LNA, (left) measured forward gain with the output port connected with an equalizer, additional equalizer will be installed in the final system to provide +-0.5dB gain flatness over 350 – 800 MHz, (right) measured noise temperature.} 
\label{fig:LNA_measurement}
\end{figure}

Received signals pass through the front-end electronics (FEE) before arriving at the digitizer.  We will use a sampling rate of 1600 MHz, so our passband lies in the first Nyquist zone. This means no mixers or local oscillators are needed in the RF system. In order to minimize noise temperature due to transmission line losses, the first-stage low-noise amplifier (LNA) is closely integrated into the feed point of the LDPA as shown in Figure \ref{fig:LNA_chassis}.  The external second-stage FEE module consists of two equalizers cascaded with a second-stage amplifiers (LNAs) and filter modules to eliminate RFI at low frequencies and to prevent aliased detection of signals and noise above 800 MHz. Figure \ref{fig:LNA_measurement} shows the measured forward gain and the noise temperature (20-30 K) of a first-stage LNA we are developing.  Comparing this to typical sky brightness temperature values, which fall with frequency from $\sim$50 K at 300 MHz to $\sim$6 K at 800 MHz, we see that the further effort to reduce LNA noise temperature can improve sensitivity, particularly at the high end of our frequency range.
The required system gain is 65 dB, which is obtained by combining the first-stage LNA gain (25-30 dB) the second-stage amplifier assembly (35-40 dB) and slope-equalizer losses (-5 dB). We will continue to optimize the LNA design and plan to install an in-house developed low-noise amplifier in the 2048-antenna array.

\begin{figure}[!tbp]
\epsscale{0.8}
\plotone{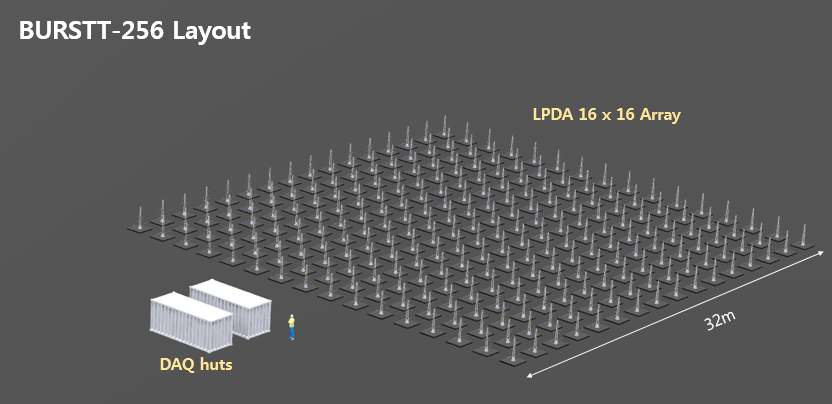}
\caption{BURSTT 256-antenna array station layout. 
\label{fig:256_layout}}
\end{figure}

The antenna array of BURSTT-256 station has a footprint of 32 m $\times$ 32 m, and the distance to the data acquisition (DAQ) hut where the digitizer is located is up to 50m, and which requires cable lengths $\sim$60 m (Figure \ref{fig:256_layout}). When using the most popular and reliable coaxial cable, LMR-400, the transfer losses are -4 dB and -8 dB for 200 MHz and 900 MHz respectively, which can be corrected in the RF slope equalizer or in the correlator after channelization.

\subsubsection{The beam pattern}

\begin{figure}[htp]
\centering
\includegraphics[width=.45\textwidth]{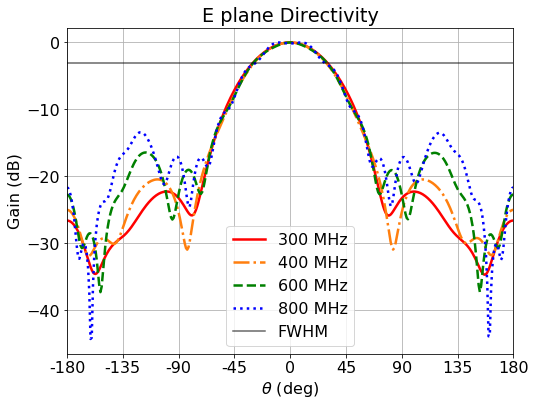}\hfill
\includegraphics[width=.45\textwidth]{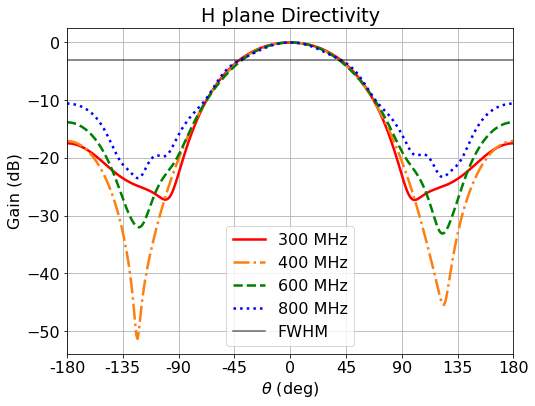}
\caption{Normalized directivity of the prototype antenna simulated using HFSS, (left) the E plane directivity and (right) H plane directivity.} 
\label{fig:beam_pattern}
\end{figure}

Figure \ref{fig:beam_pattern} shows simulated far field radiation patterns of the LPDA antenna calculated using (i.e. the Ansys high-frequency structure simulator (HFSS) software).  With the HFSS software, we utilize the finite element method to compute the EM field distribution of components. The 3 dB beam widths (FWHM) of the prototype LPDA antenna are around 80 degrees on the H plane and 60 degrees on the E plane from 300 MHz to 800 MHz (Table \ref{table: FWHM_beam}).

\begin{deluxetable}{lcccc} [!tbp]
  \tablecaption{The FWHM of the beam pattern. \label{table: FWHM_beam}}
  \tablewidth{\textwidth}
  \tablehead{
     & 300 MHz & 400 MHz & 600 MHz & 800 MHz }
  \startdata
  E plane & 58.0$^{\circ}$ & 56.0$^{\circ}$ & 58.0$^{\circ}$ & 54.0$^{\circ}$\\
  H plane & 80.0$^{\circ}$ & 78.0$^{\circ}$ & 77.0$^{\circ}$ & 80.0$^{\circ}$\\ 
  \enddata
\end{deluxetable} 

\subsection{BACK-END DESIGN AND DEVELOPMENT }\label{subsec: back-end}

When the BURSTT main station detects a candidate FRB it sends a SAVE message to the outriggers, which then transfer recently-buffered data from RAM to disk. The outriggers and main station transmit the data to a VLBI processing facility for localization processing. This process synthesizes an instrument with a nominal VLBI resolution $\lambda$/(2D) of a 7,800 km $\times$ 270 km baselines with 13 mas and 370 mas resolution in E-W and N-S respectively, allowing not only unique identification of each FRB's host galaxy, but also $\sim 100$ parsec-scale localization of the FRB within that galaxy. 

\subsubsection{Digitizer and Engines} \label{subsubsec: digitizer}

\begin{figure}[!tbp]
\epsscale{1.0}
\plotone{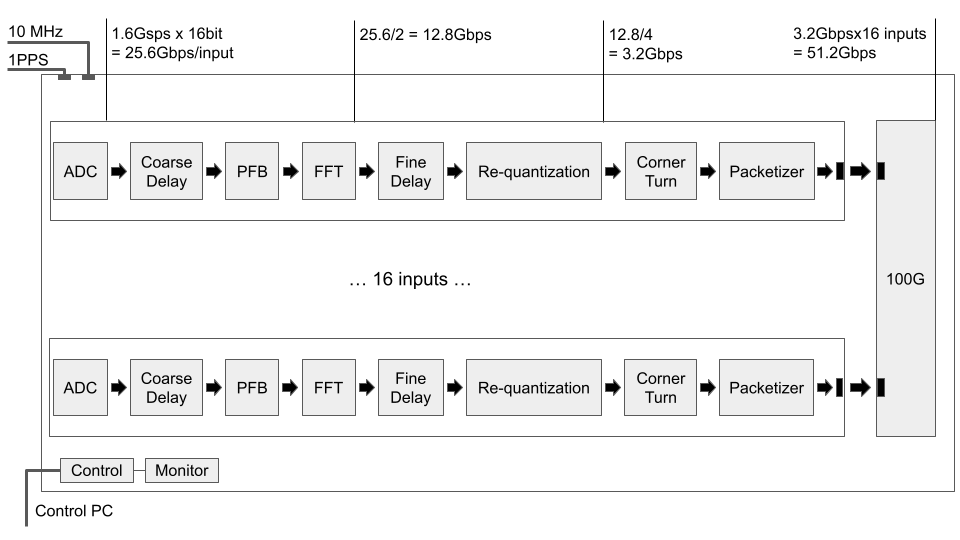}
\caption{The functional block diagram of the BURSTT F-engine. Each FPGA-based RFSoC board will process 16 inputs with an initial bandwidth of 0--800 MHz. After channelization, a choice of 400 MHz bandwidth is retained. The re-quantization block further reduces the data rate by a factor of 4. Each board will send out a total of 51.2 Gbps of data over the 100G ethernet interface to a GPU processing node.
\label{fig:FPGA_diagram}}
\end{figure}

We are employing the Xilinx ZCU216 field programmable gate array (FPGA) boards as the platform of our digitizer and channelizer. The ZCU216 is equipped with a radio frequency system on chip (RFSoC) FPGA that can process 16 RF inputs.  RFSoC systems include an embedded analog to digital converter (ADC) inside the FPGA chip, which simplifies wiring and saves physical space. This is the highest bandwidth RFSoC currently available.   Figure \ref{fig:FPGA_diagram} summarizes the functional block diagram of the RFSoC F-engine. We will be clocking the ADC at sampling rate 1600MHz for a Nyquist bandwidth of 800MHz.  A poly-phase filter bank will channelize the data, forming 2048 bands of width $\sim$390 kHz, and we will select 1024 of these for further analysis, choosing RFI-clear bands. We then re-quantized to resolution 4+4i bits to reduce the data rate. Gathering all the 16 input frequency domain data streams, a 100G Ethernet system will transmit the 100G UDP packets to the server for further data processing. The ADCs have 14-bit resolution substantially higher than the 8 bits sampling commonly used to date.

\subsubsection{Preprocessing and RFI Mitigation}

We discuss the steps of preprocessing in this section, including the RFI-removal techniques, beam-forming, and frequency channelization.

For BURSTT-256, the antennas will be organized into 16 columns, each with 16 antennas. The 16 antennas in a N-S column will be connected to one FPGA F-engine. It is thus efficient to perform 1D beamforming within the FPGA  \cite{2017ursi.confE...4N}. Each beam from the 16 F-engines is then similarly combined in the CPU-array to form the E-W direction beams, This process occurs independently for each frequency channel. This approach is similar in spirit to the 2D-FFT telescope \citep{2009PhRvD..79h3530T} although with an implementation that is tailored to our F-engine design.
For BURSTT-2048 a second layer of 2d spatial FFT will be carried out in the CPU array to synthesize all full-resolution beams.

The bandpass calibration and the system SEFD will be derived from observations of the Sun during the day. A pulsed broadband noise source will be used to monitor the gain variation during the night. 
Phase calibration will be achieved by using the many radio sources continuously available within the large FoV.

RFI Surveys in the Fushan candidate site shows that about 60\% of the 300--800~MHz frequency range is free of noticeable RFI, while an additional 15\% is affected by weak RFI. The FPGA F-engine offers a possibility of mitigating the  weak  RFI contamination using the  using a spatial filtering technique applied to time streams from the antennas. \citep{2010AJ....140.2086K}.
The basic idea is that for a spatially fixed RFI source, the cross-correlation between any two antennas will record a distinct phase relation. The covariance matrix of all the antennas contains all the phase information one can measure with the array. Any source, not just the RFI sources, will contribute to the covariance matrix. However, for signals that are above the thermal noise, one can identify them by solving the eigenvalues and eigenvectors of the covariance matrix. This procedure is performed on each spectral channel independently. The antennas voltage data can be projected onto the eigenvectors to form the eigenmodes. The eigenmodes with the strongest variances (i.e. the largest eigenvalues) are assumed to be from the RFI source. It is therefore possible to remove the RFI contribution by nulling or zeroing out the RFI eigenmodes and deprojecting them back to the antenna data space. The antenna data after the RFI removal are passed onto the beamforming stage. 

\begin{figure}[!tbp]
\epsscale{1.0}
\plotone{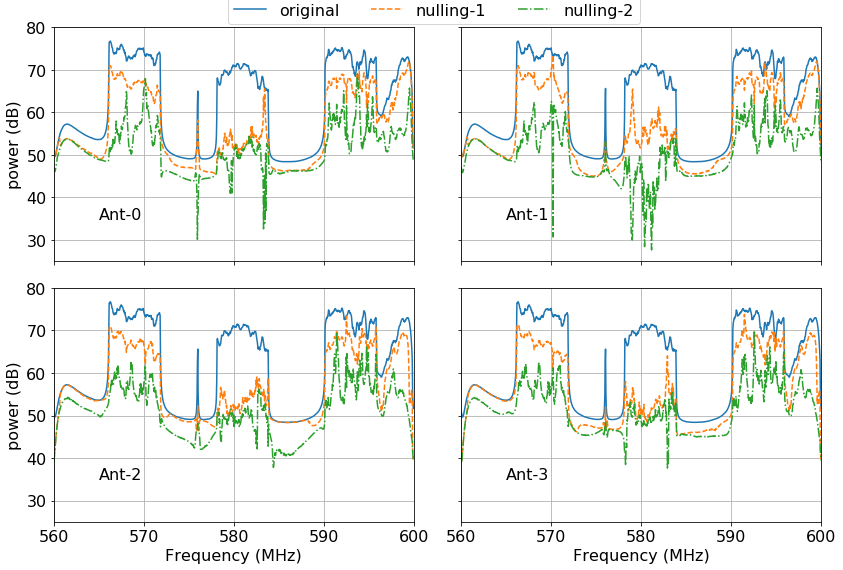}
\caption{The RFI removal testing at the Fushan site. The four panels correspond to the four antennas. The blue solid line shows the originally received power as a function of observing frequency. The orange dotted line shows the received power after nulling of the strongest eigenmode. The green dashed line corresponds to the received power after nulling of strongest two eigenmodes.
\label{fig:RFI_removal}}
\end{figure}

Figure \ref{fig:RFI_removal} is an example of this technique applied to the TV station signal that was recorded in Fushan Botanical Garden (the main station site) with a 4-antenna test system. Three TV stations, each with a 6 MHz bandwidth, are within the recorded 40 MHz bandwidth. We note here that the eigenmode decomposition works better when the variance of the antennas are comparable. For this example, the four antennas have been normalized to the mean of the four antennas at each spectral channel. It is shown that for the TV signal centered on 581MHz, the signal is reduced by about 20 dB after nulling out the strongest eigenmode. The TV signal centered on 569 MHz is only reduced by a similar amount after nulling two of the strongest eigenmodes. It is possible that at Fushan, we are receiving the TV signals from multiple towers or that multi-path propagation is an important factor. As the project expands the testing to more sites and with more antennas, the RFI removal technique will be refined.

The performance of the RFI mitigation improves as the number of antennas is increased. Using the entire array in principle gives the best RFI nulling, but at substantial increase in the  cost of the computation. To keep this system simple, we plan to update the RFI nulling solutions just once several hours or once a day. For the 16 antennas connected to one F-engine, applying the nulling solution is accomplished simply by multiplying the 16 antenna time streams for each frequency channel by a $16\times16$ complex matrix, similar to the beamforming operation. In fact, the two operations can be combined. Therefore, the RFI mitigation can be applied for each column of antennas at no additional real-time computational cost. Additional test will determine if 16-antenna RFI mitigation is sufficient or if more antennas should be combined to achieve the desired improvement.

\subsubsection{Dedispersion, search, and follow-up}

Searching for FRBs with unknown dispersion measure (DM) presents a challenge. This search has high computational cost and for BURSTT the search must be carried out real-time \citep{2019A&ARv..27....4P}, since the raw data rate is too high for disk storage. Search algorithms such as the Fast Dispersion Measure Transform (FDMT) \citep{2017ApJ...835...11Z} or the tree-dedispersion algorithm \citep{1974A&AS...15..367T, 2015Natur.528..523M, 2018ApJ...863...48C} convert the intensity versus frequency and time into intensity versus DM and time data, along with the S/N estimates. Peaks in this data with S/N higher than 10 are initial FRB candidates. These dedipsersion algroithms are well tested: the tree-dedispersion algorithm in real-time has been used by CHIME/FRB \citep{2018ApJ...863...48C}, the upgraded Molonglo Observatory Synthesis Telescope (UTMOST) \citep{2019MNRAS.488.2989F}, the Commensal Radio
Astronomy FAST Survey (CRAFTS) \citep{2021ApJ...909L...8N}, and  Deep Synoptic Array 10 (DSA-10) \citep{2019Natur.572..352R}, and the FDMT algorithm in real-time has been implemented in ASKAP \citep{2017ApJ...841L..12B}. 

Since the DM for the majority of FRBs is less than 1,000 pc cm$^{-3}$ \citep{2021ApJS..255....5C} and the BURSTT will be searching for nearby FRBs, we will initially set the  BURSTT DM range to 0 to 1,000 pc cm$^{-3}$. 

In  addition to transmitting automated SAVE messages to out outriggers, BURSTT will also send out real-time alert of the FRB events with an initial localization shortly after the detection to the community \citep{2017arXiv171008155P, 2018ApJ...863...48C}, to allow rapid or retrospective multi-frequency and multi-messenger follow-up observations. 

\section{DISCUSSION and SUMMARY}\label{section: SUMMARY}

Over 600 FRBs have been published since the first report in 2007 \citep{2007Sci...318..777L, 2019A&ARv..27....4P, 2021ApJS..257...59C}, while the complex phenomena cannot be totally explained by theories \citep{2018NatAs...2..842P, 2019PhR...821....1P}. The presence (or lack) of counterparts in other bands will provide critical constraints on both the emission mechanism and progenitor systems of FRBs \citep{2016MNRAS.462L..16P, 2019ApJ...879...40C}. Candidate models such as compact object mergers and young magnetar flares yield very different predictions for multi-wavelength counterparts \citep{2013PASJ...65L..12T, 2013arXiv1307.4924P, 2018NatAs...2..842P} yet progress has been slow due to a lack of precise and rapid FRB localizations to date. 

BURSTT with a large FoV of $\sim$10$^{4}$ deg$^{2}$ will monitor the whole visible sky for detecting and localizing $\sim$100 bright and nearby FRBs per year. First, the large FoV yields a high cadence monitoring of FRBs, which is crucial to statistically understanding the repeater and the apparent one-off events. Second, the detection and localization of nearby FRBs offers the best chance for counterpart identification, which allows for timely follow-up observations at other wavelengths, including X-ray, infrared, etc. A large sample of $\sim$100 bright FRBs per year will bring unprecedented new evidence about the nature of FRBs. 

Understanding the physical processes through nearby and bright FRBs opens the possibility of further cosmological applications. Determining FRB host galaxies, and hence distances, moreover has the potential to enable large scale structure studies and unique determinations of the content of the intergalactic medium (IGM). This includes constraints on the location of the “missing baryons” in the Universe, whether in the IGM property, or in the halos of galaxies — the  circumgalactic medium (e.g., \cite{2014ApJ...780L..33M, 2019ApJ...872...88R, 2018PhRvD..98j3518M}). Some studies suggest that the structures of galaxy halos can be uniquely constrained by FRBs \citep{2019MNRAS.485..648P, 2021ApJ...911..102O}, for comparison with models of galaxy evolution. Faraday rotation of polarized FRBs could constrain the origin and evolution of cosmic magnetism \citep{2016ApJ...824..105A}. Uncertainties in FRBs’ dispersion measures have been proposed to constrain Einstein’s weak equivalence principle \citep{2021PhRvD.104l4026H}.

The BURSTT design is scalable. Initially, 256 antennas will be deployed, which will help us to understand and improve the system. Because of its modular design, the telescope can be expanded simply by adding more main-station antennas, outriggers and correlation processors, with a corresponding increase in collecting area, localization precision, and sensitivity. In the BURSTT design we have eliminated the dishes, cylindrical reflectors, or RF-summed phased-arrays, used in other radio-telescopes. The BURSTT design therefore concentrates cost in the real-time signal-processing hardware. Most of this hardware is Commercial-Off-The-Shelf equipment, and is used for a wide range of general computing applications. Computing costs have fallen with Moore's law for decades and continue to fall. BURSTT is therefore a technology pathfinder in an area that can be expected to grow both for scientific and financial reason. If computing costs continue to fall there is no reason to stop the expansion of BURSTT at 2048 elements, since further increases in collecting area allow an increase the volume and the redshift range of the survey.

Finally, in addition to deciphering the nature of the mysterious FRBs, BURSTT will open up a vast new discovery space for non-FRB surveys due to its unique all-sky collecting area. BURSTT can contribute to the study of self-triggering and reconstruction of cosmic-ray induced extensive air showers \citep{2017PrPNP..93....1S}, as well as other extreme energy phenomena, such as Ultra-High Energy Cosmic Rays \citep{2022arXiv220308096A}, and Ultra-High Energy neutrinos \citep{2022arXiv220505845C}. Moreover, BURSTT could also potentially contribute to LIGO counterparts \citep{2022arXiv220312038T}, Extreme Scattering Events (ESEs) \citep{2018MNRAS.474.4637K}, 21cm absorbers \citep{2016MNRAS.460L..25M}, pulsar searches \citep{2021ApJS..255....5C}, interplanetary scintillation \citep{1964Natur.203.1214H}, etc. BURSTT will be a promising frontier of radio astronomy in the foreseeable future. 

\section{acknowledgments}
The BURSTT project is supported by the National Science and Technology Council (NSTC) of Taiwan through Science Vanguard Research Program 111-2123-M-001-008-. U.L.P. receives support from Ontario Research Fund—research Excellence Program (ORF-RE), Natural Sciences and Engineering Research Council of Canada (NSERC) [funding reference number RGPIN-2019-067, CRD 523638-18, 555585-20], Canadian Institute for Advanced Research (CIFAR), Canadian Foundation for Innovation (CFI), the National Science Foundation of China (Grants No. 11929301), Thoth Technology Inc, Alexander von Humboldt Foundation, and the National Science and Technology Council (NSTC) of Taiwan (111-2123-M-001 -008 -, and 111-2811-M-001 -040 -). Computations were performed on the SOSCIP Consortium’s [Blue Gene/Q, Cloud Data Analytics, Agile and/or Large Memory System] computing platform(s). SOSCIP is funded by the Federal Economic Development Agency of Southern Ontario, the Province of Ontario, IBM Canada Ltd., Ontario Centres of Excellence, Mitacs and 15 Ontario academic member institutions. P.C. acknowledges the support by the National Science and Technology Council (NSTC) of Taiwan through the grant 111-2112-M-002-029- and by Leung Center for Cosmology and Particle Astrophysics (LeCosPA), National Taiwan University. TG acknowledges the supports by the National Science and Technology Council of Taiwan through grants 108-2628-M-007-004-MY3 and 111-2112-M-007 -021. T.H. acknowledges the support of the National Science and Technology Council of Taiwan through grants 110-2112-M-005-013-MY3 and 110-2112-M-007-034-. Y.C. acknowledges the support of the National Science and Technology Council of Taiwan through grant NSTC 111-2112-M-001-090-MY3. CPH acknowledges support from the National Science and Technology Council of Taiwan through grant MOST 109-2112-M-018-009-MY3. A.Y.L.O. is supported by the National Science and Technology Council (NSTC) of Taiwan (ROC) through the grants 111-2811-M-007-009 (PI: Prof. Ray-Kuang Lee, NTHU) and 110-2112-M-005-013-MY3 (PI: Prof. Tetsuya Hashimoto, NCHU).

\bibliography{proofread}

\begin{thebibliography}{}
\expandafter\ifx\csname natexlab\endcsname\relax\def\natexlab#1{#1}\fi
\providecommand{\url}[1]{\href{#1}{#1}}
\providecommand{\dodoi}[1]{doi:~\href{http://doi.org/#1}{\nolinkurl{#1}}}
\providecommand{\doeprint}[1]{\href{http://ascl.net/#1}{\nolinkurl{http://ascl.net/#1}}}
\providecommand{\doarXiv}[1]{\href{https://arxiv.org/abs/#1}{\nolinkurl{https://arxiv.org/abs/#1}}}

\bibitem[{{Aartsen} {et~al.}(2017){Aartsen}, {Ackermann}, {Adams}, {Aguilar},
  {Ahlers}, {Ahrens}, {Altmann}, {Andeen}, {Anderson}, {Ansseau}, {Anton},
  {Archinger}, {Arg{\"u}elles}, {Auer}, {Auffenberg}, {Axani}, {Baccus}, {Bai},
  {Barnet}, {Barwick}, {Baum}, {Bay}, {Beattie}, {Beatty}, {Becker Tjus},
  {Becker}, {Bendfelt}, {BenZvi}, {Berley}, {Bernardini}, {Bernhard}, {Besson},
  {Binder}, {Bindig}, {Bissok}, {Blaufuss}, {Blot}, {Boersma}, {Bohm},
  {B{\"o}rner}, {Bos}, {Bose}, {B{\"o}ser}, {Botner}, {Bouchta}, {Braun},
  {Brayeur}, {Bretz}, {Bron}, {Burgman}, {Burreson}, {Carver}, {Casier},
  {Cheung}, {Chirkin}, {Christov}, {Clark}, {Classen}, {Coenders}, {Collin},
  {Conrad}, {Cowen}, {Cross}, {Day}, {Day}, {de Andr{\'e}}, {De Clercq}, {del
  Pino Rosendo}, {Dembinski}, {De Ridder}, {Descamps}, {Desiati}, {de Vries},
  {de Wasseige}, {de With}, {DeYoung}, {D{\'\i}az-V{\'e}lez}, {di Lorenzo},
  {Dujmovic}, {Dumm}, {Dunkman}, {Eberhardt}, {Edwards}, {Ehrhardt},
  {Eichmann}, {Eller}, {Euler}, {Evenson}, {Fahey}, {Fazely}, {Feintzeig},
  {Felde}, {Filimonov}, {Finley}, {Flis}, {F{\"o}sig}, {Franckowiak},
  {Fr{\`e}re}, {Friedman}, {Fuchs}, {Gaisser}, {Gallagher}, {Gerhardt},
  {Ghorbani}, {Giang}, {Gladstone}, {Glauch}, {Glowacki}, {Gl{\"u}senkamp},
  {Goldschmidt}, {Gonzalez}, {Grant}, {Griffith}, {Gustafsson}, {Haack},
  {Hallgren}, {Halzen}, {Hansen}, {Hansmann}, {Hanson}, {Haugen}, {Hebecker},
  {Heereman}, {Helbing}, {Hellauer}, {Heller}, {Hickford}, {Hignight}, {Hill},
  {Hoffman}, {Hoffmann}, {Hoshina}, {Huang}, {Huber}, {Hulth}, {Hultqvist},
  {In}, {Inaba}, {Ishihara}, {Jacobi}, {Jacobsen}, {Japaridze}, {Jeong},
  {Jero}, {Jones}, {Jones}, {Joseph}, {Kang}, {Kappes}, {Karg}, {Karle},
  {Katz}, {Kauer}, {Keivani}, {Kelley}, {Kemp}, {Kheirandish}, {Kim}, {Kim},
  {Kintscher}, {Kiryluk}, {Kitamura}, {Kittler}, {Klein}, {Kleinfelder},
  {Kleist}, {Kohnen}, {Koirala}, {Kolanoski}, {Konietz}, {K{\"o}pke}, {Kopper},
  {Kopper}, {Koskinen}, {Kowalski}, {Krasberg}, {Krings}, {Kroll},
  {Kr{\"u}ckl}, {Kr{\"u}ger}, {Kunnen}, {Kunwar}, {Kurahashi}, {Kuwabara},
  {Labare}, {Laihem}, {Landsman}, {Lanfranchi}, {Larson}, {Lauber}, {Laundrie},
  {Lennarz}, {Leich}, {Lesiak-Bzdak}, {Leuermann}, {Lu}, {Ludwig},
  {L{\"u}nemann}, {Mackenzie}, {Madsen}, {Maggi}, {Mahn}, {Mancina},
  {Mandelartz}, {Maruyama}, {Mase}, {Matis}, {Maunu}, {McNally}, {McParland},
  {Meade}, {Meagher}, {Medici}, {Meier}, {Meli}, {Menne}, {Merino}, {Meures},
  {Miarecki}, {Minor}, {Montaruli}, {Moulai}, {Murray}, {Nahnhauer}, {Naumann},
  {Neer}, {Newcomb}, {Niederhausen}, {Nowicki}, {Nygren}, {Obertacke Pollmann},
  {Olivas}, {O'Murchadha}, {Palczewski}, {Pandya}, {Pankova}, {Patton},
  {Peiffer}, {Penek}, {Pepper}, {P{\'e}rez de los Heros}, {Pettersen},
  {Pieloth}, {Pinat}, {Price}, {Przybylski}, {Quinnan}, {Raab}, {R{\"a}del},
  {Rameez}, {Rawlins}, {Reimann}, {Relethford}, {Relich}, {Resconi}, {Rhode},
  {Richman}, {Riedel}, {Robertson}, {Rongen}, {Roucelle}, {Rott}, {Ruhe},
  {Ryckbosch}, {Rysewyk}, {Sabbatini}, {Sanchez Herrera}, {Sandrock},
  {Sandroos}, {Sandstrom}, {Sarkar}, {Satalecka}, {Schlunder}, {Schmidt},
  {Schoenen}, {Sch{\"o}neberg}, {Schukraft}, {Schumacher}, {Seckel},
  {Seunarine}, {Solarz}, {Soldin}, {Song}, {Spiczak}, {Spiering}, {Stanev},
  {Stasik}, {Stettner}, {Steuer}, {Stezelberger}, {Stokstad}, {St{\"o}{\ss}l},
  {Str{\"o}m}, {Strotjohann}, {Sulanke}, {Sullivan}, {Sutherland}, {Taavola},
  {Taboada}, {Tatar}, {Tenholt}, {Ter-Antonyan}, {Terliuk}, {Te{\v{s}}i{\'c}},
  {Thollander}, {Tilav}, {Toale}, {Tobin}, {Toscano}, {Tosi}, {Tselengidou},
  {Turcati}, {Unger}, {Usner}, {Vandenbroucke}, {van Eijndhoven}, {Vanheule},
  {van Rossem}, {van Santen}, {Vehring}, {Voge}, {Vogel}, {Vraeghe}, {Wahl},
  {Walck}, {Wallace}, {Wallraff}, {Wandkowsky}, {Weaver}, {Weiss}, {Wendt},
  {Westerhoff}, {Wharton}, {Whelan}, {Wickmann}, {Wiebe}, {Wiebusch}, {Wille},
  {Williams}, {Wills}, {Wisniewski}, {Wolf}, {Wood}, {Woolsey}, {Woschnagg},
  {Xu}, {Xu}, {Xu}, {Yanez}, {Yodh}, {Yoshida}, \&
  {Zoll}}]{2017JInst..12P3012A}
{Aartsen}, M.~G., {Ackermann}, M., {Adams}, J., {et~al.} 2017, Journal of
  Instrumentation, 12, P03012, \dodoi{10.1088/1748-0221/12/03/P03012}

\bibitem[{{Aartsen} {et~al.}(2018){Aartsen}, {Ackermann}, {Adams}, {Aguilar},
  {Ahlers}, {Ahrens}, {Samarai}, {Altmann}, {Andeen}, {Anderson}, {Ansseau},
  {Anton}, {Arg{\"u}elles}, {Auffenberg}, {Axani}, {Bagherpour}, {Bai},
  {Barron}, {Barwick}, {Baum}, {Bay}, {Beatty}, {Becker Tjus}, {Becker},
  {BenZvi}, {Berley}, {Bernardini}, {Besson}, {Binder}, {Bindig}, {Blaufuss},
  {Blot}, {Bohm}, {B{\"o}rner}, {Bos}, {B{\"o}ser}, {Botner}, {Bourbeau},
  {Bourbeau}, {Bradascio}, {Braun}, {Brenzke}, {Bretz}, {Bron},
  {Brostean-Kaiser}, {Burgman}, {Busse}, {Carver}, {Cheung}, {Chirkin},
  {Christov}, {Clark}, {Classen}, {Collin}, {Conrad}, {Coppin}, {Correa},
  {Cowen}, {Cross}, {Dave}, {Day}, {de Andr{\'e}}, {De Clercq}, {DeLaunay},
  {Dembinski}, {De Ridder}, {Desiati}, {de Vries}, {de Wasseige}, {de With},
  {DeYoung}, {D{\'\i}az-V{\'e}lez}, {di Lorenzo}, {Dujmovic}, {Dumm},
  {Dunkman}, {Dvorak}, {Eberhardt}, {Ehrhardt}, {Eichmann}, {Eller}, {Evenson},
  {Fahey}, {Fazely}, {Felde}, {Filimonov}, {Finley}, {Flis}, {Franckowiak},
  {Friedman}, {Fritz}, {Gaisser}, {Gallagher}, {Gerhardt}, {Ghorbani}, {Giang},
  {Glauch}, {Gl{\"u}senkamp}, {Goldschmidt}, {Gonzalez}, {Grant}, {Griffith},
  {Haack}, {Hallgren}, {Halzen}, {Hanson}, {Hebecker}, {Heereman}, {Helbing},
  {Hellauer}, {Hickford}, {Hignight}, {Hill}, {Hoffman}, {Hoffmann}, {Hoinka},
  {Hokanson-Fasig}, {Hoshina}, {Huang}, {Huber}, {Hultqvist}, {H{\"u}nnefeld},
  {Hussain}, {In}, {Iovine}, {Ishihara}, {Jacobi}, {Japaridze}, {Jeong},
  {Jero}, {Jones}, {Kalaczynski}, {Kang}, {Kappes}, {Kappesser}, {Karg},
  {Karle}, {Katz}, {Kauer}, {Keivani}, {Kelley}, {Kheirandish}, {Kim}, {Kim},
  {Kintscher}, {Kiryluk}, {Kittler}, {Klein}, {Koirala}, {Kolanoski},
  {K{\"o}pke}, {Kopper}, {Kopper}, {Koschinsky}, {Koskinen}, {Kowalski},
  {Krings}, {Kroll}, {Kr{\"u}ckl}, {Kunwar}, {Kurahashi}, {Kuwabara},
  {Kyriacou}, {Labare}, {Lanfranchi}, {Larson}, {Lauber}, {Leonard},
  {Lesiak-Bzdak}, {Leuermann}, {Liu}, {Lozano Mariscal}, {Lu}, {L{\"u}nemann},
  {Luszczak}, {Madsen}, {Maggi}, {Mahn}, {Mancina}, {Maruyama}, {Mase},
  {Maunu}, {Meagher}, {Medici}, {Meier}, {Menne}, {Merino}, {Meures},
  {Miarecki}, {Micallef}, {Moment{\'e}}, {Montaruli}, {Moore}, {Moulai},
  {Nahnhauer}, {Nakarmi}, {Naumann}, {Neer}, {Niederhausen}, {Nowicki},
  {Nygren}, {Obertacke Pollmann}, {Olivas}, {O'Murchadha}, {O'Sullivan},
  {Palczewski}, {Pandya}, {Pankova}, {Peiffer}, {Pepper}, {P{\'e}rez de los
  Heros}, {Pieloth}, {Pinat}, {Plum}, {Price}, {Przybylski}, {Raab},
  {R{\"a}del}, {Rameez}, {Rauch}, {Rawlins}, {Rea}, {Reimann}, {Relethford},
  {Relich}, {Resconi}, {Rhode}, {Richman}, {Robertson}, {Rongen}, {Rott},
  {Ruhe}, {Ryckbosch}, {Rysewyk}, {Safa}, {S{\"a}lzer}, {Sanchez Herrera},
  {Sandrock}, {Sandroos}, {Santander}, {Sarkar}, {Sarkar}, {Satalecka},
  {Schlunder}, {Schmidt}, {Schneider}, {Schoenen}, {Sch{\"o}neberg},
  {Schumacher}, {Sclafani}, {Seckel}, {Seunarine}, {Soedingrekso}, {Soldin},
  {Song}, {Spiczak}, {Spiering}, {Stachurska}, {Stamatikos}, {Stanev},
  {Stasik}, {Stein}, {Stettner}, {Steuer}, {Stezelberger}, {Stokstad},
  {St{\"o}{\ss}l}, {Strotjohann}, {Stuttard}, {Sullivan}, {Sutherland},
  {Taboada}, {Tatar}, {Tenholt}, {Ter-Antonyan}, {Terliuk}, {Tilav}, {Toale},
  {Tobin}, {T{\"o}nnis}, {Toscano}, {Tosi}, {Tselengidou}, {Tung}, {Turcati},
  {Turley}, {Ty}, {Unger}, {Usner}, {Vandenbroucke}, {Van Driessche}, {van
  Eijk}, {van Eijndhoven}, {Vanheule}, {van Santen}, {Vogel}, {Vraeghe},
  {Walck}, {Wallace}, {Wallraff}, {Wandler}, {Wandkowsky}, {Waza}, {Weaver},
  {Weiss}, {Wendt}, {Werthebach}, {Westerhoff}, {Whelan}, {Wiebe}, {Wiebusch},
  {Wille}, {Williams}, {Wills}, {Wolf}, {Wood}, {Wood}, {Woolsey}, {Woschnagg},
  {Xu}, {Xu}, {Xu}, {Yanez}, {Yodh}, {Yoshida}, {Yuan}, \& {IceCube
  Collaboration}}]{2018ApJ...857..117A}
---. 2018, \apj, 857, 117, \dodoi{10.3847/1538-4357/aab4f8}

\bibitem[{{Aartsen} {et~al.}(2020){Aartsen}, {Ackermann}, {Adams}, {Aguilar},
  {Ahlers}, {Ahrens}, {Alispach}, {Andeen}, {Anderson}, {Ansseau}, {Anton},
  {Arg{\"u}elles}, {Auffenberg}, {Axani}, {Backes}, {Bagherpour}, {Bai},
  {Balagopal}, {Barbano}, {Barwick}, {Bastian}, {Baum}, {Baur}, {Bay},
  {Beatty}, {Becker}, {Becker Tjus}, {BenZvi}, {Berley}, {Bernardini},
  {Besson}, {Binder}, {Bindig}, {Blaufuss}, {Blot}, {Bohm}, {B{\"o}rner},
  {B{\"o}ser}, {Botner}, {B{\"o}ttcher}, {Bourbeau}, {Bourbeau}, {Bradascio},
  {Braun}, {Bron}, {Brostean-Kaiser}, {Burgman}, {Buscher}, {Busse}, {Carver},
  {Chen}, {Cheung}, {Chirkin}, {Choi}, {Clark}, {Classen}, {Coleman}, {Collin},
  {Conrad}, {Coppin}, {Correa}, {Cowen}, {Cross}, {Dave}, {De Clercq},
  {DeLaunay}, {Dembinski}, {Deoskar}, {De Ridder}, {Desiati}, {de Vries}, {de
  Wasseige}, {de With}, {DeYoung}, {Diaz}, {D{\'\i}az-V{\'e}lez}, {Dujmovic},
  {Dunkman}, {Dvorak}, {Eberhardt}, {Ehrhardt}, {Eller}, {Engel}, {Evenson},
  {Fahey}, {Fazely}, {Felde}, {Filimonov}, {Finley}, {Franckowiak}, {Friedman},
  {Fritz}, {Gaisser}, {Gallagher}, {Ganster}, {Garrappa}, {Gerhardt},
  {Ghorbani}, {Glauch}, {Gl{\"u}senkamp}, {Goldschmidt}, {Gonzalez}, {Grant},
  {Griffith}, {Griswold}, {G{\"u}nder}, {G{\"u}nd{\"u}z}, {Haack}, {Hallgren},
  {Halve}, {Halzen}, {Hanson}, {Haungs}, {Hebecker}, {Heereman}, {Heix},
  {Helbing}, {Hellauer}, {Henningsen}, {Hickford}, {Hignight}, {Hill},
  {Hoffman}, {Hoffmann}, {Hoinka}, {Hokanson-Fasig}, {Hoshina}, {Huang},
  {Huber}, {Huber}, {Hultqvist}, {H{\"u}nnefeld}, {Hussain}, {In}, {Iovine},
  {Ishihara}, {Japaridze}, {Jeong}, {Jero}, {Jones}, {Jonske}, {Joppe}, {Kang},
  {Kang}, {Kappes}, {Kappesser}, {Karg}, {Karl}, {Karle}, {Katz}, {Kauer},
  {Kelley}, {Kheirandish}, {Kim}, {Kintscher}, {Kiryluk}, {Kittler}, {Klein},
  {Koirala}, {Kolanoski}, {K{\"o}pke}, {Kopper}, {Kopper}, {Koskinen},
  {Kowalski}, {Krings}, {Kr{\"u}ckl}, {Kulacz}, {Kurahashi}, {Kyriacou},
  {Labare}, {Lanfranchi}, {Larson}, {Lauber}, {Lazar}, {Leonard},
  {Leszczy{\'n}ska}, {Leuermann}, {Liu}, {Lohfink}, {Mariscal}, {Lu},
  {Lucarelli}, {L{\"u}nemann}, {Luszczak}, {Lyu}, {Ma}, {Madsen}, {Maggi},
  {Mahn}, {Makino}, {Mallik}, {Mallot}, {Mancina}, {Mari{\c{s}}}, {Maruyama},
  {Mase}, {Maunu}, {McNally}, {Meagher}, {Medici}, {Medina}, {Meier},
  {Meighen-Berger}, {Menne}, {Merino}, {Meures}, {Micallef}, {Mockler},
  {Moment{\'e}}, {Montaruli}, {Moore}, {Morse}, {Moulai}, {Muth}, {Nagai},
  {Naumann}, {Neer}, {Niederhausen}, {Nowicki}, {Nygren}, {Pollmann}, {Oehler},
  {Olivas}, {O'Murchadha}, {O'Sullivan}, {Palczewski}, {Pandya}, {Pankova},
  {Park}, {Peiffer}, {de los Heros}, {Philippen}, {Pieloth}, {Pinat},
  {Pizzuto}, {Plum}, {Porcelli}, {Price}, {Przybylski}, {Raab}, {Raissi},
  {Rameez}, {Rauch}, {Rawlins}, {Rea}, {Reimann}, {Relethford}, {Renschler},
  {Renzi}, {Resconi}, {Rhode}, {Richman}, {Robertson}, {Rongen}, {Rott},
  {Ruhe}, {Ryckbosch}, {Rysewyk}, {Safa}, {Herrera}, {Sandrock}, {Sandroos},
  {Santander}, {Sarkar}, {Sarkar}, {Satalecka}, {Schaufel}, {Schieler},
  {Schlunder}, {Schmidt}, {Schneider}, {Schneider}, {Schr{\"o}der},
  {Schumacher}, {Sclafani}, {Seckel}, {Seunarine}, {Shefali}, {Silva},
  {Snihur}, {Soedingrekso}, {Soldin}, {Song}, {Spiczak}, {Spiering},
  {Stachurska}, {Stamatikos}, {Stanev}, {Stein}, {Steinm{\"u}ller}, {Stettner},
  {Steuer}, {Stezelberger}, {Stokstad}, {St{\"o}{\ss}l}, {Strotjohann},
  {St{\"u}rwald}, {Stuttard}, {Sullivan}, {Taboada}, {Tenholt}, {Ter-Antonyan},
  {Terliuk}, {Tilav}, {Tollefson}, {Tomankova}, {T{\"o}nnis}, {Toscano},
  {Tosi}, {Trettin}, {Tselengidou}, {Tung}, {Turcati}, {Turcotte}, {Turley},
  {Ty}, {Unger}, {Unland Elorrieta}, {Usner}, {Vandenbroucke}, {Van Driessche},
  {van Eijk}, {van Eijndhoven}, {Vanheule}, {van Santen}, {Vraeghe}, {Walck},
  {Wallace}, {Wallraff}, {Wandkowsky}, {Watson}, {Weaver}, {Weindl}, {Weiss},
  {Weldert}, {Wendt}, {Werthebach}, {Whelan}, {Whitehorn}, {Wiebe}, {Wiebusch},
  {Wille}, {Williams}, {Wills}, {Wolf}, {Wood}, {Wood}, {Woschnagg}, {Wrede},
  {Xu}, {Xu}, {Xu}, {Yanez}, {Yodh}, {Yoshida}, {Yuan}, \&
  {Z{\"o}cklein}}]{2020ApJ...890..111A}
---. 2020, \apj, 890, 111, \dodoi{10.3847/1538-4357/ab564b}

\bibitem[{{Abbott} {et~al.}(2009){Abbott}, {Abbott}, {Adhikari}, {Ajith},
  {Allen}, {Allen}, {Amin}, {Anderson}, {Anderson}, {Arain}, {Araya},
  {Armandula}, {Armor}, {Aso}, {Aston}, {Aufmuth}, {Aulbert}, {Babak}, {Baker},
  {Ballmer}, {Barker}, {Barker}, {Barr}, {Barriga}, {Barsotti}, {Barton},
  {Bartos}, {Bassiri}, {Bastarrika}, {Behnke}, {Benacquista}, {Betzwieser},
  {Beyersdorf}, {Bilenko}, {Billingsley}, {Biswas}, {Black}, {Blackburn},
  {Blackburn}, {Blair}, {Bland}, {Bodiya}, {Bogue}, {Bork}, {Boschi}, {Bose},
  {Brady}, {Braginsky}, {Brau}, {Bridges}, {Brinkmann}, {Brooks}, {Brown},
  {Brummit}, {Brunet}, {Bullington}, {Buonanno}, {Burmeister}, {Byer},
  {Cadonati}, {Camp}, {Cannizzo}, {Cannon}, {Cao}, {Cardenas}, {Caride},
  {Castaldi}, {Caudill}, {Cavagli{\`a}}, {Cepeda}, {Chalermsongsak},
  {Chalkley}, {Charlton}, {Chatterji}, {Chelkowski}, {Chen}, {Christensen},
  {Chung}, {Clark}, {Clark}, {Clayton}, {Cokelaer}, {Colacino}, {Conte},
  {Cook}, {Corbitt}, {Cornish}, {Coward}, {Coyne}, {Creighton}, {Creighton},
  {Cruise}, {Culter}, {Cumming}, {Cunningham}, {Danilishin}, {Danzmann},
  {Daudert}, {Davies}, {Daw}, {DeBra}, {Degallaix}, {Dergachev}, {Desai},
  {DeSalvo}, {Dhurandhar}, {D{\'\i}az}, {Dietz}, {Donovan}, {Dooley}, {Doomes},
  {Drever}, {Dueck}, {Duke}, {Dumas}, {Dwyer}, {Echols}, {Edgar}, {Effler},
  {Ehrens}, {Espinoza}, {Etzel}, {Evans}, {Evans}, {Fairhurst}, {Faltas},
  {Fan}, {Fazi}, {Fehrmenn}, {Finn}, {Flasch}, {Foley}, {Forrest},
  {Fotopoulos}, {Franzen}, {Frede}, {Frei}, {Frei}, {Freise}, {Frey}, {Fricke},
  {Fritschel}, {Frolov}, {Fyffe}, {Galdi}, {Garofoli}, {Gholami}, {Giaime},
  {Giampanis}, {Giardina}, {Goda}, {Goetz}, {Goggin}, {Gonz{\'a}lez},
  {Gorodetsky}, {Go{\ss}ler}, {Gouaty}, {Grant}, {Gras}, {Gray}, {Gray},
  {Greenhalgh}, {Gretarsson}, {Grimaldi}, {Grosso}, {Grote}, {Grunewald},
  {Guenther}, {Gustafson}, {Gustafson}, {Hage}, {Hallam}, {Hammer}, {Hammond},
  {Hanna}, {Hanson}, {Harms}, {Harry}, {Harry}, {Harstad}, {Haughian},
  {Hayama}, {Heefner}, {Heng}, {Heptonstall}, {Hewitson}, {Hild}, {Hirose},
  {Hoak}, {Hodge}, {Holt}, {Hosken}, {Hough}, {Hoyland}, {Hughey}, {Huttner},
  {Ingram}, {Isogai}, {Ito}, {Ivanov}, {Johnson}, {Johnson}, {Jones}, {Jones},
  {Jones}, {Ju}, {Kalmus}, {Kalogera}, {Kandhasamy}, {Kanner}, {Kasprzyk},
  {Katsavounidis}, {Kawabe}, {Kawamura}, {Kawazoe}, {Kells}, {Keppel},
  {Khalaidovski}, {Khalili}, {Khan}, {Khazanov}, {King}, {Kissel}, {Klimenko},
  {Kokeyama}, {Kondrashov}, {Kopparapu}, {Koranda}, {Kozak}, {Krishnan},
  {Kumar}, {Kwee}, {Lam}, {Landry}, {Lantz}, {Lazzarini}, {Lei}, {Lei},
  {Leindecker}, {Leonor}, {Li}, {Lin}, {Lindquist}, {Littenberg}, {Lockerbie},
  {Lodhia}, {Longo}, {Lormand}, {Lu}, {Lubi{\'n}ski}, {Lucianetti}, {L{\"u}ck},
  {Machenschalk}, {MacInnis}, {Mageswaran}, {Mailand}, {Mandel}, {Mandic},
  {M{\'a}rka}, {M{\'a}rka}, {Markosyan}, {Markowitz}, {Maros}, {Martin},
  {Martin}, {Marx}, {Mason}, {Matichard}, {Matone}, {Matzner}, {Mavalvala},
  {McCarthy}, {McClelland}, {McGuire}, {McHugh}, {McIntyre}, {McKechan},
  {McKenzie}, {Mehmet}, {Melatos}, {Melissinos}, {Men{\'e}ndez}, {Mendell},
  {Mercer}, {Meshkov}, {Messenger}, {Meyer}, {Miller}, {Minelli}, {Mino},
  {Mitrofanov}, {Mitselmakher}, {Mittleman}, {Miyakawa}, {Moe}, {Mohanty},
  {Mohapatra}, {Moreno}, {Morioka}, {Mors}, {Mossavi}, {Mow Lowry}, {Mueller},
  {M{\"u}ller-Ebhardt}, {Muhammad}, {Mukherjee}, {Mukhopadhyay}, {Mullavey},
  {Munch}, {Murray}, {Myers}, {Myers}, {Nash}, {Nelson}, {Newton}, {Nishizawa},
  {Numata}, {O'Dell}, {O'Reilly}, {O'Shaughnessy}, {Ochsner}, {Ogin},
  {Ottaway}, {Ottens}, {Overmier}, {Owen}, {Pan}, {Pankow}, {Papa},
  {Parameshwaraiah}, {Patel}, {Pedraza}, {Penn}, {Perraca}, {Pierro}, {Pinto},
  {Pitkin}, {Pletsch}, {Plissi}, {Postiglione}, {Principe}, {Prix},
  {Prokhorov}, {Punken}, {Quetschke}, {Raab}, {Rabeling}, {Radkins}, {Raffai},
  {Raics}, {Rainer}, {Rakhmanov}, {Raymond}, {Reed}, {Reed}, {Rehbein}, {Reid},
  {Reitze}, {Riesen}, {Riles}, {Rivera}, {Roberts}, {Robertson}, {Robinson},
  {Robinson}, {Roddy}, {R{\"o}ver}, {Rollins}, {Romano}, {Romie}, {Rowan},
  {R{\"u}diger}, {Russell}, {Ryan}, {Sakata}, {de la Jordana}, {Sandberg},
  {Sannibale}, {Santamar{\'\i}a}, {Saraf}, {Sarin}, {Sathyaprakash}, {Sato},
  {Satterthwaite}, {Saulson}, {Savage}, {Savov}, {Scanlan}, {Schilling},
  {Schnabel}, {Schofield}, {Schulz}, {Schutz}, {Schwinberg}, {Scott}, {Scott},
  {Searle}, {Sears}, {Seifert}, {Sellers}, {Sengupta}, {Sergeev}, {Shapiro},
  {Shawhan}, {Shoemaker}, {Sibley}, {Siemens}, {Sigg}, {Sinha}, {Sintes},
  {Slagmolen}, {Slutsky}, {Smith}, {Smith}, {Smith}, {Somiya}, {Sorazu},
  {Stein}, {Stein}, {Steplewski}, {Stochino}, {Stone}, {Strain}, {Strigin},
  {Stroeer}, {Stuver}, {Summerscales}, {Sun}, {Sung}, {Sutton}, {Szokoly},
  {Talukder}, {Tang}, {Tanner}, {Tarabrin}, {Taylor}, {Taylor}, {Thacker},
  {Thorne}, {Th{\"u}ring}, {Tokmakov}, {Torres}, {Torrie}, {Traylor}, {Trias},
  {Ugolini}, {Ulmen}, {Urbanek}, {Vahlbruch}, {Vallisneri}, {van den Broeck},
  {van der Sluys}, {van Veggel}, {Vass}, {Vaulin}, {Vecchio}, {Veitch},
  {Veitch}, {Veltkamp}, {Villar}, {Vorvick}, {Vyachanin}, {Waldman}, {Wallace},
  {Ward}, {Weidner}, {Weinert}, {Weinstein}, {Weiss}, {Wen}, {Wen}, {Wette},
  {Whelan}, {Whitcomb}, {Whiting}, {Wilkinson}, {Willems}, {Williams},
  {Williams}, {Willke}, {Wilmut}, {Winkelmann}, {Winkler}, {Wipf}, {Wiseman},
  {Woan}, {Wooley}, {Worden}, {Wu}, {Yakushin}, {Yamamoto}, {Yan}, {Yoshida},
  {Zanolin}, {Zhang}, {Zhang}, {Zhao}, {Zotov}, {Zucker}, {M{\"u}hlen}, \&
  {Zweizig}}]{2009RPPh...72g6901A}
{Abbott}, B.~P., {Abbott}, R., {Adhikari}, R., {et~al.} 2009, Reports on
  Progress in Physics, 72, 076901, \dodoi{10.1088/0034-4885/72/7/076901}

\bibitem[{{Abbott} {et~al.}(2016){Abbott}, {Abbott}, {Abbott}, {Abernathy},
  {Acernese}, {Ackley}, {Adams}, {Adams}, {Addesso}, {Adhikari}, {Adya},
  {Affeldt}, {Agathos}, {Agatsuma}, {Aggarwal}, {Aguiar}, {Aiello}, {Ain},
  {Ajith}, {Allen}, {Allocca}, {Altin}, {Anderson}, {Anderson}, {Arai},
  {Araya}, {Arceneaux}, {Areeda}, {Arnaud}, {Arun}, {Ascenzi}, {Ashton}, {Ast},
  {Aston}, {Astone}, {Aufmuth}, {Aulbert}, {Babak}, {Bacon}, {Bader}, {Baker},
  {Baldaccini}, {Ballardin}, {Ballmer}, {Barayoga}, {Barclay}, {Barish},
  {Barker}, {Barone}, {Barr}, {Barsotti}, {Barsuglia}, {Barta}, {Bartlett},
  {Bartos}, {Bassiri}, {Basti}, {Batch}, {Baune}, {Bavigadda}, {Bazzan},
  {Behnke}, {Bejger}, {Bell}, {Bell}, {Berger}, {Bergman}, {Bergmann}, {Berry},
  {Bersanetti}, {Bertolini}, {Betzwieser}, {Bhagwat}, {Bhandare}, {Bilenko},
  {Billingsley}, {Birch}, {Birney}, {Biscans}, {Bisht}, {Bitossi}, {Biwer},
  {Bizouard}, {Blackburn}, {Blair}, {Blair}, {Blair}, {Bloemen}, {Bock},
  {Bodiya}, {Boer}, {Bogaert}, {Bogan}, {Bohe}, {Boh{\'e}mier}, {Bojtos},
  {Bond}, {Bondu}, {Bonnand}, {Boom}, {Bork}, {Boschi}, {Bose}, {Bouffanais},
  {Bozzi}, {Bradaschia}, {Brady}, {Braginsky}, {Branchesi}, {Brau}, {Briant},
  {Brillet}, {Brinkmann}, {Brisson}, {Brockill}, {Brooks}, {Brown}, {Brown},
  {Brown}, {Buchanan}, {Buikema}, {Bulik}, {Bulten}, {Buonanno}, {Buskulic},
  {Buy}, {Byer}, {Cabero}, {Cadonati}, {Cagnoli}, {Cahillane}, {Calder{\'o}n
  Bustillo}, {Callister}, {Calloni}, {Camp}, {Cannon}, {Cao}, {Capano},
  {Capocasa}, {Carbognani}, {Caride}, {Casanueva Diaz}, {Casentini}, {Caudill},
  {Cavagli{\`a}}, {Cavalier}, {Cavalieri}, {Cella}, {Cepeda}, {Cerboni
  Baiardi}, {Cerretani}, {Cesarini}, {Chakraborty}, {Chalermsongsak},
  {Chamberlin}, {Chan}, {Chao}, {Charlton}, {Chassande-Mottin}, {Chen}, {Chen},
  {Cheng}, {Chincarini}, {Chiummo}, {Cho}, {Cho}, {Chow}, {Christensen}, {Chu},
  {Chua}, {Chung}, {Ciani}, {Clara}, {Clark}, {Clayton}, {Cleva}, {Coccia},
  {Cohadon}, {Cokelaer}, {Colla}, {Collette}, {Cominsky}, {Constancio},
  {Conte}, {Conti}, {Cook}, {Corbitt}, {Cornish}, {Corsi}, {Cortese}, {Costa},
  {Coughlin}, {Coughlin}, {Coulon}, {Countryman}, {Couvares}, {Cowan},
  {Coward}, {Cowart}, {Coyne}, {Coyne}, {Craig}, {Creighton}, {Creighton},
  {Cripe}, {Crowder}, {Cumming}, {Cunningham}, {Cuoco}, {Dal Canton},
  {Danilishin}, {D'Antonio}, {Danzmann}, {Darman}, {Dattilo}, {Dave},
  {Daveloza}, {Davier}, {Davies}, {Daw}, {Day}, {De}, {DeBra}, {Debreczeni},
  {Degallaix}, {De Laurentis}, {Del{\'e}glise}, {Del Pozzo}, {Denker}, {Dent},
  {Dereli}, {Dergachev}, {DeRosa}, {De Rosa}, {DeSalvo}, {Dhurandhar},
  {D{\'\i}az}, {Dietz}, {Di Fiore}, {Di Giovanni}, {Di Lieto}, {Di Pace}, {Di
  Palma}, {Di Virgilio}, {Dojcinoski}, {Dolique}, {Donovan}, {Dooley},
  {Doravari}, {Douglas}, {Downes}, {Drago}, {Drever}, {Driggers}, {Du},
  {Ducrot}, {Dwyer}, {Edo}, {Edwards}, {Effler}, {Eggenstein}, {Ehrens},
  {Eichholz}, {Eikenberry}, {Engels}, {Essick}, {Etzel}, {Evans}, {Evans},
  {Everett}, {Factourovich}, {Fafone}, {Fair}, {Fairhurst}, {Fan}, {Fang},
  {Farinon}, {Farr}, {Farr}, {Favata}, {Fays}, {Fehrmann}, {Fejer}, {Ferrante},
  {Ferreira}, {Ferrini}, {Fidecaro}, {Fiori}, {Fiorucci}, {Fisher}, {Flaminio},
  {Fletcher}, {Fotopoulos}, {Fournier}, {Franco}, {Frasca}, {Frasconi}, {Frei},
  {Frei}, {Freise}, {Frey}, {Frey}, {Fricke}, {Fritschel}, {Frolov}, {Fulda},
  {Fyffe}, {Gabbard}, {Gair}, {Gammaitoni}, {Gaonkar}, {Garufi}, {Gatto},
  {Gaur}, {Gehrels}, {Gemme}, {Gendre}, {Genin}, {Gennai}, {George}, {Gergely},
  {Germain}, {Ghosh}, {Ghosh}, {Giaime}, {Giardina}, {Giazotto}, {Gill},
  {Glaefke}, {Goetz}, {Goetz}, {Goggin}, {Gondan}, {Gonz{\'a}lez}, {Gonzalez
  Castro}, {Gopakumar}, {Gordon}, {Gorodetsky}, {Gossan}, {Gosselin}, {Gouaty},
  {Graef}, {Graff}, {Granata}, {Grant}, {Gras}, {Gray}, {Greco}, {Green},
  {Groot}, {Grote}, {Grunewald}, {Guidi}, {Guo}, {Gupta}, {Gupta}, {Gushwa},
  {Gustafson}, {Gustafson}, {Hacker}, {Hall}, {Hall}, {Hammond}, {Haney},
  {Hanke}, {Hanks}, {Hanna}, {Hannam}, {Hanson}, {Hardwick}, {Harms}, {Harry},
  {Harry}, {Hart}, {Hartman}, {Haster}, {Haughian}, {Heidmann}, {Heintze},
  {Heitmann}, {Hello}, {Hemming}, {Hendry}, {Heng}, {Hennig}, {Heptonstall},
  {Heurs}, {Hild}, {Hoak}, {Hodge}, {Hofman}, {Hollitt}, {Holt}, {Holz},
  {Hopkins}, {Hosken}, {Hough}, {Houston}, {Howell}, {Hu}, {Huang}, {Huerta},
  {Huet}, {Hughey}, {Husa}, {Huttner}, {Huynh-Dinh}, {Idrisy}, {Indik},
  {Ingram}, {Inta}, {Isa}, {Isac}, {Isi}, {Islas}, {Isogai}, {Iyer}, {Izumi},
  {Jacqmin}, {Jang}, {Jani}, {Jaranowski}, {Jawahar}, {Jim{\'e}nez-Forteza},
  {Johnson}, {Jones}, {Jones}, {Jones}, {Jonker}, {Ju}, {Haris}, {Kalaghatgi},
  {Kalogera}, {Kandhasamy}, {Kang}, {Kanner}, {Karki}, {Kasprzack},
  {Katsavounidis}, {Katzman}, {Kaufer}, {Kaur}, {Kawabe}, {Kawazoe},
  {K{\'e}f{\'e}lian}, {Kehl}, {Keitel}, {Kelley}, {Kells}, {Keppel}, {Kennedy},
  {Key}, {Khalaidovski}, {Khalili}, {Khan}, {Khan}, {Khan}, {Khazanov},
  {Kijbunchoo}, {Kim}, {Kim}, {Kim}, {Kim}, {Kim}, {Kim}, {King}, {King},
  {Kinzel}, {Kissel}, {Kleybolte}, {Klimenko}, {Koehlenbeck}, {Kokeyama},
  {Koley}, {Kondrashov}, {Kontos}, {Korobko}, {Korth}, {Kowalska}, {Kozak},
  {Kringel}, {Krishnan}, {Kr{\'o}lak}, {Krueger}, {Kuehn}, {Kumar}, {Kuo},
  {Kutynia}, {Lackey}, {Landry}, {Lange}, {Lantz}, {Lasky}, {Lazzarini},
  {Lazzaro}, {Leaci}, {Leavey}, {Lebigot}, {Lee}, {Lee}, {Lee}, {Lee}, {Lenon},
  {Leonardi}, {Leong}, {Leroy}, {Letendre}, {Levin}, {Levine}, {Li}, {Libson},
  {Littenberg}, {Lockerbie}, {Logue}, {Lombardi}, {Lord}, {Lorenzini},
  {Loriette}, {Lormand}, {Losurdo}, {Lough}, {L{\"u}ck}, {Lundgren}, {Luo},
  {Lynch}, {Ma}, {MacDonald}, {Machenschalk}, {MacInnis}, {Macleod},
  {Maga{\~n}a-Sandoval}, {Magee}, {Mageswaran}, {Majorana}, {Maksimovic},
  {Malvezzi}, {Man}, {Mandel}, {Mandic}, {Mangano}, {Mansell}, {Manske},
  {Mantovani}, {Marchesoni}, {Marion}, {M{\'a}rka}, {M{\'a}rka}, {Markosyan},
  {Maros}, {Martelli}, {Martellini}, {Martin}, {Martin}, {Martynov}, {Marx},
  {Mason}, {Masserot}, {Massinger}, {Masso-Reid}, {Matichard}, {Matone},
  {Mavalvala}, {Mazumder}, {Mazzolo}, {McCarthy}, {McClelland}, {McCormick},
  {McGuire}, {McIntyre}, {McIver}, {McKechan}, {McManus}, {McWilliams},
  {Meacher}, {Meadors}, {Meidam}, {Melatos}, {Mendell}, {Mendoza-Gandara},
  {Mercer}, {Merilh}, {Merzougui}, {Meshkov}, {Messaritaki}, {Messenger},
  {Messick}, {Meyers}, {Mezzani}, {Miao}, {Michel}, {Middleton}, {Mikhailov},
  {Milano}, {Miller}, {Millhouse}, {Minenkov}, {Ming}, {Mirshekari}, {Mishra},
  {Mitra}, {Mitrofanov}, {Mitselmakher}, {Mittleman}, {Moggi}, {Mohan},
  {Mohapatra}, {Montani}, {Moore}, {Moore}, {Moraru}, {Moreno}, {Morriss},
  {Mossavi}, {Mours}, {Mow-Lowry}, {Mueller}, {Mueller}, {Muir}, {Mukherjee},
  {Mukherjee}, {Mukherjee}, {Mukund}, {Mullavey}, {Munch}, {Murphy}, {Murray},
  {Mytidis}, {Nardecchia}, {Naticchioni}, {Nayak}, {Necula}, {Nedkova},
  {Nelemans}, {Neri}, {Neunzert}, {Newton}, {Nguyen}, {Nielsen}, {Nissanke},
  {Nitz}, {Nocera}, {Nolting}, {Normandin}, {Nuttall}, {Oberling}, {Ochsner},
  {O'Dell}, {Oelker}, {Ogin}, {Oh}, {Oh}, {Ohme}, {Oliver}, {Oppermann},
  {Oram}, {O'Reilly}, {O'Shaughnessy}, {Ottaway}, {Ottens}, {Overmier}, {Owen},
  {Pai}, {Pai}, {Palamos}, {Palashov}, {Palomba}, {Pal-Singh}, {Pan}, {Pan},
  {Pankow}, {Pannarale}, {Pant}, {Paoletti}, {Paoli}, {Papa}, {Paris},
  {Parker}, {Pascucci}, {Pasqualetti}, {Passaquieti}, {Passuello},
  {Patricelli}, {Patrick}, {Pearlstone}, {Pedraza}, {Pedurand}, {Pekowsky},
  {Pele}, {Penn}, {Perreca}, {Phelps}, {Piccinni}, {Pichot}, {Piergiovanni},
  {Pierro}, {Pillant}, {Pinard}, {Pinto}, {Pitkin}, {Poggiani}, {Popolizio},
  {Post}, {Powell}, {Prasad}, {Predoi}, {Premachandra}, {Prestegard}, {Price},
  {Prijatelj}, {Principe}, {Privitera}, {Prodi}, {Prokhorov}, {Puncken},
  {Punturo}, {Puppo}, {P{\"u}rrer}, {Qi}, {Qin}, {Quetschke}, {Quintero},
  {Quitzow-James}, {Raab}, {Rabeling}, {Radkins}, {Raffai}, {Raja},
  {Rakhmanov}, {Rapagnani}, {Raymond}, {Razzano}, {Re}, {Read}, {Reed},
  {Regimbau}, {Rei}, {Reid}, {Reitze}, {Rew}, {Reyes}, {Ricci}, {Riles},
  {Robertson}, {Robie}, {Robinet}, {Robinson}, {Rocchi}, {Rodriguez},
  {Rolland}, {Rollins}, {Roma}, {Romano}, {Romanov}, {Romie}, {Rosi{\'n}ska},
  {Rowan}, {R{\"u}diger}, {Ruggi}, {Ryan}, {Sachdev}, {Sadecki}, {Sadeghian},
  {Salconi}, {Saleem}, {Salemi}, {Samajdar}, {Sammut}, {Sanchez}, {Sandberg},
  {Sandeen}, {Sanders}, {Santamar{\'\i}a}, {Sassolas}, {Sathyaprakash},
  {Saulson}, {Sauter}, {Savage}, {Sawadsky}, {Schale}, {Schilling}, {Schmidt},
  {Schmidt}, {Schnabel}, {Schofield}, {Sch{\"o}nbeck}, {Schreiber}, {Schuette},
  {Schutz}, {Scott}, {Scott}, {Sellers}, {Sengupta}, {Sentenac}, {Sequino},
  {Sergeev}, {Serna}, {Setyawati}, {Sevigny}, {Shaddock}, {Shah}, {Shahriar},
  {Shaltev}, {Shao}, {Shapiro}, {Shawhan}, {Sheperd}, {Shoemaker}, {Shoemaker},
  {Siellez}, {Siemens}, {Sigg}, {Silva}, {Simakov}, {Singer}, {Singer},
  {Singh}, {Singh}, {Singhal}, {Sintes}, {Slagmolen}, {Smith}, {Smith},
  {Smith}, {Son}, {Sorazu}, {Sorrentino}, {Souradeep}, {Srivastava}, {Staley},
  {Steinke}, {Steinlechner}, {Steinlechner}, {Steinmeyer}, {Stephens}, {Stone},
  {Strain}, {Straniero}, {Stratta}, {Strauss}, {Strigin}, {Sturani}, {Stuver},
  {Summerscales}, {Sun}, {Sutton}, {Swinkels}, {Szczepa{\'n}czyk}, {Tacca},
  {Talukder}, {Tanner}, {T{\'a}pai}, {Tarabrin}, {Taracchini}, {Taylor},
  {Theeg}, {Thirugnanasambandam}, {Thomas}, {Thomas}, {Thomas}, {Thorne},
  {Thorne}, {Thrane}, {Tiwari}, {Tiwari}, {Tokmakov}, {Tomlinson}, {Tonelli},
  {Torres}, {Torrie}, {T{\"o}yr{\"a}}, {Travasso}, {Traylor}, {Trifir{\`o}},
  {Tringali}, {Trozzo}, {Tse}, {Turconi}, {Tuyenbayev}, {Ugolini},
  {Unnikrishnan}, {Urban}, {Usman}, {Vahlbruch}, {Vajente}, {Valdes}, {van
  Bakel}, {van Beuzekom}, {van den Brand}, {Van Den Broeck}, {Vander-Hyde},
  {van der Schaaf}, {van Heijningen}, {van Veggel}, {Vardaro}, {Vass},
  {Vas{\'u}th}, {Vaulin}, {Vecchio}, {Vedovato}, {Veitch}, {Veitch},
  {Venkateswara}, {Verkindt}, {Vetrano}, {Vicer{\'e}}, {Vinciguerra}, {Vine},
  {Vinet}, {Vitale}, {Vo}, {Vocca}, {Vorvick}, {Voss}, {Vousden}, {Vyatchanin},
  {Wade}, {Wade}, {Wade}, {Walker}, {Wallace}, {Walsh}, {Wang}, {Wang}, {Wang},
  {Wang}, {Wang}, {Ward}, {Warner}, {Was}, {Weaver}, {Wei}, {Weinert},
  {Weinstein}, {Weiss}, {Welborn}, {Wen}, {We{\ss}els}, {West}, {Westphal},
  {Wette}, {Whelan}, {White}, {Whiting}, {Wiesner}, {Williams}, {Williamson},
  {Willis}, {Willke}, {Wimmer}, {Winkler}, {Wipf}, {Wiseman}, {Wittel}, {Woan},
  {Worden}, {Wright}, {Wu}, {Yablon}, {Yam}, {Yamamoto}, {Yancey}, {Yap}, {Yu},
  {Yvert}, {Zadro{\.Z}ny}, {Zangrando}, {Zanolin}, {Zendri}, {Zevin}, {Zhang},
  {Zhang}, {Zhang}, {Zhang}, {Zhao}, {Zhou}, {Zhou}, {Zhu}, {Zucker}, {Zuraw},
  {Zweizig}, {LIGO Scientific Collaboration}, \& {Virgo
  Collaboration}}]{2016PhRvD..93l2003A}
{Abbott}, B.~P., {Abbott}, R., {Abbott}, T.~D., {et~al.} 2016, \prd, 93,
  122003, \dodoi{10.1103/PhysRevD.93.122003}

\bibitem[{{Abbott} {et~al.}(2017){Abbott}, {Abbott}, {Abbott}, {Acernese},
  {Ackley}, {Adams}, {Adams}, {Addesso}, {Adhikari}, {Adya}, {Affeldt},
  {Afrough}, {Agarwal}, {Agathos}, {Agatsuma}, {Aggarwal}, {Aguiar}, {Aiello},
  {Ain}, {Ajith}, {Allen}, {Allen}, {Allocca}, {Altin}, {Amato}, {Ananyeva},
  {Anderson}, {Anderson}, {Angelova}, {Antier}, {Appert}, {Arai}, {Araya},
  {Areeda}, {Arnaud}, {Arun}, {Ascenzi}, {Ashton}, {Ast}, {Aston}, {Astone},
  {Atallah}, {Aufmuth}, {Aulbert}, {AultONeal}, {Austin}, {Avila-Alvarez},
  {Babak}, {Bacon}, {Bader}, {Bae}, {Baker}, {Baldaccini}, {Ballardin},
  {Ballmer}, {Banagiri}, {Barayoga}, {Barclay}, {Barish}, {Barker}, {Barkett},
  {Barone}, {Barr}, {Barsotti}, {Barsuglia}, {Barta}, {Barthelmy}, {Bartlett},
  {Bartos}, {Bassiri}, {Basti}, {Batch}, {Bawaj}, {Bayley}, {Bazzan},
  {B{\'e}csy}, {Beer}, {Bejger}, {Belahcene}, {Bell}, {Berger}, {Bergmann},
  {Bero}, {Berry}, {Bersanetti}, {Bertolini}, {Betzwieser}, {Bhagwat},
  {Bhandare}, {Bilenko}, {Billingsley}, {Billman}, {Birch}, {Birney},
  {Birnholtz}, {Biscans}, {Biscoveanu}, {Bisht}, {Bitossi}, {Biwer},
  {Bizouard}, {Blackburn}, {Blackman}, {Blair}, {Blair}, {Blair}, {Bloemen},
  {Bock}, {Bode}, {Boer}, {Bogaert}, {Bohe}, {Bondu}, {Bonilla}, {Bonnand},
  {Boom}, {Bork}, {Boschi}, {Bose}, {Bossie}, {Bouffanais}, {Bozzi},
  {Bradaschia}, {Brady}, {Branchesi}, {Brau}, {Briant}, {Brillet}, {Brinkmann},
  {Brisson}, {Brockill}, {Broida}, {Brooks}, {Brown}, {Brown}, {Brunett},
  {Buchanan}, {Buikema}, {Bulik}, {Bulten}, {Buonanno}, {Buskulic}, {Buy},
  {Byer}, {Cabero}, {Cadonati}, {Cagnoli}, {Cahillane}, {Calder{\'o}n
  Bustillo}, {Callister}, {Calloni}, {Camp}, {Canepa}, {Canizares}, {Cannon},
  {Cao}, {Cao}, {Capano}, {Capocasa}, {Carbognani}, {Caride}, {Carney},
  {Casanueva Diaz}, {Casentini}, {Caudill}, {Cavagli{\`a}}, {Cavalier},
  {Cavalieri}, {Cella}, {Cepeda}, {Cerd{\'a}-Dur{\'a}n}, {Cerretani},
  {Cesarini}, {Chamberlin}, {Chan}, {Chao}, {Charlton}, {Chase},
  {Chassande-Mottin}, {Chatterjee}, {Chatziioannou}, {Cheeseboro}, {Chen},
  {Chen}, {Chen}, {Cheng}, {Chia}, {Chincarini}, {Chiummo}, {Chmiel}, {Cho},
  {Cho}, {Chow}, {Christensen}, {Chu}, {Chua}, {Chua}, {Chung}, {Chung},
  {Ciani}, {Ciolfi}, {Cirelli}, {Cirone}, {Clara}, {Clark}, {Clearwater},
  {Cleva}, {Cocchieri}, {Coccia}, {Cohadon}, {Cohen}, {Colla}, {Collette},
  {Cominsky}, {Constancio}, {Conti}, {Cooper}, {Corban}, {Corbitt},
  {Cordero-Carri{\'o}n}, {Corley}, {Cornish}, {Corsi}, {Cortese}, {Costa},
  {Coughlin}, {Coughlin}, {Coulon}, {Countryman}, {Couvares}, {Covas}, {Cowan},
  {Coward}, {Cowart}, {Coyne}, {Coyne}, {Creighton}, {Creighton}, {Cripe},
  {Crowder}, {Cullen}, {Cumming}, {Cunningham}, {Cuoco}, {Dal Canton},
  {D{\'a}lya}, {Danilishin}, {D'Antonio}, {Danzmann}, {Dasgupta}, {Da Silva
  Costa}, {Dattilo}, {Dave}, {Davier}, {Davis}, {Daw}, {Day}, {De}, {DeBra},
  {Degallaix}, {De Laurentis}, {Del{\'e}glise}, {Del Pozzo}, {Demos}, {Denker},
  {Dent}, {De Pietri}, {Dergachev}, {De Rosa}, {DeRosa}, {De Rossi}, {DeSalvo},
  {de Varona}, {Devenson}, {Dhurandhar}, {D{\'\i}az}, {Di Fiore}, {Di
  Giovanni}, {Di Girolamo}, {Di Lieto}, {Di Pace}, {Di Palma}, {Di Renzo},
  {Doctor}, {Dolique}, {Donovan}, {Dooley}, {Doravari}, {Dorrington},
  {Douglas}, {Dovale {\'A}lvarez}, {Downes}, {Drago}, {Dreissigacker},
  {Driggers}, {Du}, {Ducrot}, {Dupej}, {Dwyer}, {Edo}, {Edwards}, {Effler},
  {Ehrens}, {Eichholz}, {Eikenberry}, {Eisenstein}, {Essick}, {Estevez},
  {Etienne}, {Etzel}, {Evans}, {Evans}, {Factourovich}, {Fafone}, {Fair},
  {Fairhurst}, {Fan}, {Farinon}, {Farr}, {Farr}, {Fauchon-Jones}, {Favata},
  {Fays}, {Fee}, {Fehrmann}, {Feicht}, {Fejer}, {Fernandez-Galiana},
  {Ferrante}, {Ferreira}, {Ferrini}, {Fidecaro}, {Finstad}, {Fiori},
  {Fiorucci}, {Fishbach}, {Fisher}, {Fitz-Axen}, {Flaminio}, {Fletcher},
  {Fong}, {Font}, {Forsyth}, {Forsyth}, {Fournier}, {Frasca}, {Frasconi},
  {Frei}, {Freise}, {Frey}, {Frey}, {Fries}, {Fritschel}, {Frolov}, {Fulda},
  {Fyffe}, {Gabbard}, {Gadre}, {Gaebel}, {Gair}, {Gammaitoni}, {Ganija},
  {Gaonkar}, {Garcia-Quiros}, {Garufi}, {Gateley}, {Gaudio}, {Gaur},
  {Gayathri}, {Gehrels}, {Gemme}, {Genin}, {Gennai}, {George}, {George},
  {Gergely}, {Germain}, {Ghonge}, {Ghosh}, {Ghosh}, {Ghosh}, {Giaime},
  {Giardina}, {Giazotto}, {Gill}, {Glover}, {Goetz}, {Goetz}, {Gomes},
  {Goncharov}, {Gonz{\'a}lez}, {Gonzalez Castro}, {Gopakumar}, {Gorodetsky},
  {Gossan}, {Gosselin}, {Gouaty}, {Grado}, {Graef}, {Granata}, {Grant}, {Gras},
  {Gray}, {Greco}, {Green}, {Gretarsson}, {Griswold}, {Groot}, {Grote},
  {Grunewald}, {Gruning}, {Guidi}, {Guo}, {Gupta}, {Gupta}, {Gushwa},
  {Gustafson}, {Gustafson}, {Halim}, {Hall}, {Hall}, {Hamilton}, {Hammond},
  {Haney}, {Hanke}, {Hanks}, {Hanna}, {Hannam}, {Hannuksela}, {Hanson},
  {Hardwick}, {Harms}, {Harry}, {Harry}, {Hart}, {Haster}, {Haughian}, {Healy},
  {Heidmann}, {Heintze}, {Heitmann}, {Hello}, {Hemming}, {Hendry}, {Heng},
  {Hennig}, {Heptonstall}, {Heurs}, {Hild}, {Hinderer}, {Hoak}, {Hofman},
  {Holt}, {Holz}, {Hopkins}, {Horst}, {Hough}, {Houston}, {Howell}, {Hreibi},
  {Hu}, {Huerta}, {Huet}, {Hughey}, {Husa}, {Huttner}, {Huynh-Dinh}, {Indik},
  {Inta}, {Intini}, {Isa}, {Isac}, {Isi}, {Iyer}, {Izumi}, {Jacqmin}, {Jani},
  {Jaranowski}, {Jawahar}, {Jim{\'e}nez-Forteza}, {Johnson}, {Jones}, {Jones},
  {Jonker}, {Ju}, {Junker}, {Kalaghatgi}, {Kalogera}, {Kamai}, {Kandhasamy},
  {Kang}, {Kanner}, {Kapadia}, {Karki}, {Karvinen}, {Kasprzack}, {Katolik},
  {Katsavounidis}, {Katzman}, {Kaufer}, {Kawabe}, {K{\'e}f{\'e}lian}, {Keitel},
  {Kemball}, {Kennedy}, {Kent}, {Key}, {Khalili}, {Khan}, {Khan}, {Khan},
  {Khazanov}, {Kijbunchoo}, {Kim}, {Kim}, {Kim}, {Kim}, {Kim}, {Kim},
  {Kimbrell}, {King}, {King}, {Kinley-Hanlon}, {Kirchhoff}, {Kissel},
  {Kleybolte}, {Klimenko}, {Knowles}, {Koch}, {Koehlenbeck}, {Koley},
  {Kondrashov}, {Kontos}, {Korobko}, {Korth}, {Kowalska}, {Kozak},
  {Kr{\"a}mer}, {Kringel}, {Krishnan}, {Kr{\'o}lak}, {Kuehn}, {Kumar}, {Kumar},
  {Kumar}, {Kuo}, {Kutynia}, {Kwang}, {Lackey}, {Lai}, {Landry}, {Lang},
  {Lange}, {Lantz}, {Lanza}, {Larson}, {Lartaux-Vollard}, {Lasky}, {Laxen},
  {Lazzarini}, {Lazzaro}, {Leaci}, {Leavey}, {Lee}, {Lee}, {Lee}, {Lee}, {Lee},
  {Lehmann}, {Lenon}, {Leonardi}, {Leroy}, {Letendre}, {Levin}, {Li}, {Linker},
  {Littenberg}, {Liu}, {Lo}, {Lockerbie}, {London}, {Lord}, {Lorenzini},
  {Loriette}, {Lormand}, {Losurdo}, {Lough}, {Lousto}, {Lovelace}, {L{\"u}ck},
  {Lumaca}, {Lundgren}, {Lynch}, {Ma}, {Macas}, {Macfoy}, {Machenschalk},
  {MacInnis}, {Macleod}, {Maga{\~n}a Hernandez}, {Maga{\~n}a-Sandoval},
  {Maga{\~n}a Zertuche}, {Magee}, {Majorana}, {Maksimovic}, {Man}, {Mandic},
  {Mangano}, {Mansell}, {Manske}, {Mantovani}, {Marchesoni}, {Marion},
  {M{\'a}rka}, {M{\'a}rka}, {Markakis}, {Markosyan}, {Markowitz}, {Maros},
  {Marquina}, {Marsh}, {Martelli}, {Martellini}, {Martin}, {Martin},
  {Martynov}, {Mason}, {Massera}, {Masserot}, {Massinger}, {Masso-Reid},
  {Mastrogiovanni}, {Matas}, {Matichard}, {Matone}, {Mavalvala}, {Mazumder},
  {McCarthy}, {McClelland}, {McCormick}, {McCuller}, {McGuire}, {McIntyre},
  {McIver}, {McManus}, {McNeill}, {McRae}, {McWilliams}, {Meacher}, {Meadors},
  {Mehmet}, {Meidam}, {Mejuto-Villa}, {Melatos}, {Mendell}, {Mercer}, {Merilh},
  {Merzougui}, {Meshkov}, {Messenger}, {Messick}, {Metzdorff}, {Meyers},
  {Miao}, {Michel}, {Middleton}, {Mikhailov}, {Milano}, {Miller}, {Miller},
  {Miller}, {Millhouse}, {Milovich-Goff}, {Minazzoli}, {Minenkov}, {Ming},
  {Mishra}, {Mitra}, {Mitrofanov}, {Mitselmakher}, {Mittleman}, {Moffa},
  {Moggi}, {Mogushi}, {Mohan}, {Mohapatra}, {Montani}, {Moore}, {Moraru},
  {Moreno}, {Morriss}, {Mours}, {Mow-Lowry}, {Mueller}, {Muir}, {Mukherjee},
  {Mukherjee}, {Mukherjee}, {Mukund}, {Mullavey}, {Munch}, {Mu{\~n}iz},
  {Muratore}, {Murray}, {Napier}, {Nardecchia}, {Naticchioni}, {Nayak},
  {Neilson}, {Nelemans}, {Nelson}, {Nery}, {Neunzert}, {Nevin}, {Newport},
  {Newton}, {Ng}, {Nguyen}, {Nguyen}, {Nichols}, {Nielsen}, {Nissanke}, {Nitz},
  {Noack}, {Nocera}, {Nolting}, {North}, {Nuttall}, {Oberling}, {O'Dea},
  {Ogin}, {Oh}, {Oh}, {Ohme}, {Okada}, {Oliver}, {Oppermann}, {Oram},
  {O'Reilly}, {Ormiston}, {Ortega}, {O'Shaughnessy}, {Ossokine}, {Ottaway},
  {Overmier}, {Owen}, {Pace}, {Page}, {Page}, {Pai}, {Pai}, {Palamos},
  {Palashov}, {Palomba}, {Pal-Singh}, {Pan}, {Pan}, {Pang}, {Pang}, {Pankow},
  {Pannarale}, {Pant}, {Paoletti}, {Paoli}, {Papa}, {Parida}, {Parker},
  {Pascucci}, {Pasqualetti}, {Passaquieti}, {Passuello}, {Patil}, {Patricelli},
  {Pearlstone}, {Pedraza}, {Pedurand}, {Pekowsky}, {Pele}, {Penn}, {Perez},
  {Perreca}, {Perri}, {Pfeiffer}, {Phelps}, {Piccinni}, {Pichot},
  {Piergiovanni}, {Pierro}, {Pillant}, {Pinard}, {Pinto}, {Pirello}, {Pitkin},
  {Poe}, {Poggiani}, {Popolizio}, {Porter}, {Post}, {Powell}, {Prasad},
  {Pratt}, {Pratten}, {Predoi}, {Prestegard}, {Price}, {Prijatelj}, {Principe},
  {Privitera}, {Prodi}, {Prokhorov}, {Puncken}, {Punturo}, {Puppo},
  {P{\"u}rrer}, {Qi}, {Quetschke}, {Quintero}, {Quitzow-James}, {Raab},
  {Rabeling}, {Radkins}, {Raffai}, {Raja}, {Rajan}, {Rajbhandari}, {Rakhmanov},
  {Ramirez}, {Ramos-Buades}, {Rapagnani}, {Raymond}, {Razzano}, {Read},
  {Regimbau}, {Rei}, {Reid}, {Reitze}, {Ren}, {Reyes}, {Ricci}, {Ricker},
  {Rieger}, {Riles}, {Rizzo}, {Robertson}, {Robie}, {Robinet}, {Rocchi},
  {Rolland}, {Rollins}, {Roma}, {Romano}, {Romel}, {Romie}, {Rosi{\'n}ska},
  {Ross}, {Rowan}, {R{\"u}diger}, {Ruggi}, {Rutins}, {Ryan}, {Sachdev},
  {Sadecki}, {Sadeghian}, {Sakellariadou}, {Salconi}, {Saleem}, {Salemi},
  {Samajdar}, {Sammut}, {Sampson}, {Sanchez}, {Sanchez}, {Sanchis-Gual},
  {Sandberg}, {Sanders}, {Sassolas}, {Sathyaprakash}, {Saulson}, {Sauter},
  {Savage}, {Sawadsky}, {Schale}, {Scheel}, {Scheuer}, {Schmidt}, {Schmidt},
  {Schnabel}, {Schofield}, {Sch{\"o}nbeck}, {Schreiber}, {Schuette}, {Schulte},
  {Schutz}, {Schwalbe}, {Scott}, {Scott}, {Seidel}, {Sellers}, {Sengupta},
  {Sentenac}, {Sequino}, {Sergeev}, {Shaddock}, {Shaffer}, {Shah}, {Shahriar},
  {Shaner}, {Shao}, {Shapiro}, {Shawhan}, {Sheperd}, {Shoemaker}, {Shoemaker},
  {Siellez}, {Siemens}, {Sieniawska}, {Sigg}, {Silva}, {Singer}, {Singh},
  {Singhal}, {Sintes}, {Slagmolen}, {Smith}, {Smith}, {Smith}, {Somala}, {Son},
  {Sonnenberg}, {Sorazu}, {Sorrentino}, {Souradeep}, {Spencer}, {Srivastava},
  {Staats}, {Staley}, {Steinke}, {Steinlechner}, {Steinlechner}, {Steinmeyer},
  {Stevenson}, {Stone}, {Stops}, {Strain}, {Stratta}, {Strigin}, {Strunk},
  {Sturani}, {Stuver}, {Summerscales}, {Sun}, {Sunil}, {Suresh}, {Sutton},
  {Swinkels}, {Szczepa{\'n}czyk}, {Tacca}, {Tait}, {Talbot}, {Talukder},
  {Tanner}, {T{\'a}pai}, {Taracchini}, {Tasson}, {Taylor}, {Taylor}, {Tewari},
  {Theeg}, {Thies}, {Thomas}, {Thomas}, {Thomas}, {Thorne}, {Thorne}, {Thrane},
  {Tiwari}, {Tiwari}, {Tokmakov}, {Toland}, {Tonelli}, {Tornasi},
  {Torres-Forn{\'e}}, {Torrie}, {T{\"o}yr{\"a}}, {Travasso}, {Traylor},
  {Trinastic}, {Tringali}, {Trozzo}, {Tsang}, {Tse}, {Tso}, {Tsukada}, {Tsuna},
  {Tuyenbayev}, {Ueno}, {Ugolini}, {Unnikrishnan}, {Urban}, {Usman},
  {Vahlbruch}, {Vajente}, {Valdes}, {van Bakel}, {van Beuzekom}, {van den
  Brand}, {Van Den Broeck}, {Vander-Hyde}, {van der Schaaf}, {van Heijningen},
  {van Veggel}, {Vardaro}, {Varma}, {Vass}, {Vas{\'u}th}, {Vecchio},
  {Vedovato}, {Veitch}, {Veitch}, {Venkateswara}, {Venugopalan}, {Verkindt},
  {Vetrano}, {Vicer{\'e}}, {Viets}, {Vinciguerra}, {Vine}, {Vinet}, {Vitale},
  {Vo}, {Vocca}, {Vorvick}, {Vyatchanin}, {Wade}, {Wade}, {Wade}, {Walet},
  {Walker}, {Wallace}, {Walsh}, {Wang}, {Wang}, {Wang}, {Wang}, {Wang}, {Ward},
  {Warner}, {Was}, {Watchi}, {Weaver}, {Wei}, {Weinert}, {Weinstein}, {Weiss},
  {Wen}, {Wessel}, {Wessels}, {Westerweck}, {Westphal}, {Wette}, {Whelan},
  {Whitcomb}, {Whiting}, {Whittle}, {Wilken}, {Williams}, {Williams},
  {Williamson}, {Willis}, {Willke}, {Wimmer}, {Winkler}, {Wipf}, {Wittel},
  {Woan}, {Woehler}, {Wofford}, {Wong}, {Worden}, {Wright}, {Wu}, {Wysocki},
  {Xiao}, {Yamamoto}, {Yancey}, {Yang}, {Yap}, {Yazback}, {Yu}, {Yu}, {Yvert},
  {Zadro{\.z}ny}, {Zanolin}, {Zelenova}, {Zendri}, {Zevin}, {Zhang}, {Zhang},
  {Zhang}, {Zhang}, {Zhao}, {Zhou}, {Zhou}, {Zhu}, {Zhu}, {Zimmerman},
  {Zucker}, {Zweizig}, {LIGO Scientific Collaboration}, {Virgo Collaboration},
  {Wilson-Hodge}, {Bissaldi}, {Blackburn}, {Briggs}, {Burns}, {Cleveland},
  {Connaughton}, {Gibby}, {Giles}, {Goldstein}, {Hamburg}, {Jenke}, {Hui},
  {Kippen}, {Kocevski}, {McBreen}, {Meegan}, {Paciesas}, {Poolakkil}, {Preece},
  {Racusin}, {Roberts}, {Stanbro}, {Veres}, {von Kienlin}, {GBM}, {Savchenko},
  {Ferrigno}, {Kuulkers}, {Bazzano}, {Bozzo}, {Brandt}, {Chenevez},
  {Courvoisier}, {Diehl}, {Domingo}, {Hanlon}, {Jourdain}, {Laurent}, {Lebrun},
  {Lutovinov}, {Martin-Carrillo}, {Mereghetti}, {Natalucci}, {Rodi}, {Roques},
  {Sunyaev}, {Ubertini}, {INTEGRAL}, {Aartsen}, {Ackermann}, {Adams},
  {Aguilar}, {Ahlers}, {Ahrens}, {Samarai}, {Altmann}, {Andeen}, {Anderson},
  {Ansseau}, {Anton}, {Arg{\"u}elles}, {Auffenberg}, {Axani}, {Bagherpour},
  {Bai}, {Barron}, {Barwick}, {Baum}, {Bay}, {Beatty}, {Becker Tjus},
  {Bernardini}, {Besson}, {Binder}, {Bindig}, {Blaufuss}, {Blot}, {Bohm},
  {B{\"o}rner}, {Bos}, {Bose}, {B{\"o}ser}, {Botner}, {Bourbeau}, {Bourbeau},
  {Bradascio}, {Braun}, {Brayeur}, {Brenzke}, {Bretz}, {Bron},
  {Brostean-Kaiser}, {Burgman}, {Carver}, {Casey}, {Casier}, {Cheung},
  {Chirkin}, {Christov}, {Clark}, {Classen}, {Coenders}, {Collin}, {Conrad},
  {Cowen}, {Cross}, {Day}, {de Andr{\'e}}, {De Clercq}, {DeLaunay},
  {Dembinski}, {De Ridder}, {Desiati}, {de Vries}, {de Wasseige}, {de With},
  {DeYoung}, {D{\'\i}az-V{\'e}lez}, {di Lorenzo}, {Dujmovic}, {Dumm},
  {Dunkman}, {Dvorak}, {Eberhardt}, {Ehrhardt}, {Eichmann}, {Eller}, {Evenson},
  {Fahey}, {Fazely}, {Felde}, {Filimonov}, {Finley}, {Flis}, {Franckowiak},
  {Friedman}, {Fuchs}, {Gaisser}, {Gallagher}, {Gerhardt}, {Ghorbani}, {Giang},
  {Glauch}, {Gl{\"u}senkamp}, {Goldschmidt}, {Gonzalez}, {Grant}, {Griffith},
  {Haack}, {Hallgren}, {Halzen}, {Hanson}, {Hebecker}, {Heereman}, {Helbing},
  {Hellauer}, {Hickford}, {Hignight}, {Hill}, {Hoffman}, {Hoffmann},
  {Hokanson-Fasig}, {Hoshina}, {Huang}, {Huber}, {Hultqvist}, {H{\"u}nnefeld},
  {In}, {Ishihara}, {Jacobi}, {Japaridze}, {Jeong}, {Jero}, {Jones},
  {Kalaczynski}, {Kang}, {Kappes}, {Karg}, {Karle}, {Kauer}, {Keivani},
  {Kelley}, {Kheirandish}, {Kim}, {Kim}, {Kintscher}, {Kiryluk}, {Kittler},
  {Klein}, {Kohnen}, {Koirala}, {Kolanoski}, {K{\"o}pke}, {Kopper}, {Kopper},
  {Koschinsky}, {Koskinen}, {Kowalski}, {Krings}, {Kroll}, {Kr{\"u}ckl},
  {Kunnen}, {Kunwar}, {Kurahashi}, {Kuwabara}, {Kyriacou}, {Labare},
  {Lanfranchi}, {Larson}, {Lauber}, {Lesiak-Bzdak}, {Leuermann}, {Liu}, {Lu},
  {L{\"u}nemann}, {Luszczak}, {Madsen}, {Maggi}, {Mahn}, {Mancina}, {Maruyama},
  {Mase}, {Maunu}, {McNally}, {Meagher}, {Medici}, {Meier}, {Menne}, {Merino},
  {Meures}, {Miarecki}, {Micallef}, {Moment{\'e}}, {Montaruli}, {Moore},
  {Moulai}, {Nahnhauer}, {Nakarmi}, {Naumann}, {Neer}, {Niederhausen},
  {Nowicki}, {Nygren}, {Obertacke Pollmann}, {Olivas}, {O'Murchadha},
  {Palczewski}, {Pandya}, {Pankova}, {Peiffer}, {Pepper}, {P{\'e}rez de los
  Heros}, {Pieloth}, {Pinat}, {Price}, {Przybylski}, {Raab}, {R{\"a}del},
  {Rameez}, {Rawlins}, {Rea}, {Reimann}, {Relethford}, {Relich}, {Resconi},
  {Rhode}, {Richman}, {Robertson}, {Rongen}, {Rott}, {Ruhe}, {Ryckbosch},
  {Rysewyk}, {S{\"a}lzer}, {Sanchez Herrera}, {Sandrock}, {Sandroos},
  {Santander}, {Sarkar}, {Sarkar}, {Satalecka}, {Schlunder}, {Schmidt},
  {Schneider}, {Schoenen}, {Sch{\"o}neberg}, {Schumacher}, {Seckel},
  {Seunarine}, {Soedingrekso}, {Soldin}, {Song}, {Spiczak}, {Spiering},
  {Stachurska}, {Stamatikos}, {Stanev}, {Stasik}, {Stettner}, {Steuer},
  {Stezelberger}, {Stokstad}, {St{\"o}ssl}, {Strotjohann}, {Stuttard},
  {Sullivan}, {Sutherland}, {Taboada}, {Tatar}, {Tenholt}, {Ter-Antonyan},
  {Terliuk}, {Te{\v{s}}i{\'c}}, {Tilav}, {Toale}, {Tobin}, {Toscano}, {Tosi},
  {Tselengidou}, {Tung}, {Turcati}, {Turley}, {Ty}, {Unger}, {Usner},
  {Vandenbroucke}, {Van Driessche}, {van Eijndhoven}, {Vanheule}, {van Santen},
  {Vehring}, {Vogel}, {Vraeghe}, {Walck}, {Wallace}, {Wallraff}, {Wandler},
  {Wandkowsky}, {Waza}, {Weaver}, {Weiss}, {Wendt}, {Werthebach}, {Whelan},
  {Wiebe}, {Wiebusch}, {Wille}, {Williams}, {Wills}, {Wolf}, {Wood}, {Woolsey},
  {Woschnagg}, {Xu}, {Xu}, {Xu}, {Yanez}, {Yodh}, {Yoshida}, {Yuan}, {Zoll},
  {IceCube Collaboration}, {Balasubramanian}, {Mate}, {Bhalerao},
  {Bhattacharya}, {Vibhute}, {Dewangan}, {Rao}, {Vadawale}, {AstroSat Cadmium
  Zinc Telluride Imager Team}, {Svinkin}, {Hurley}, {Aptekar}, {Frederiks},
  {Golenetskii}, {Kozlova}, {Lysenko}, {Oleynik}, {Tsvetkova}, {Ulanov},
  {Cline}, {IPN Collaboration}, {Li}, {Xiong}, {Zhang}, {Lu}, {Song}, {Cao},
  {Chang}, {Chen}, {Chen}, {Chen}, {Chen}, {Chen}, {Chen}, {Cui}, {Cui},
  {Deng}, {Dong}, {Du}, {Fu}, {Gao}, {Gao}, {Gao}, {Ge}, {Gu}, {Guan}, {Guo},
  {Han}, {Hu}, {Huang}, {Huo}, {Jia}, {Jiang}, {Jiang}, {Jin}, {Jin}, {Li},
  {Li}, {Li}, {Li}, {Li}, {Li}, {Li}, {Li}, {Li}, {Li}, {Li}, {Liang}, {Liao},
  {Liu}, {Liu}, {Liu}, {Liu}, {Liu}, {Liu}, {Liu}, {Lu}, {Lu}, {Luo}, {Ma},
  {Meng}, {Nang}, {Nie}, {Ou}, {Qu}, {Sai}, {Sun}, {Tan}, {Tao}, {Tao}, {Tuo},
  {Wang}, {Wang}, {Wang}, {Wang}, {Wang}, {Wen}, {Wu}, {Wu}, {Xiao}, {Xu},
  {Xu}, {Yan}, {Yang}, {Yang}, {Yang}, {Zhang}, {Zhang}, {Zhang}, {Zhang},
  {Zhang}, {Zhang}, {Zhang}, {Zhang}, {Zhang}, {Zhang}, {Zhang}, {Zhang},
  {Zhang}, {Zhang}, {Zhang}, {Zhang}, {Zhang}, {Zhang}, {Zhao}, {Zhao}, {Zhao},
  {Zheng}, {Zhu}, {Zhu}, {Zou}, {Insight-HXMT Collaboration}, {Albert},
  {Andr{\'e}}, {Anghinolfi}, {Ardid}, {Aubert}, {Aublin}, {Avgitas}, {Baret},
  {Barrios-Mart{\'\i}}, {Basa}, {Belhorma}, {Bertin}, {Biagi}, {Bormuth},
  {Bourret}, {Bouwhuis}, {Br{\^a}nza{\c{s}}}, {Bruijn}, {Brunner}, {Busto},
  {Capone}, {Caramete}, {Carr}, {Celli}, {Cherkaoui El Moursli}, {Chiarusi},
  {Circella}, {Coelho}, {Coleiro}, {Coniglione}, {Costantini}, {Coyle},
  {Creusot}, {D{\'\i}az}, {Deschamps}, {De Bonis}, {Distefano}, {Di Palma},
  {Domi}, {Donzaud}, {Dornic}, {Drouhin}, {Eberl}, {El Bojaddaini}, {El
  Khayati}, {Els{\"a}sser}, {Enzenh{\"o}fer}, {Ettahiri}, {Fassi}, {Felis},
  {Fusco}, {Gay}, {Giordano}, {Glotin}, {Gr{\'e}goire}, {Ruiz}, {Graf},
  {Hallmann}, {van Haren}, {Heijboer}, {Hello}, {Hern{\'a}ndez-Rey},
  {H{\"o}ssl}, {Hofest{\"a}dt}, {Hugon}, {Illuminati}, {James}, {de Jong},
  {Jongen}, {Kadler}, {Kalekin}, {Katz}, {Kiessling}, {Kouchner}, {Kreter},
  {Kreykenbohm}, {Kulikovskiy}, {Lachaud}, {Lahmann}, {Lef{\`e}vre}, {Leonora},
  {Lotze}, {Loucatos}, {Marcelin}, {Margiotta}, {Marinelli},
  {Mart{\'\i}nez-Mora}, {Mele}, {Melis}, {Michael}, {Migliozzi}, {Moussa},
  {Navas}, {Nezri}, {Organokov}, {P{\u{a}}v{\u{a}}la{\c{s}}}, {Pellegrino},
  {Perrina}, {Piattelli}, {Popa}, {Pradier}, {Quinn}, {Racca}, {Riccobene},
  {S{\'a}nchez-Losa}, {Salda{\~n}a}, {Salvadori}, {Samtleben}, {Sanguineti},
  {Sapienza}, {Sieger}, {Spurio}, {Stolarczyk}, {Taiuti}, {Tayalati},
  {Trovato}, {Turpin}, {T{\"o}nnis}, {Vallage}, {Van Elewyck}, {Versari},
  {Vivolo}, {Vizzoca}, {Wilms}, {Zornoza}, {Z{\'u}{\~n}iga}, {ANTARES
  Collaboration}, {Beardmore}, {Breeveld}, {Burrows}, {Cenko}, {Cusumano},
  {D'A{\`\i}}, {de Pasquale}, {Emery}, {Evans}, {Giommi}, {Gronwall}, {Kennea},
  {Krimm}, {Kuin}, {Lien}, {Marshall}, {Melandri}, {Nousek}, {Oates},
  {Osborne}, {Pagani}, {Page}, {Palmer}, {Perri}, {Siegel}, {Sbarufatti},
  {Tagliaferri}, {Tohuvavohu}, {Swift Collaboration}, {Tavani}, {Verrecchia},
  {Bulgarelli}, {Evangelista}, {Pacciani}, {Feroci}, {Pittori}, {Giuliani},
  {Del Monte}, {Donnarumma}, {Argan}, {Trois}, {Ursi}, {Cardillo}, {Piano},
  {Longo}, {Lucarelli}, {Munar-Adrover}, {Fuschino}, {Labanti}, {Marisaldi},
  {Minervini}, {Fioretti}, {Parmiggiani}, {Gianotti}, {Trifoglio}, {Di Persio},
  {Antonelli}, {Barbiellini}, {Caraveo}, {Cattaneo}, {Costa}, {Colafrancesco},
  {D'Amico}, {Ferrari}, {Morselli}, {Paoletti}, {Picozza}, {Pilia}, {Rappoldi},
  {Soffitta}, {Vercellone}, {AGILE Team}, {Foley}, {Coulter}, {Kilpatrick},
  {Drout}, {Piro}, {Shappee}, {Siebert}, {Simon}, {Ulloa}, {Kasen}, {Madore},
  {Murguia-Berthier}, {Pan}, {Prochaska}, {Ramirez-Ruiz}, {Rest},
  {Rojas-Bravo}, {1M2H Team}, {Berger}, {Soares-Santos}, {Annis}, {Alexander},
  {Allam}, {Balbinot}, {Blanchard}, {Brout}, {Butler}, {Chornock}, {Cook},
  {Cowperthwaite}, {Diehl}, {Drlica-Wagner}, {Drout}, {Durret}, {Eftekhari},
  {Finley}, {Fong}, {Frieman}, {Fryer}, {Garc{\'\i}a-Bellido}, {Gruendl},
  {Hartley}, {Herner}, {Kessler}, {Lin}, {Lopes}, {Louren{\c{c}}o}, {Margutti},
  {Marshall}, {Matheson}, {Medina}, {Metzger}, {Mu{\~n}oz}, {Muir}, {Nicholl},
  {Nugent}, {Palmese}, {Paz-Chinch{\'o}n}, {Quataert}, {Sako}, {Sauseda},
  {Schlegel}, {Scolnic}, {Secco}, {Smith}, {Sobreira}, {Villar}, {Vivas},
  {Wester}, {Williams}, {Yanny}, {Zenteno}, {Zhang}, {Abbott}, {Banerji},
  {Bechtol}, {Benoit-L{\'e}vy}, {Bertin}, {Brooks}, {Buckley-Geer}, {Burke},
  {Capozzi}, {Carnero Rosell}, {Carrasco Kind}, {Castander}, {Crocce}, {Cunha},
  {D'Andrea}, {da Costa}, {Davis}, {DePoy}, {Desai}, {Dietrich}, {Eifler},
  {Fernandez}, {Flaugher}, {Fosalba}, {Gaztanaga}, {Gerdes}, {Giannantonio},
  {Goldstein}, {Gruen}, {Gschwend}, {Gutierrez}, {Honscheid}, {James},
  {Jeltema}, {Johnson}, {Johnson}, {Kent}, {Krause}, {Kron}, {Kuehn}, {Lahav},
  {Lima}, {Maia}, {March}, {Martini}, {McMahon}, {Menanteau}, {Miller},
  {Miquel}, {Mohr}, {Nichol}, {Ogando}, {Plazas}, {Romer}, {Roodman}, {Rykoff},
  {Sanchez}, {Scarpine}, {Schindler}, {Schubnell}, {Sevilla-Noarbe}, {Sheldon},
  {Smith}, {Smith}, {Stebbins}, {Suchyta}, {Swanson}, {Tarle}, {Thomas},
  {Troxel}, {Tucker}, {Vikram}, {Walker}, {Wechsler}, {Weller}, {Carlin},
  {Gill}, {Li}, {Marriner}, {Neilsen}, {Dark Energy Camera GW-EM
  Collaboration}, {DES Collaboration}, {Haislip}, {Kouprianov}, {Reichart},
  {Sand}, {Tartaglia}, {Valenti}, {Yang}, {DLT40 Collaboration}, {Benetti},
  {Brocato}, {Campana}, {Cappellaro}, {Covino}, {D'Avanzo}, {D'Elia}, {Getman},
  {Ghirlanda}, {Ghisellini}, {Limatola}, {Nicastro}, {Palazzi}, {Pian},
  {Piranomonte}, {Possenti}, {Rossi}, {Salafia}, {Tomasella}, {Amati},
  {Antonelli}, {Bernardini}, {Bufano}, {Capaccioli}, {Casella}, {Dadina}, {De
  Cesare}, {Di Paola}, {Giuffrida}, {Giunta}, {Israel}, {Lisi}, {Maiorano},
  {Mapelli}, {Masetti}, {Pescalli}, {Pulone}, {Salvaterra}, {Schipani},
  {Spera}, {Stamerra}, {Stella}, {Testa}, {Turatto}, {Vergani}, {Aresu},
  {Bachetti}, {Buffa}, {Burgay}, {Buttu}, {Caria}, {Carretti}, {Casasola},
  {Castangia}, {Carboni}, {Casu}, {Concu}, {Corongiu}, {Deiana}, {Egron},
  {Fara}, {Gaudiomonte}, {Gusai}, {Ladu}, {Loru}, {Leurini}, {Marongiu},
  {Melis}, {Melis}, {Migoni}, {Milia}, {Navarrini}, {Orlati}, {Ortu}, {Palmas},
  {Pellizzoni}, {Perrodin}, {Pisanu}, {Poppi}, {Righini}, {Saba}, {Serra},
  {Serrau}, {Stagni}, {Surcis}, {Vacca}, {Vargiu}, {Hunt}, {Jin}, {Klose},
  {Kouveliotou}, {Mazzali}, {M{\o}ller}, {Nava}, {Piran}, {Selsing}, {Vergani},
  {Wiersema}, {Toma}, {Higgins}, {Mundell}, {di Serego Alighieri}, {G{\'o}tz},
  {Gao}, {Gomboc}, {Kaper}, {Kobayashi}, {Kopac}, {Mao}, {Starling}, {Steele},
  {van der Horst}, {GRAWITA: GRAvitational Wave Inaf TeAm}, {Acero}, {Atwood},
  {Baldini}, {Barbiellini}, {Bastieri}, {Berenji}, {Bellazzini}, {Bissaldi},
  {Blandford}, {Bloom}, {Bonino}, {Bottacini}, {Bregeon}, {Buehler}, {Buson},
  {Cameron}, {Caputo}, {Caraveo}, {Cavazzuti}, {Chekhtman}, {Cheung}, {Chiang},
  {Ciprini}, {Cohen-Tanugi}, {Cominsky}, {Costantin}, {Cuoco}, {D'Ammando}, {de
  Palma}, {Digel}, {Di Lalla}, {Di Mauro}, {Di Venere}, {Dubois}, {Fegan},
  {Focke}, {Franckowiak}, {Fukazawa}, {Funk}, {Fusco}, {Gargano}, {Gasparrini},
  {Giglietto}, {Giordano}, {Giroletti}, {Glanzman}, {Green}, {Grondin},
  {Guillemot}, {Guiriec}, {Harding}, {Horan}, {J{\'o}hannesson}, {Kamae},
  {Kensei}, {Kuss}, {La Mura}, {Latronico}, {Lemoine-Goumard}, {Longo},
  {Loparco}, {Lovellette}, {Lubrano}, {Magill}, {Maldera}, {Manfreda},
  {Mazziotta}, {McEnery}, {Meyer}, {Michelson}, {Mirabal}, {Monzani},
  {Moretti}, {Morselli}, {Moskalenko}, {Negro}, {Nuss}, {Ojha}, {Omodei},
  {Orienti}, {Orlando}, {Palatiello}, {Paliya}, {Paneque}, {Pesce-Rollins},
  {Piron}, {Porter}, {Principe}, {Rain{\`o}}, {Rando}, {Razzano}, {Razzaque},
  {Reimer}, {Reimer}, {Reposeur}, {Rochester}, {Saz Parkinson}, {Sgr{\`o}},
  {Siskind}, {Spada}, {Spandre}, {Suson}, {Takahashi}, {Tanaka}, {Thayer},
  {Thayer}, {Thompson}, {Tibaldo}, {Torres}, {Torresi}, {Troja}, {Venters},
  {Vianello}, {Zaharijas}, {Fermi Large Area Telescope Collaboration},
  {Allison}, {Bannister}, {Dobie}, {Kaplan}, {Lenc}, {Lynch}, {Murphy},
  {Sadler}, {Australia Telescope Compact Array}, {Hotan}, {James}, {Oslowski},
  {Raja}, {Shannon}, {Whiting}, {Australian SKA Pathfinder}, {Arcavi},
  {Howell}, {McCully}, {Hosseinzadeh}, {Hiramatsu}, {Poznanski}, {Barnes},
  {Zaltzman}, {Vasylyev}, {Maoz}, {Las Cumbres Observatory Group}, {Cooke},
  {Bailes}, {Wolf}, {Deller}, {Lidman}, {Wang}, {Gendre}, {Andreoni}, {Ackley},
  {Pritchard}, {Bessell}, {Chang}, {M{\"o}ller}, {Onken}, {Scalzo},
  {Ridden-Harper}, {Sharp}, {Tucker}, {Farrell}, {Elmer}, {Johnston},
  {Venkatraman Krishnan}, {Keane}, {Green}, {Jameson}, {Hu}, {Ma}, {Sun}, {Wu},
  {Wang}, {Shang}, {Hu}, {Ashley}, {Yuan}, {Li}, {Tao}, {Zhu}, {Zhang},
  {Suntzeff}, {Zhou}, {Yang}, {Orange}, {Morris}, {Cucchiara}, {Giblin},
  {Klotz}, {Staff}, {Thierry}, {Schmidt}, {OzGrav}, {(Deeper}, {Wider},
  {program}, {AST3}, {CAASTRO Collaborations}, {Tanvir}, {Levan}, {Cano}, {de
  Ugarte-Postigo}, {Gonz{\'a}lez-Fern{\'a}ndez}, {Greiner}, {Hjorth}, {Irwin},
  {Kr{\"u}hler}, {Mandel}, {Milvang-Jensen}, {O'Brien}, {Rol}, {Rosetti},
  {Rosswog}, {Rowlinson}, {Steeghs}, {Th{\"o}ne}, {Ulaczyk}, {Watson}, {Bruun},
  {Cutter}, {Figuera Jaimes}, {Fujii}, {Fruchter}, {Gompertz}, {Jakobsson},
  {Hodosan}, {J{\`e}rgensen}, {Kangas}, {Kann}, {Rabus}, {Schr{\o}der},
  {Stanway}, {Wijers}, {VINROUGE Collaboration}, {Lipunov}, {Gorbovskoy},
  {Kornilov}, {Tyurina}, {Balanutsa}, {Kuznetsov}, {Vlasenko}, {Podesta},
  {Lopez}, {Podesta}, {Levato}, {Saffe}, {Mallamaci}, {Budnev}, {Gress},
  {Kuvshinov}, {Gorbunov}, {Vladimirov}, {Zimnukhov}, {Gabovich}, {Yurkov},
  {Sergienko}, {Rebolo}, {Serra-Ricart}, {Tlatov}, {Ishmuhametova}, {MASTER
  Collaboration}, {Abe}, {Aoki}, {Aoki}, {Asakura}, {Baar}, {Barway}, {Bond},
  {Doi}, {Finet}, {Fujiyoshi}, {Furusawa}, {Honda}, {Itoh}, {Kanda},
  {Kawabata}, {Kawabata}, {Kim}, {Koshida}, {Kuroda}, {Lee}, {Liu},
  {Matsubayashi}, {Miyazaki}, {Morihana}, {Morokuma}, {Motohara}, {Murata},
  {Nagai}, {Nagashima}, {Nagayama}, {Nakaoka}, {Nakata}, {Ohsawa}, {Ohshima},
  {Ohta}, {Okita}, {Saito}, {Saito}, {Sako}, {Sekiguchi}, {Sumi}, {Tajitsu},
  {Takahashi}, {Takayama}, {Tamura}, {Tanaka}, {Tanaka}, {Terai}, {Tominaga},
  {Tristram}, {Uemura}, {Utsumi}, {Yamaguchi}, {Yasuda}, {Yoshida}, {Zenko},
  {J-GEM}, {Adams}, {Anupama}, {Bally}, {Barway}, {Bellm}, {Blagorodnova},
  {Cannella}, {Chandra}, {Chatterjee}, {Clarke}, {Cobb}, {Cook}, {Copperwheat},
  {De}, {Emery}, {Feindt}, {Foster}, {Fox}, {Frail}, {Fremling}, {Frohmaier},
  {Garcia}, {Ghosh}, {Giacintucci}, {Goobar}, {Gottlieb}, {Grefenstette},
  {Hallinan}, {Harrison}, {Heida}, {Helou}, {Ho}, {Horesh}, {Hotokezaka}, {Ip},
  {Itoh}, {Jacobs}, {Jencson}, {Kasen}, {Kasliwal}, {Kassim}, {Kim}, {Kiran},
  {Kuin}, {Kulkarni}, {Kupfer}, {Lau}, {Madsen}, {Mazzali}, {Miller},
  {Miyasaka}, {Mooley}, {Myers}, {Nakar}, {Ngeow}, {Nugent}, {Ofek},
  {Palliyaguru}, {Pavana}, {Perley}, {Peters}, {Pike}, {Piran}, {Qi}, {Quimby},
  {Rana}, {Rosswog}, {Rusu}, {Sadler}, {Van Sistine}, {Sollerman}, {Xu}, {Yan},
  {Yatsu}, {Yu}, {Zhang}, {Zhao}, {GROWTH}, {JAGWAR}, {Caltech-NRAO},
  {TTU-NRAO}, {NuSTAR Collaborations}, {Chambers}, {Huber}, {Schultz},
  {Bulger}, {Flewelling}, {Magnier}, {Lowe}, {Wainscoat}, {Waters}, {Willman},
  {Pan-STARRS}, {Ebisawa}, {Hanyu}, {Harita}, {Hashimoto}, {Hidaka}, {Hori},
  {Ishikawa}, {Isobe}, {Iwakiri}, {Kawai}, {Kawai}, {Kawamuro}, {Kawase},
  {Kitaoka}, {Makishima}, {Matsuoka}, {Mihara}, {Morita}, {Morita}, {Nakahira},
  {Nakajima}, {Nakamura}, {Negoro}, {Oda}, {Sakamaki}, {Sasaki}, {Serino},
  {Shidatsu}, {Shimomukai}, {Sugawara}, {Sugita}, {Sugizaki}, {Tachibana},
  {Takao}, {Tanimoto}, {Tomida}, {Tsuboi}, {Tsunemi}, {Ueda}, {Ueno}, {Yamada},
  {Yamaoka}, {Yamauchi}, {Yatabe}, {Yoneyama}, {Yoshii}, {MAXI Team}, {Coward},
  {Crisp}, {Macpherson}, {Andreoni}, {Laugier}, {Noysena}, {Klotz}, {Gendre},
  {Thierry}, {Turpin}, {Consortium}, {Im}, {Choi}, {Kim}, {Yoon}, {Lim}, {Lee},
  {Lee}, {Kim}, {Ko}, {Joe}, {Kwon}, {Kim}, {Lim}, {Choi}, {KU Collaboration},
  {Fynbo}, {Malesani}, {Xu}, {Optical Telescope}, {Smartt}, {Jerkstrand},
  {Kankare}, {Sim}, {Fraser}, {Inserra}, {Maguire}, {Leloudas}, {Magee},
  {Shingles}, {Smith}, {Young}, {Kotak}, {Gal-Yam}, {Lyman}, {Homan},
  {Agliozzo}, {Anderson}, {Angus}, {Ashall}, {Barbarino}, {Bauer}, {Berton},
  {Botticella}, {Bulla}, {Cannizzaro}, {Cartier}, {Cikota}, {Clark}, {De Cia},
  {Della Valle}, {Dennefeld}, {Dessart}, {Dimitriadis}, {Elias-Rosa}, {Firth},
  {Fl{\"o}rs}, {Frohmaier}, {Galbany}, {Gonz{\'a}lez-Gait{\'a}n}, {Gromadzki},
  {Guti{\'e}rrez}, {Hamanowicz}, {Harmanen}, {Heintz}, {Hernandez}, {Hodgkin},
  {Hook}, {Izzo}, {James}, {Jonker}, {Kerzendorf}, {Kostrzewa-Rutkowska},
  {Kromer}, {Kuncarayakti}, {Lawrence}, {Manulis}, {Mattila}, {McBrien},
  {M{\"u}ller}, {Nordin}, {O'Neill}, {Onori}, {Palmerio}, {Pastorello},
  {Patat}, {Pignata}, {Podsiadlowski}, {Razza}, {Reynolds}, {Roy}, {Ruiter},
  {Rybicki}, {Salmon}, {Pumo}, {Prentice}, {Seitenzahl}, {Smith}, {Sollerman},
  {Sullivan}, {Szegedi}, {Taddia}, {Taubenberger}, {Terreran}, {Van Soelen},
  {Vos}, {Walton}, {Wright}, {Wyrzykowski}, {Yaron}, {pre=''(''>ePESSTO},
  {Chen}, {Kr{\"u}hler}, {Schady}, {Wiseman}, {Greiner}, {Rau}, {Schweyer},
  {Klose}, {Nicuesa Guelbenzu}, {GROND}, {Palliyaguru}, {Tech University},
  {Shara}, {Williams}, {Vaisanen}, {Potter}, {Romero Colmenero}, {Crawford},
  {Buckley}, {Mao}, {SALT Group}, {D{\'\i}az}, {Macri}, {Garc{\'\i}a Lambas},
  {Mendes de Oliveira}, {Nilo Castell{\'o}n}, {Ribeiro}, {S{\'a}nchez},
  {Schoenell}, {Abramo}, {Akras}, {Alcaniz}, {Artola}, {Beroiz}, {Bonoli},
  {Cabral}, {Camuccio}, {Chavushyan}, {Coelho}, {Colazo}, {Costa-Duarte},
  {Cuevas Larenas}, {Dom{\'\i}nguez Romero}, {Dultzin}, {Fern{\'a}ndez},
  {Garc{\'\i}a}, {Girardini}, {Gon{\c{c}}alves}, {Gon{\c{c}}alves}, {Gurovich},
  {Jim{\'e}nez-Teja}, {Kanaan}, {Lares}, {Lopes de Oliveira}, {L{\'o}pez-Cruz},
  {Melia}, {Molino}, {Padilla}, {Pe{\~n}uela}, {Placco}, {Qui{\~n}ones},
  {Ram{\'\i}rez Rivera}, {Renzi}, {Riguccini}, {R{\'\i}os-L{\'o}pez},
  {Rodriguez}, {Sampedro}, {Schneiter}, {Sodr{\'e}}, {Starck}, {Torres-Flores},
  {Tornatore}, {Zadro{\.z}ny}, {Castillo}, {TOROS: Transient Robotic
  Observatory of South Collaboration}, {Castro-Tirado}, {Tello}, {Hu}, {Zhang},
  {Cunniffe}, {Castell{\'o}n}, {Hiriart}, {Caballero-Garc{\'\i}a},
  {Jel{\'\i}nek}, {Kub{\'a}nek}, {P{\'e}rez del Pulgar}, {Park}, {Jeong},
  {Castro Cer{\'o}n}, {Pandey}, {Yock}, {Querel}, {Fan}, {Wang}, {BOOTES
  Collaboration}, {Beardsley}, {Brown}, {Crosse}, {Emrich}, {Franzen},
  {Gaensler}, {Horsley}, {Johnston-Hollitt}, {Kenney}, {Morales}, {Pallot},
  {Sokolowski}, {Steele}, {Tingay}, {Trott}, {Walker}, {Wayth}, {Williams},
  {Wu}, {Murchison Widefield Array}, {Yoshida}, {Sakamoto}, {Kawakubo},
  {Yamaoka}, {Takahashi}, {Asaoka}, {Ozawa}, {Torii}, {Shimizu}, {Tamura},
  {Ishizaki}, {Cherry}, {Ricciarini}, {Penacchioni}, {Marrocchesi}, {CALET
  Collaboration}, {Pozanenko}, {Volnova}, {Mazaeva}, {Minaev}, {Krugov},
  {Kusakin}, {Reva}, {Moskvitin}, {Rumyantsev}, {Inasaridze}, {Klunko},
  {Tungalag}, {Schmalz}, {Burhonov}, {IKI-GW Follow-up Collaboration},
  {Abdalla}, {Abramowski}, {Aharonian}, {Ait Benkhali}, {Ang{\"u}ner},
  {Arakawa}, {Arrieta}, {Aubert}, {Backes}, {Balzer}, {Barnard}, {Becherini},
  {Becker Tjus}, {Berge}, {Bernhard}, {Bernl{\"o}hr}, {Blackwell},
  {B{\"o}ttcher}, {Boisson}, {Bolmont}, {Bonnefoy}, {Bordas}, {Bregeon},
  {Brun}, {Brun}, {Bryan}, {B{\"u}chele}, {Bulik}, {Capasso}, {Caroff},
  {Carosi}, {Casanova}, {Cerruti}, {Chakraborty}, {Chaves}, {Chen},
  {Chevalier}, {Colafrancesco}, {Condon}, {Conrad}, {Davids}, {Decock}, {Deil},
  {Devin}, {deWilt}, {Dirson}, {Djannati-Ata{\"\i}}, {Donath}, {O'C. Drury},
  {Dutson}, {Dyks}, {Edwards}, {Egberts}, {Emery}, {Ernenwein}, {Eschbach},
  {Farnier}, {Fegan}, {Fernandes}, {Fiasson}, {Fontaine}, {Funk},
  {F{\"u}ssling}, {Gabici}, {Gallant}, {Garrigoux}, {Gat{\'e}}, {Giavitto},
  {Giebels}, {Glawion}, {Glicenstein}, {Gottschall}, {Grondin}, {Hahn},
  {Haupt}, {Hawkes}, {Heinzelmann}, {Henri}, {Hermann}, {Hinton}, {Hofmann},
  {Hoischen}, {Holch}, {Holler}, {Horns}, {Ivascenko}, {Iwasaki},
  {Jacholkowska}, {Jamrozy}, {Jankowsky}, {Jankowsky}, {Jingo}, {Jouvin},
  {Jung-Richardt}, {Kastendieck}, {Katarzy{\'n}ski}, {Katsuragawa},
  {Kerszberg}, {Khangulyan}, {Kh{\'e}lifi}, {King}, {Klepser}, {Klochkov},
  {Klu{\'z}niak}, {Komin}, {Kosack}, {Krakau}, {Kraus}, {Kr{\"u}ger}, {Laffon},
  {Lamanna}, {Lau}, {Lees}, {Lefaucheur}, {Lemi{\`e}re}, {Lemoine-Goumard},
  {Lenain}, {Leser}, {Lohse}, {Lorentz}, {Liu}, {Lypova}, {Malyshev},
  {Marandon}, {Marcowith}, {Mariaud}, {Marx}, {Maurin}, {Maxted}, {Mayer},
  {Meintjes}, {Meyer}, {Mitchell}, {Moderski}, {Mohamed}, {Mohrmann},
  {Mor{\r{a}}}, {Moulin}, {Murach}, {Nakashima}, {de Naurois}, {Ndiyavala},
  {Niederwanger}, {Niemiec}, {Oakes}, {O'Brien}, {Odaka}, {Ohm}, {Ostrowski},
  {Oya}, {Padovani}, {Panter}, {Parsons}, {Pekeur}, {Pelletier}, {Perennes},
  {Petrucci}, {Peyaud}, {Piel}, {Pita}, {Poireau}, {Poon}, {Prokhorov},
  {Prokoph}, {P{\"u}hlhofer}, {Punch}, {Quirrenbach}, {Raab}, {Rauth},
  {Reimer}, {Reimer}, {Renaud}, {de los Reyes}, {Rieger}, {Rinchiuso},
  {Romoli}, {Rowell}, {Rudak}, {Rulten}, {Sahakian}, {Saito}, {Sanchez},
  {Santangelo}, {Sasaki}, {Schlickeiser}, {Sch{\"u}ssler}, {Schulz},
  {Schwanke}, {Schwemmer}, {Seglar-Arroyo}, {Settimo}, {Seyffert}, {Shafi},
  {Shilon}, {Shiningayamwe}, {Simoni}, {Sol}, {Spanier}, {Spir-Jacob},
  {Stawarz}, {Steenkamp}, {Stegmann}, {Steppa}, {Sushch}, {Takahashi},
  {Tavernet}, {Tavernier}, {Taylor}, {Terrier}, {Tibaldo}, {Tiziani},
  {Tluczykont}, {Trichard}, {Tsirou}, {Tsuji}, {Tuffs}, {Uchiyama}, {van der
  Walt}, {van Eldik}, {van Rensburg}, {van Soelen}, {Vasileiadis}, {Veh},
  {Venter}, {Viana}, {Vincent}, {Vink}, {Voisin}, {V{\"o}lk}, {Vuillaume},
  {Wadiasingh}, {Wagner}, {Wagner}, {Wagner}, {White}, {Wierzcholska},
  {Willmann}, {W{\"o}rnlein}, {Wouters}, {Yang}, {Zaborov}, {Zacharias},
  {Zanin}, {Zdziarski}, {Zech}, {Zefi}, {Ziegler}, {Zorn}, {{\.Z}ywucka},
  {H.~E.~S.~S. Collaboration}, {Fender}, {Broderick}, {Rowlinson}, {Wijers},
  {Stewart}, {ter Veen}, {Shulevski}, {LOFAR Collaboration}, {Kavic},
  {Simonetti}, {League}, {Tsai}, {Obenberger}, {Nathaniel}, {Taylor}, {Dowell},
  {Liebling}, {Estes}, {Lippert}, {Sharma}, {Vincent}, {Farella}, {Wavelength
  Array}, {Abeysekara}, {Albert}, {Alfaro}, {Alvarez}, {Arceo},
  {Arteaga-Vel{\'a}zquez}, {Avila Rojas}, {Ayala Solares}, {Barber}, {Becerra
  Gonzalez}, {Becerril}, {Belmont-Moreno}, {BenZvi}, {Berley}, {Bernal},
  {Braun}, {Brisbois}, {Caballero-Mora}, {Capistr{\'a}n}, {Carrami{\~n}ana},
  {Casanova}, {Castillo}, {Cotti}, {Cotzomi}, {Couti{\~n}o de Le{\'o}n}, {De
  Le{\'o}n}, {De la Fuente}, {Diaz Hernandez}, {Dichiara}, {Dingus},
  {DuVernois}, {D{\'\i}az-V{\'e}lez}, {Ellsworth}, {Engel},
  {Enr{\'\i}quez-Rivera}, {Fiorino}, {Fleischhack}, {Fraija},
  {Garc{\'\i}a-Gonz{\'a}lez}, {Garfias}, {Gerhardt}, {Gonz{\~o}lez Mu{\~n}oz},
  {Gonz{\'a}lez}, {Goodman}, {Hampel-Arias}, {Harding}, {Hernandez},
  {Hernandez-Almada}, {Hona}, {H{\"u}ntemeyer}, {Iriarte}, {Jardin-Blicq},
  {Joshi}, {Kaufmann}, {Kieda}, {Lara}, {Lauer}, {Lennarz}, {Le{\'o}n Vargas},
  {Linnemann}, {Longinotti}, {Raya}, {Luna-Garc{\'\i}a}, {L{\'o}pez-Coto},
  {Malone}, {Marinelli}, {Martinez}, {Martinez-Castellanos},
  {Mart{\'\i}nez-Castro}, {Mart{\'\i}nez-Huerta}, {Matthews},
  {Miranda-Romagnoli}, {Moreno}, {Mostaf{\'a}}, {Nellen}, {Newbold}, {Nisa},
  {Noriega-Papaqui}, {Pelayo}, {Pretz}, {P{\'e}rez-P{\'e}rez}, {Ren}, {Rho},
  {Rivi{\`e}re}, {Rosa-Gonz{\'a}lez}, {Rosenberg}, {Ruiz-Velasco}, {Salazar},
  {Salesa Greus}, {Sandoval}, {Schneider}, {Schoorlemmer}, {Sinnis}, {Smith},
  {Springer}, {Surajbali}, {Tibolla}, {Tollefson}, {Torres}, {Ukwatta},
  {Weisgarber}, {Westerhoff}, {Wisher}, {Wood}, {Yapici}, {Yodh}, {Younk},
  {Zhou}, {{\'A}lvarez}, {HAWC Collaboration}, {Aab}, {Abreu}, {Aglietta},
  {Albuquerque}, {Albury}, {Allekotte}, {Almela}, {Alvarez Castillo},
  {Alvarez-Mu{\~n}iz}, {Anastasi}, {Anchordoqui}, {Andrada}, {Andringa},
  {Aramo}, {Arsene}, {Asorey}, {Assis}, {Avila}, {Badescu}, {Balaceanu},
  {Barbato}, {Barreira Luz}, {Becker}, {Bellido}, {Berat}, {Bertaina},
  {Bertou}, {Biermann}, {Biteau}, {Blaess}, {Blanco}, {Blazek}, {Bleve},
  {Boh{\'a}{\v{c}}ov{\'a}}, {Bonifazi}, {Borodai}, {Botti}, {Brack}, {Brancus},
  {Bretz}, {Bridgeman}, {Briechle}, {Buchholz}, {Bueno}, {Buitink}, {Buscemi},
  {Caballero-Mora}, {Caccianiga}, {Cancio}, {Canfora}, {Caruso}, {Castellina},
  {Catalani}, {Cataldi}, {Cazon}, {Chavez}, {Chinellato}, {Chudoba}, {Clay},
  {Cobos Cerutti}, {Colalillo}, {Coleman}, {Collica}, {Coluccia},
  {Concei{\c{c}}{\~a}o}, {Consolati}, {Contreras}, {Cooper}, {Coutu},
  {Covault}, {Cronin}, {D'Amico}, {Daniel}, {Dasso}, {Daumiller}, {Dawson},
  {Day}, {de Almeida}, {de Jong}, {De Mauro}, {de Mello Neto}, {De Mitri}, {de
  Oliveira}, {de Souza}, {Debatin}, {Deligny}, {D{\'\i}az Castro}, {Diogo},
  {Dobrigkeit}, {D'Olivo}, {Dorosti}, {Dos Anjos}, {Dova}, {Dundovic}, {Ebr},
  {Engel}, {Erdmann}, {Erfani}, {Escobar}, {Espadanal}, {Etchegoyen}, {Falcke},
  {Farmer}, {Farrar}, {Fauth}, {Fazzini}, {Feldbusch}, {Fenu}, {Fick},
  {Figueira}, {Filip{\v{c}}i{\v{c}}}, {Freire}, {Fujii}, {Fuster},
  {Ga{\"\i}or}, {Garc{\'\i}a}, {Gat{\'e}}, {Gemmeke}, {Gherghel-Lascu}, {Ghia},
  {Giaccari}, {Giammarchi}, {Giller}, {G{\l}as}, {Glaser}, {Golup}, {G{\'o}mez
  Berisso}, {G{\'o}mez Vitale}, {Gonz{\'a}lez}, {Gorgi}, {Gottowik}, {Grillo},
  {Grubb}, {Guarino}, {Guedes}, {Halliday}, {Hampel}, {Hansen}, {Harari},
  {Harrison}, {Harvey}, {Haungs}, {Hebbeker}, {Heck}, {Heimann}, {Herve},
  {Hill}, {Hojvat}, {Holt}, {Homola}, {H{\"o}randel}, {Horvath},
  {Hrabovsk{\'y}}, {Huege}, {Hulsman}, {Insolia}, {Isar}, {Jandt}, {Johnsen},
  {Josebachuili}, {Jurysek}, {K{\"a}{\"a}p{\"a}}, {Kampert}, {Keilhauer},
  {Kemmerich}, {Kemp}, {Kieckhafer}, {Klages}, {Kleifges}, {Kleinfeller},
  {Krause}, {Krohm}, {Kuempel}, {Kukec Mezek}, {Kunka}, {Kuotb Awad}, {Lago},
  {LaHurd}, {Lang}, {Lauscher}, {Legumina}, {Leigui de Oliveira},
  {Letessier-Selvon}, {Lhenry-Yvon}, {Link}, {Lo Presti}, {Lopes}, {L{\'o}pez},
  {L{\'o}pez Casado}, {Lorek}, {Luce}, {Lucero}, {Malacari}, {Mallamaci},
  {Mandat}, {Mantsch}, {Mariazzi}, {Maris}, {Marsella}, {Martello}, {Martinez},
  {Mart{\'\i}nez Bravo}, {Mas{\'\i}as Meza}, {Mathes}, {Mathys}, {Matthews},
  {Matthiae}, {Mayotte}, {Mazur}, {Medina}, {Medina-Tanco}, {Melo},
  {Menshikov}, {Merenda}, {Michal}, {Micheletti}, {Middendorf}, {Miramonti},
  {Mitrica}, {Mockler}, {Mollerach}, {Montanet}, {Morello}, {Morlino},
  {M{\"u}ller}, {M{\"u}ller}, {Muller}, {M{\"u}ller}, {Mussa}, {Naranjo},
  {Nguyen}, {Niculescu-Oglinzanu}, {Niechciol}, {Niemietz}, {Niggemann},
  {Nitz}, {Nosek}, {Novotny}, {No{\v{z}}ka}, {N{\'u}{\~n}ez}, {Oikonomou},
  {Olinto}, {Palatka}, {Pallotta}, {Papenbreer}, {Parente}, {Parra}, {Paul},
  {Pech}, {Pedreira}, {P{\c{e}}kala}, {Pe{\~n}a-Rodriguez}, {Pereira},
  {Perlin}, {Perrone}, {Peters}, {Petrera}, {Phuntsok}, {Pierog}, {Pimenta},
  {Pirronello}, {Platino}, {Plum}, {Poh}, {Porowski}, {Prado}, {Privitera},
  {Prouza}, {Quel}, {Querchfeld}, {Quinn}, {Ramos-Pollan}, {Rautenberg},
  {Ravignani}, {Ridky}, {Riehn}, {Risse}, {Ristori}, {Rizi}, {Rodrigues de
  Carvalho}, {Rodriguez Fernandez}, {Rodriguez Rojo}, {Roncoroni}, {Roth},
  {Roulet}, {Rovero}, {Ruehl}, {Saffi}, {Saftoiu}, {Salamida}, {Salazar},
  {Saleh}, {Salina}, {S{\'a}nchez}, {Sanchez-Lucas}, {Santos}, {Santos},
  {Sarazin}, {Sarmento}, {Sarmiento-Cano}, {Sato}, {Schauer}, {Scherini},
  {Schieler}, {Schimp}, {Schmidt}, {Scholten}, {Schov{\'a}nek}, {Schr{\"o}der},
  {Schr{\"o}der}, {Schulz}, {Schumacher}, {Sciutto}, {Segreto}, {Shadkam},
  {Shellard}, {Sigl}, {Silli}, {{\v{S}}m{\'\i}da}, {Snow}, {Sommers},
  {Sonntag}, {Soriano}, {Squartini}, {Stanca}, {Stani{\v{c}}}, {Stasielak},
  {Stassi}, {Stolpovskiy}, {Strafella}, {Streich}, {Suarez},
  {Suarez-Dur{\'a}n}, {Sudholz}, {Suomij{\"a}rvi}, {Supanitsky},
  {{\v{S}}up{\'\i}k}, {Swain}, {Szadkowski}, {Taboada}, {Taborda},
  {Timmermans}, {Todero Peixoto}, {Tomankova}, {Tom{\'e}}, {Torralba Elipe},
  {Travnicek}, {Trini}, {Tueros}, {Ulrich}, {Unger}, {Urban}, {Vald{\'e}s
  Galicia}, {Vali{\~n}o}, {Valore}, {van Aar}, {van Bodegom}, {van den Berg},
  {van Vliet}, {Varela}, {Vargas C{\'a}rdenas}, {V{\'a}zquez}, {Veberi{\v{c}}},
  {Ventura}, {Vergara Quispe}, {Verzi}, {Vicha}, {Villase{\~n}or}, {Vorobiov},
  {Wahlberg}, {Wainberg}, {Walz}, {Watson}, {Weber}, {Weindl}, {Wiede{\'n}ski},
  {Wiencke}, {Wilczy{\'n}ski}, {Wirtz}, {Wittkowski}, {Wundheiler}, {Yang},
  {Yushkov}, {Zas}, {Zavrtanik}, {Zavrtanik}, {Zepeda}, {Zimmermann},
  {Ziolkowski}, {Zong}, {Zuccarello}, {Pierre Auger Collaboration}, {Kim},
  {Schulze}, {Bauer}, {Corral-Santana}, {de Gregorio-Monsalvo},
  {Gonz{\'a}lez-L{\'o}pez}, {Hartmann}, {Ishwara-Chandra}, {Mart{\'\i}n},
  {Mehner}, {Misra}, {Micha{\l}owski}, {Resmi}, {ALMA Collaboration}, {Paragi},
  {Agudo}, {An}, {Beswick}, {Casadio}, {Frey}, {Jonker}, {Kettenis}, {Marcote},
  {Moldon}, {Szomoru}, {van Langevelde}, {Yang}, {Euro VLBI Team}, {Cwiek},
  {Cwiok}, {Czyrkowski}, {Dabrowski}, {Kasprowicz}, {Mankiewicz}, {Nawrocki},
  {Opiela}, {Piotrowski}, {Wrochna}, {Zaremba}, {{\.Z}arnecki}, {Pi of Sky
  Collaboration}, {Haggard}, {Nynka}, {Ruan}, {Chandra Team at McGill
  University}, {Bland}, {Booler}, {Devillepoix}, {de Gois}, {Hancock}, {Howie},
  {Paxman}, {Sansom}, {Towner}, {Desert Fireball Network}, {Tonry}, {Coughlin},
  {Stubbs}, {Denneau}, {Heinze}, {Stalder}, {Weiland}, {ATLAS}, {Eatough},
  {Kramer}, {Kraus}, {Time Resolution Universe Survey}, {Troja}, {Piro},
  {Becerra Gonz{\'a}lez}, {Butler}, {Fox}, {Khandrika}, {Kutyrev}, {Lee},
  {Ricci}, {Ryan}, {S{\'a}nchez-Ram{\'\i}rez}, {Veilleux}, {Watson},
  {Wieringa}, {Burgess}, {van Eerten}, {Fontes}, {Fryer}, {Korobkin},
  {Wollaeger}, {RIMAS}, {RATIR}, {Camilo}, {Foley}, {Goedhart}, {Makhathini},
  {Oozeer}, {Smirnov}, {Fender}, {Woudt}, \& {South
  Africa/MeerKAT}}]{2017ApJ...848L..12A}
---. 2017, \apjl, 848, L12, \dodoi{10.3847/2041-8213/aa91c9}

\bibitem[{{Ackermann} {et~al.}(2022){Ackermann}, {Agarwalla},
  {Alvarez-Mu{\~n}iz}, {Arg{\"u}elles}, {Bustamante}, {Clark}, {Cummings},
  {Decoene}, {Denton}, {Dornic}, {Dzhilkibaev}, {Farzan}, {Garcia}, {Vittoria
  Garzelli}, {Glaser}, {Heijboer}, {H{\"o}randel}, {Illuminati}, {Jeong},
  {Kelley}, {Kelly}, {Kheirandish}, {Klein}, {Krizmanic}, {Larson}, {Lu},
  {Murase}, {Narang}, {Otte}, {Prechelt}, {Prohira}, {Hall Reno}, {Resconi},
  {Santander}, {Vasil'evna Suvorova}, {Vandenbroucke}, {Wiencke}, {Wissel},
  {Yoshida}, {Yuan}, {Zas}, {Zhelnin}, \& {Zhou}}]{2022arXiv220308096A}
{Ackermann}, M., {Agarwalla}, S.~K., {Alvarez-Mu{\~n}iz}, J., {et~al.} 2022,
  arXiv e-prints, arXiv:2203.08096.
\newblock \doarXiv{2203.08096}

\bibitem[{{Adri{\'a}n-Mart{\'\i}nez} {et~al.}(2016){Adri{\'a}n-Mart{\'\i}nez},
  {Ageron}, {Aharonian}, {Aiello}, {Albert}, {Ameli}, {Anassontzis}, {Andre},
  {Androulakis}, {Anghinolfi}, {Anton}, {Ardid}, {Avgitas}, {Barbarino},
  {Barbarito}, {Baret}, {Barrios-Mart{\'\i}}, {Belhorma}, {Belias}, {Berbee},
  {van den Berg}, {Bertin}, {Beurthey}, {van Beveren}, {Beverini}, {Biagi},
  {Biagioni}, {Billault}, {Bond{\`\i}}, {Bormuth}, {Bouhadef}, {Bourlis},
  {Bourret}, {Boutonnet}, {Bouwhuis}, {Bozza}, {Bruijn}, {Brunner}, {Buis},
  {Busto}, {Cacopardo}, {Caillat}, {Calamai}, {Calvo}, {Capone}, {Caramete},
  {Cecchini}, {Celli}, {Champion}, {Cherkaoui El Moursli}, {Cherubini},
  {Chiarusi}, {Circella}, {Classen}, {Cocimano}, {Coelho}, {Coleiro},
  {Colonges}, {Coniglione}, {Cordelli}, {Cosquer}, {Coyle}, {Creusot},
  {Cuttone}, {D'Amico}, {De Bonis}, {De Rosa}, {De Sio}, {Di Capua}, {Di
  Palma}, {D{\'\i}az Garc{\'\i}a}, {Distefano}, {Donzaud}, {Dornic},
  {Dorosti-Hasankiadeh}, {Drakopoulou}, {Drouhin}, {Drury}, {Durocher},
  {Eberl}, {Eichie}, {van Eijk}, {El Bojaddaini}, {El Khayati}, {Elsaesser},
  {Enzenh{\"o}fer}, {Fassi}, {Favali}, {Fermani}, {Ferrara}, {Filippidis},
  {Frascadore}, {Fusco}, {Gal}, {Galat{\`a}}, {Garufi}, {Gay}, {Gebyehu},
  {Giordano}, {Gizani}, {Gracia}, {Graf}, {Gr{\'e}goire}, {Grella}, {Habel},
  {Hallmann}, {van Haren}, {Harissopulos}, {Heid}, {Heijboer}, {Heine},
  {Henry}, {Hern{\'a}ndez-Rey}, {Hevinga}, {Hofest{\"a}dt}, {Hugon},
  {Illuminati}, {James}, {Jansweijer}, {Jongen}, {de Jong}, {Kadler},
  {Kalekin}, {Kappes}, {Katz}, {Keller}, {Kieft}, {Kie{\ss}ling}, {Koffeman},
  {Kooijman}, {Kouchner}, {Kulikovskiy}, {Lahmann}, {Lamare}, {Leisos},
  {Leonora}, {Clark}, {Liolios}, {Llorens Alvarez}, {Lo Presti}, {L{\"o}hner},
  {Lonardo}, {Lotze}, {Loucatos}, {Maccioni}, {Mannheim}, {Margiotta},
  {Marinelli}, {Mari{\c{s}}}, {Markou}, {Mart{\'\i}nez-Mora}, {Martini},
  {Mele}, {Melis}, {Michael}, {Migliozzi}, {Migneco}, {Mijakowski}, {Miraglia},
  {Mollo}, {Mongelli}, {Morganti}, {Moussa}, {Musico}, {Musumeci}, {Navas},
  {Nicolau}, {Olcina}, {Olivetto}, {Orlando}, {Papaikonomou}, {Papaleo},
  {P{\u{a}}v{\u{a}}la{\c{s}}}, {Peek}, {Pellegrino}, {Perrina}, {Pfutzner},
  {Piattelli}, {Pikounis}, {Poma}, {Popa}, {Pradier}, {Pratolongo},
  {P{\"u}hlhofer}, {Pulvirenti}, {Quinn}, {Racca}, {Raffaelli}, {Randazzo},
  {Rapidis}, {Razis}, {Real}, {Resvanis}, {Reubelt}, {Riccobene}, {Rossi},
  {Rovelli}, {Salda{\~n}a}, {Salvadori}, {Samtleben}, {S{\'a}nchez
  Garc{\'\i}a}, {S{\'a}nchez Losa}, {Sanguineti}, {Santangelo}, {Santonocito},
  {Sapienza}, {Schimmel}, {Schmelling}, {Sciacca}, {Sedita}, {Seitz}, {Sgura},
  {Simeone}, {Siotis}, {Sipala}, {Spisso}, {Spurio}, {Stavropoulos},
  {Steijger}, {Stellacci}, {Stransky}, {Taiuti}, {Tayalati}, {T{\'e}zier},
  {Theraube}, {Thompson}, {Timmer}, {T{\"o}nnis}, {Trasatti}, {Trovato},
  {Tsirigotis}, {Tzamarias}, {Tzamariudaki}, {Vallage}, {Van Elewyck},
  {Vermeulen}, {Vicini}, {Viola}, {Vivolo}, {Volkert}, {Voulgaris}, {Wiggers},
  {Wilms}, {de Wolf}, {Zachariadou}, {Zornoza}, \&
  {Z{\'u}{\~n}iga}}]{2016JPhG...43h4001A}
{Adri{\'a}n-Mart{\'\i}nez}, S., {Ageron}, M., {Aharonian}, F., {et~al.} 2016,
  Journal of Physics G Nuclear Physics, 43, 084001,
  \dodoi{10.1088/0954-3899/43/8/084001}

\bibitem[{{Ai} {et~al.}(2021){Ai}, {Gao}, \& {Zhang}}]{2021ApJ...906L...5A}
{Ai}, S., {Gao}, H., \& {Zhang}, B. 2021, \apjl, 906, L5,
  \dodoi{10.3847/2041-8213/abcec9}

\bibitem[{{Akahori} {et~al.}(2016){Akahori}, {Ryu}, \&
  {Gaensler}}]{2016ApJ...824..105A}
{Akahori}, T., {Ryu}, D., \& {Gaensler}, B.~M. 2016, \apj, 824, 105,
  \dodoi{10.3847/0004-637X/824/2/105}

\bibitem[{{Anna-Thomas} {et~al.}(2022){Anna-Thomas}, {Connor}, {Burke-Spolaor},
  {Beniamini}, {Aggarwal}, {Law}, {Lynch}, {Li}, {Feng}, {Ocker}, {Cruces},
  {Chatterjee}, {Yu}, {Niu}, \& {Xue}}]{2022arXiv220211112A}
{Anna-Thomas}, R., {Connor}, L., {Burke-Spolaor}, S., {et~al.} 2022, arXiv
  e-prints, arXiv:2202.11112.
\newblock \doarXiv{2202.11112}

\bibitem[{{Avrorin} {et~al.}(2022){Avrorin}, {Avrorin}, {Ayinutdinov},
  {Allakhverdyan}, {Banash}, {Bardachova}, {Belolaptikov}, {Borina},
  {Brudanin}, {Budnev}, {Gafarov}, {Golubkov}, {Gorshkov}, {Gres'},
  {Dwornitski}, {Dzhilkibaev}, {Dik}, {Domogatskii}, {Doroshenko}, {Dyachok},
  {Elzhov}, {Zaborov}, {Katulin}, {Kebkal}, {Kebkal}, {Kozhin}, {Kolbin},
  {Konishchev}, {Kopanski}, {Korobchenko}, {Koshechkin}, {Kruglov}, {Kryukov},
  {Kulepov}, {Maletski}, {Malyshkin}, {Milenin}, {Mirgazov}, {Nazari},
  {Naumov}, {Noga}, {Petukhov}, {Pliskovskii}, {Rozanov}, {Rushay}, {Ryabov},
  {Safronov}, {Sirenko}, {Skurikhin}, {Solovjev}, {Sorokovikov}, {Stromakov},
  {Suvorova}, {Sushenok}, {Tabolenko}, {Tarashchanskii}, {Fait}, {Fialkovskii},
  {Khramov}, {Shaibonov}, {Shelepov}, {{\v{S}}imkovic}, {Stekl}, {Etskerova},
  {Yablokova}, \& {Yakovlev}}]{2022JETP..134..399A}
{Avrorin}, A.~V., {Avrorin}, A.~D., {Ayinutdinov}, V.~M., {et~al.} 2022, Soviet
  Journal of Experimental and Theoretical Physics, 134, 399,
  \dodoi{10.1134/S1063776122040148}

\bibitem[{{Bannister} {et~al.}(2017){Bannister}, {Shannon}, {Macquart},
  {Flynn}, {Edwards}, {O'Neill}, {Os{\l}owski}, {Bailes}, {Zackay}, {Clarke},
  {D'Addario}, {Dodson}, {Hall}, {Jameson}, {Jones}, {Navarro}, {Trinh},
  {Allison}, {Anderson}, {Bell}, {Chippendale}, {Collier}, {Heald}, {Heywood},
  {Hotan}, {Lee-Waddell}, {Madrid}, {Marvil}, {McConnell}, {Popping},
  {Voronkov}, {Whiting}, {Allen}, {Bock}, {Brodrick}, {Cooray}, {DeBoer},
  {Diamond}, {Ekers}, {Gough}, {Hampson}, {Harvey-Smith}, {Hay}, {Hayman},
  {Jackson}, {Johnston}, {Koribalski}, {McClure-Griffiths}, {Mirtschin}, {Ng},
  {Norris}, {Pearce}, {Phillips}, {Roxby}, {Troup}, \&
  {Westmeier}}]{2017ApJ...841L..12B}
{Bannister}, K.~W., {Shannon}, R.~M., {Macquart}, J.~P., {et~al.} 2017, \apjl,
  841, L12, \dodoi{10.3847/2041-8213/aa71ff}

\bibitem[{{Bassa} {et~al.}(2017){Bassa}, {Tendulkar}, {Adams}, {Maddox},
  {Bogdanov}, {Bower}, {Burke-Spolaor}, {Butler}, {Chatterjee}, {Cordes},
  {Hessels}, {Kaspi}, {Law}, {Marcote}, {Paragi}, {Ransom}, {Scholz},
  {Spitler}, \& {van Langevelde}}]{2017ApJ...843L...8B}
{Bassa}, C.~G., {Tendulkar}, S.~P., {Adams}, E.~A.~K., {et~al.} 2017, \apjl,
  843, L8, \dodoi{10.3847/2041-8213/aa7a0c}

\bibitem[{{Bhandari} {et~al.}(2018){Bhandari}, {Keane}, {Barr}, {Jameson},
  {Petroff}, {Johnston}, {Bailes}, {Bhat}, {Burgay}, {Burke-Spolaor}, {Caleb},
  {Eatough}, {Flynn}, {Green}, {Jankowski}, {Kramer}, {Krishnan}, {Morello},
  {Possenti}, {Stappers}, {Tiburzi}, {van Straten}, {Andreoni}, {Butterley},
  {Chandra}, {Cooke}, {Corongiu}, {Coward}, {Dhillon}, {Dodson}, {Hardy},
  {Howell}, {Jaroenjittichai}, {Klotz}, {Littlefair}, {Marsh}, {Mickaliger},
  {Muxlow}, {Perrodin}, {Pritchard}, {Sawangwit}, {Terai}, {Tominaga}, {Torne},
  {Totani}, {Trois}, {Turpin}, {Niino}, {Wilson}, {Albert}, {Andr{\'e}},
  {Anghinolfi}, {Anton}, {Ardid}, {Aubert}, {Avgitas}, {Baret},
  {Barrios-Mart{\'\i}}, {Basa}, {Belhorma}, {Bertin}, {Biagi}, {Bormuth},
  {Bourret}, {Bouwhuis}, {Br{\^a}nza{\c{s}}}, {Bruijn}, {Brunner}, {Busto},
  {Capone}, {Caramete}, {Carr}, {Celli}, {Moursli}, {Chiarusi}, {Circella},
  {Coelho}, {Coleiro}, {Coniglione}, {Costantini}, {Coyle}, {Creusot},
  {D{\'\i}az}, {Deschamps}, {De Bonis}, {Distefano}, {Palma}, {Domi},
  {Donzaud}, {Dornic}, {Drouhin}, {Eberl}, {Bojaddaini}, {Khayati},
  {Els{\"a}sser}, {Enzenh{\"o}fer}, {Ettahiri}, {Fassi}, {Felis}, {Fusco},
  {Gay}, {Giordano}, {Glotin}, {Gregoire}, {Gracia-Ruiz}, {Graf}, {Hallmann},
  {van Haren}, {Heijboer}, {Hello}, {Hern{\'a}ndez-Rey}, {H{\"o}{\ss}l},
  {Hofest{\"a}dt}, {Hugon}, {Illuminati}, {James}, {de Jong}, {Jongen},
  {Kadler}, {Kalekin}, {Katz}, {Kie{\ss}ling}, {Kouchner}, {Kreter},
  {Kreykenbohm}, {Kulikovskiy}, {Lachaud}, {Lahmann}, {Lef{\`e}vre}, {Leonora},
  {Loucatos}, {Marcelin}, {Margiotta}, {Marinelli}, {Mart{\'\i}nez-Mora},
  {Mele}, {Melis}, {Michael}, {Migliozzi}, {Moussa}, {Navas}, {Nezri},
  {Organokov}, {P{\v{a}}v{\v{a}}la{\c{s}}}, {Pellegrino}, {Perrina},
  {Piattelli}, {Popa}, {Pradier}, {Quinn}, {Racca}, {Riccobene},
  {S{\'a}nchez-Losa}, {Salda{\~n}a}, {Salvadori}, {Samtleben}, {Sanguineti},
  {Sapienza}, {Sch{\"u}ssler}, {Sieger}, {Spurio}, {Stolarczyk}, {Taiuti},
  {Tayalati}, {Trovato}, {Turpin}, {T{\"o}nnis}, {Vallage}, {Van Elewyck},
  {Versari}, {Vivolo}, {Vizzocca}, {Wilms}, {Zornoza}, \&
  {Z{\'u}{\~n}iga}}]{2018MNRAS.475.1427B}
{Bhandari}, S., {Keane}, E.~F., {Barr}, E.~D., {et~al.} 2018, \mnras, 475,
  1427, \dodoi{10.1093/mnras/stx3074}

\bibitem[{{Bhandari} {et~al.}(2020){Bhandari}, {Sadler}, {Prochaska}, {Simha},
  {Ryder}, {Marnoch}, {Bannister}, {Macquart}, {Flynn}, {Shannon}, {Tejos},
  {Corro-Guerra}, {Day}, {Deller}, {Ekers}, {Lopez}, {Mahony}, {Nu{\~n}ez}, \&
  {Phillips}}]{2020ApJ...895L..37B}
{Bhandari}, S., {Sadler}, E.~M., {Prochaska}, J.~X., {et~al.} 2020, \apjl, 895,
  L37, \dodoi{10.3847/2041-8213/ab672e}

\bibitem[{{Bhandari} {et~al.}(2022){Bhandari}, {Heintz}, {Aggarwal}, {Marnoch},
  {Day}, {Sydnor}, {Burke-Spolaor}, {Law}, {Xavier Prochaska}, {Tejos},
  {Bannister}, {Butler}, {Deller}, {Ekers}, {Flynn}, {Fong}, {James}, {Lazio},
  {Luo}, {Mahony}, {Ryder}, {Sadler}, {Shannon}, {Han}, {Lee}, \&
  {Zhang}}]{2022AJ....163...69B}
{Bhandari}, S., {Heintz}, K.~E., {Aggarwal}, K., {et~al.} 2022, \aj, 163, 69,
  \dodoi{10.3847/1538-3881/ac3aec}

\bibitem[{{Bij} {et~al.}(2021){Bij}, {Lin}, {Li}, {van Kerkwijk}, {Pen}, {Lu},
  {Main}, {Peterson}, {Quine}, \& {Vanderlinde}}]{2021ApJ...920...38B}
{Bij}, A., {Lin}, H.-H., {Li}, D., {et~al.} 2021, \apj, 920, 38,
  \dodoi{10.3847/1538-4357/ac1589}

\bibitem[{{Bilicki} {et~al.}(2014){Bilicki}, {Jarrett}, {Peacock}, {Cluver}, \&
  {Steward}}]{2014ApJS..210....9B}
{Bilicki}, M., {Jarrett}, T.~H., {Peacock}, J.~A., {Cluver}, M.~E., \&
  {Steward}, L. 2014, \apjs, 210, 9, \dodoi{10.1088/0067-0049/210/1/9}

\bibitem[{{Bilicki} {et~al.}(2016){Bilicki}, {Peacock}, {Jarrett}, {Cluver},
  {Maddox}, {Brown}, {Taylor}, {Hambly}, {Solarz}, {Holwerda}, {Baldry},
  {Loveday}, {Moffett}, {Hopkins}, {Driver}, {Alpaslan}, \&
  {Bland-Hawthorn}}]{2016ApJS..225....5B}
{Bilicki}, M., {Peacock}, J.~A., {Jarrett}, T.~H., {et~al.} 2016, \apjs, 225,
  5, \dodoi{10.3847/0067-0049/225/1/5}

\bibitem[{{Bochenek} {et~al.}(2020{\natexlab{a}}){Bochenek}, {McKenna},
  {Belov}, {Kocz}, {Kulkarni}, {Lamb}, {Ravi}, \&
  {Woody}}]{2020PASP..132c4202B}
{Bochenek}, C.~D., {McKenna}, D.~L., {Belov}, K.~V., {et~al.}
  2020{\natexlab{a}}, \pasp, 132, 034202, \dodoi{10.1088/1538-3873/ab63b3}

\bibitem[{{Bochenek} {et~al.}(2020{\natexlab{b}}){Bochenek}, {Ravi}, {Belov},
  {Hallinan}, {Kocz}, {Kulkarni}, \& {McKenna}}]{2020Natur.587...59B}
{Bochenek}, C.~D., {Ravi}, V., {Belov}, K.~V., {et~al.} 2020{\natexlab{b}},
  \nat, 587, 59, \dodoi{10.1038/s41586-020-2872-x}

\bibitem[{{Braun} {et~al.}(2019){Braun}, {Bonaldi}, {Bourke}, {Keane}, \&
  {Wagg}}]{2019arXiv191212699B}
{Braun}, R., {Bonaldi}, A., {Bourke}, T., {Keane}, E., \& {Wagg}, J. 2019,
  arXiv e-prints, arXiv:1912.12699.
\newblock \doarXiv{1912.12699}

\bibitem[{{Cassanelli} {et~al.}(2022){Cassanelli}, {Leung}, {Rahman},
  {Vanderlinde}, {Mena-Parra}, {Cary}, {Masui}, {Luo}, {Lin}, {Bij}, {Gill},
  {Baker}, {Bandura}, {Berger}, {Boyle}, {Brar}, {Chatterjee}, {Cubranic},
  {Dobbs}, {Fonseca}, {Good}, {Kaczmarek}, {Kaspi}, {Landecker}, {Lanman},
  {Li}, {McKee}, {Meyers}, {Michilli}, {Naidu}, {Ng}, {Patel}, {Pearlman},
  {Pen}, {Pleunis}, {Quine}, {Renard}, {Sanghavi}, {Smith}, {Stairs}, \&
  {Tendulkar}}]{2022AJ....163...65C}
{Cassanelli}, T., {Leung}, C., {Rahman}, M., {et~al.} 2022, \aj, 163, 65,
  \dodoi{10.3847/1538-3881/ac3d2f}

\bibitem[{{Chatterjee} {et~al.}(2017){Chatterjee}, {Law}, {Wharton},
  {Burke-Spolaor}, {Hessels}, {Bower}, {Cordes}, {Tendulkar}, {Bassa},
  {Demorest}, {Butler}, {Seymour}, {Scholz}, {Abruzzo}, {Bogdanov}, {Kaspi},
  {Keimpema}, {Lazio}, {Marcote}, {McLaughlin}, {Paragi}, {Ransom}, {Rupen},
  {Spitler}, \& {van Langevelde}}]{2017Natur.541...58C}
{Chatterjee}, S., {Law}, C.~J., {Wharton}, R.~S., {et~al.} 2017, \nat, 541, 58,
  \dodoi{10.1038/nature20797}

\bibitem[{{Chen} {et~al.}(2022){Chen}, {Hashimoto}, {Goto}, {Kim}, {Santos},
  {On}, {Lu}, \& {Hsiao}}]{2022MNRAS.509.1227C}
{Chen}, B.~H., {Hashimoto}, T., {Goto}, T., {et~al.} 2022, \mnras, 509, 1227,
  \dodoi{10.1093/mnras/stab2994}

\bibitem[{{CHIME Collaboration} {et~al.}(2022){CHIME Collaboration}, {Amiri},
  {Bandura}, {Boskovic}, {Chen}, {Cliche}, {Deng}, {Denman}, {Dobbs},
  {Fandino}, {Foreman}, {Halpern}, {Hanna}, {Hill}, {Hinshaw}, {H{\"o}fer},
  {Kania}, {Klages}, {Landecker}, {MacEachern}, {Masui}, {Mena-Parra},
  {Milutinovic}, {Mirhosseini}, {Newburgh}, {Nitsche}, {Ordog}, {Pen},
  {Pinsonneault-Marotte}, {Polzin}, {Reda}, {Renard}, {Shaw}, {Siegel},
  {Singh}, {Smegal}, {Tretyakov}, {van Gassen}, {Vanderlinde}, {Wang}, {Wiebe},
  {Willis}, \& {Wulf}}]{2022ApJS..261...29C}
{CHIME Collaboration}, {Amiri}, M., {Bandura}, K., {et~al.} 2022, \apjs, 261,
  29, \dodoi{10.3847/1538-4365/ac6fd9}

\bibitem[{{CHIME/FRB Collaboration} {et~al.}(2018){CHIME/FRB Collaboration},
  {Amiri}, {Bandura}, {Berger}, {Bhardwaj}, {Boyce}, {Boyle}, {Brar},
  {Burhanpurkar}, {Chawla}, {Chowdhury}, {Cliche}, {Cranmer}, {Cubranic},
  {Deng}, {Denman}, {Dobbs}, {Fandino}, {Fonseca}, {Gaensler}, {Giri},
  {Gilbert}, {Good}, {Guliani}, {Halpern}, {Hinshaw}, {H{\"o}fer}, {Josephy},
  {Kaspi}, {Landecker}, {Lang}, {Liao}, {Masui}, {Mena-Parra}, {Naidu},
  {Newburgh}, {Ng}, {Patel}, {Pen}, {Pinsonneault-Marotte}, {Pleunis}, {Rafiei
  Ravandi}, {Ransom}, {Renard}, {Scholz}, {Sigurdson}, {Siegel}, {Smith},
  {Stairs}, {Tendulkar}, {Vanderlinde}, \& {Wiebe}}]{2018ApJ...863...48C}
{CHIME/FRB Collaboration}, {Amiri}, M., {Bandura}, K., {et~al.} 2018, \apj,
  863, 48, \dodoi{10.3847/1538-4357/aad188}

\bibitem[{{Chime/Frb Collaboration} {et~al.}(2020){Chime/Frb Collaboration},
  {Amiri}, {Andersen}, {Bandura}, {Bhardwaj}, {Boyle}, {Brar}, {Chawla},
  {Chen}, {Cliche}, {Cubranic}, {Deng}, {Denman}, {Dobbs}, {Dong}, {Fandino},
  {Fonseca}, {Gaensler}, {Giri}, {Good}, {Halpern}, {Hessels}, {Hill},
  {H{\"o}fer}, {Josephy}, {Kania}, {Karuppusamy}, {Kaspi}, {Keimpema},
  {Kirsten}, {Landecker}, {Lang}, {Leung}, {Li}, {Lin}, {Marcote}, {Masui},
  {McKinven}, {Mena-Parra}, {Merryfield}, {Michilli}, {Milutinovic},
  {Mirhosseini}, {Naidu}, {Newburgh}, {Ng}, {Nimmo}, {Paragi}, {Patel}, {Pen},
  {Pinsonneault-Marotte}, {Pleunis}, {Rafiei-Ravandi}, {Rahman}, {Ransom},
  {Renard}, {Sanghavi}, {Scholz}, {Shaw}, {Shin}, {Siegel}, {Singh}, {Smegal},
  {Smith}, {Stairs}, {Tendulkar}, {Tretyakov}, {Vanderlinde}, {Wang}, {Wang},
  {Wulf}, {Yadav}, \& {Zwaniga}}]{2020Natur.582..351C}
{Chime/Frb Collaboration}, {Amiri}, M., {Andersen}, B.~C., {et~al.} 2020, \nat,
  582, 351, \dodoi{10.1038/s41586-020-2398-2}

\bibitem[{{CHIME/FRB Collaboration} {et~al.}(2020){CHIME/FRB Collaboration},
  {Andersen}, {Bandura}, {Bhardwaj}, {Bij}, {Boyce}, {Boyle}, {Brar},
  {Cassanelli}, {Chawla}, {Chen}, {Cliche}, {Cook}, {Cubranic}, {Curtin},
  {Denman}, {Dobbs}, {Dong}, {Fandino}, {Fonseca}, {Gaensler}, {Giri}, {Good},
  {Halpern}, {Hill}, {Hinshaw}, {H{\"o}fer}, {Josephy}, {Kania}, {Kaspi},
  {Landecker}, {Leung}, {Li}, {Lin}, {Masui}, {McKinven}, {Mena-Parra},
  {Merryfield}, {Meyers}, {Michilli}, {Milutinovic}, {Mirhosseini},
  {M{\"u}nchmeyer}, {Naidu}, {Newburgh}, {Ng}, {Patel}, {Pen},
  {Pinsonneault-Marotte}, {Pleunis}, {Quine}, {Rafiei-Ravandi}, {Rahman},
  {Ransom}, {Renard}, {Sanghavi}, {Scholz}, {Shaw}, {Shin}, {Siegel}, {Singh},
  {Smegal}, {Smith}, {Stairs}, {Tan}, {Tendulkar}, {Tretyakov}, {Vanderlinde},
  {Wang}, {Wulf}, \& {Zwaniga}}]{2020Natur.587...54C}
{CHIME/FRB Collaboration}, {Andersen}, B.~C., {Bandura}, K.~M., {et~al.} 2020,
  \nat, 587, 54, \dodoi{10.1038/s41586-020-2863-y}

\bibitem[{{CHIME/FRB Collaboration} {et~al.}(2021){CHIME/FRB Collaboration},
  {Amiri}, {Andersen}, {Bandura}, {Berger}, {Bhardwaj}, {Boyce}, {Boyle},
  {Brar}, {Breitman}, {Cassanelli}, {Chawla}, {Chen}, {Cliche}, {Cook},
  {Cubranic}, {Curtin}, {Deng}, {Dobbs}, {Dong}, {Eadie}, {Fandino}, {Fonseca},
  {Gaensler}, {Giri}, {Good}, {Halpern}, {Hill}, {Hinshaw}, {Josephy},
  {Kaczmarek}, {Kader}, {Kania}, {Kaspi}, {Landecker}, {Lang}, {Leung}, {Li},
  {Lin}, {Masui}, {McKinven}, {Mena-Parra}, {Merryfield}, {Meyers}, {Michilli},
  {Milutinovic}, {Mirhosseini}, {M{\"u}nchmeyer}, {Naidu}, {Newburgh}, {Ng},
  {Patel}, {Pen}, {Petroff}, {Pinsonneault-Marotte}, {Pleunis},
  {Rafiei-Ravandi}, {Rahman}, {Ransom}, {Renard}, {Sanghavi}, {Scholz}, {Shaw},
  {Shin}, {Siegel}, {Sikora}, {Singh}, {Smith}, {Stairs}, {Tan}, {Tendulkar},
  {Vanderlinde}, {Wang}, {Wulf}, \& {Zwaniga}}]{2021ApJS..257...59C}
{CHIME/FRB Collaboration}, {Amiri}, M., {Andersen}, B.~C., {et~al.} 2021,
  \apjs, 257, 59, \dodoi{10.3847/1538-4365/ac33ab}

\bibitem[{{CHIME/FRB Collaboration} {et~al.}(2022){CHIME/FRB Collaboration},
  {Bandura}, {Bhardwaj}, {Boyle}, {Brar}, {Breitman}, {Cassanelli},
  {Chatterjee}, {Chawla}, {Cliche}, {Cubranic}, {Curtin}, {Deng}, {Dobbs},
  {Dong}, {Fonseca}, {Gaensler}, {Giri}, {Good}, {Hill}, {Josephy},
  {Kaczmarek}, {Kader}, {Kania}, {Kaspi}, {Leung}, {Li}, {Lin}, {Masui},
  {McKinven}, {Mena-Parra}, {Merryfield}, {Meyers}, {Michilli}, {Naidu},
  {Newburgh}, {Ng}, {Ordog}, {Patel}, {Pearlman}, {Pen}, {Petroff}, {Pleunis},
  {Rafiei-Ravandi}, {Rahman}, {Ransom}, {Renard}, {Sanghavi}, {Scholz}, {Shaw},
  {Shin}, {Siegel}, {Singh}, {Smith}, {Stairs}, {Tan}, {Tendulkar},
  {Vanderlinde}, {Wiebe}, {Wulf}, \& {Zwaniga}}]{2022Natur.607..256C}
{CHIME/FRB Collaboration}, Andersen, B.~C., {Bandura}, K., {Bhardwaj}, M.,
  {et~al.} 2022, \nat, 607, 256, \dodoi{10.1038/s41586-022-04841-8}

\bibitem[{{CHIME/Pulsar Collaboration} {et~al.}(2021){CHIME/Pulsar
  Collaboration}, {Amiri}, {Bandura}, {Boyle}, {Brar}, {Cliche}, {Crowter},
  {Cubranic}, {Demorest}, {Denman}, {Dobbs}, {Dong}, {Fandino}, {Fonseca},
  {Good}, {Halpern}, {Hill}, {H{\"o}fer}, {Kaspi}, {Landecker}, {Leung}, {Lin},
  {Luo}, {Masui}, {McKee}, {Mena-Parra}, {Meyers}, {Michilli}, {Naidu},
  {Newburgh}, {Ng}, {Patel}, {Pinsonneault-Marotte}, {Ransom}, {Renard},
  {Scholz}, {Shaw}, {Sikora}, {Stairs}, {Tan}, {Tendulkar}, {Tretyakov},
  {Vanderlinde}, {Wang}, \& {Wang}}]{2021ApJS..255....5C}
{CHIME/Pulsar Collaboration}, {Amiri}, M., {Bandura}, K.~M., {et~al.} 2021,
  \apjs, 255, 5, \dodoi{10.3847/1538-4365/abfdcb}

\bibitem[{{Coleman} {et~al.}(2022){Coleman}, {Eser}, {Mayotte}, {Sarazin},
  {Schr{\"o}der}, {Soldin}, {Venters}, {Aloisio}, {Alvarez-Mu{\~n}iz}, {Alves
  Batista}, {Bergman}, {Bertaina}, {Caccianiga}, {Deligny}, {Dembinski},
  {Denton}, {di Matteo}, {Globus}, {Glombitza}, {Golup}, {Haungs},
  {H{\"o}randel}, {Jaffe}, {Kelley}, {Krizmanic}, {Lu}, {Matthews},
  {Mari{\c{s}}}, {Mussa}, {Oikonomou}, {Pierog}, {Santos}, {Tinyakov},
  {Tsunesada}, {Unger}, {Yushkov}, {Albrow}, {Anchordoqui}, {Andeen}, {Arnone},
  {Barghini}, {Bechtol}, {Bellido}, {Casolino}, {Castellina}, {Cazon},
  {Concei{\c{c}}{\~a}o}, {Cremonini}, {Dujmovic}, {Engel}, {Farrar}, {Fenu},
  {Ferrarese}, {Fujii}, {Gardiol}, {Gritsevich}, {Homola}, {Huege}, {Kampert},
  {Kang}, {Kido}, {Klimov}, {Kotera}, {Kozelov}, {Leszczy{\'n}ska}, {Madsen},
  {Marcelli}, {Marisaldi}, {Martineau-Huynh}, {Mayotte}, {Mulrey}, {Murase},
  {Muzio}, {Ogio}, {Olinto}, {Onel}, {Paul}, {Piotrowski}, {Plum}, {Pont},
  {Reininghaus}, {Riedel}, {Riehn}, {Roth}, {Sako}, {Schl{\"u}ter},
  {Shoemaker}, {Sidhu}, {Sidelnik}, {Timmermans}, {Tkachenko}, {Veberi{\v{c}}},
  {Verpoest}, {Verzi}, {V{\'\i}cha}, {Winn}, {Zas}, \&
  {Zotov}}]{2022arXiv220505845C}
{Coleman}, A., {Eser}, J., {Mayotte}, E., {et~al.} 2022, arXiv e-prints,
  arXiv:2205.05845.
\newblock \doarXiv{2205.05845}

\bibitem[{{Connor} {et~al.}(2016){Connor}, {Lin}, {Masui}, {Oppermann}, {Pen},
  {Peterson}, {Roman}, \& {Sievers}}]{2016MNRAS.460.1054C}
{Connor}, L., {Lin}, H.-H., {Masui}, K., {et~al.} 2016, \mnras, 460, 1054,
  \dodoi{10.1093/mnras/stw907}

\bibitem[{{Connor} {et~al.}(2021){Connor}, {Shila}, {Kulkarni}, {Flygare},
  {Hallinan}, {Li}, {Lu}, {Ravi}, \& {Weinreb}}]{2021PASP..133g5001C}
{Connor}, L., {Shila}, K.~A., {Kulkarni}, S.~R., {et~al.} 2021, \pasp, 133,
  075001, \dodoi{10.1088/1538-3873/ac0bcc}

\bibitem[{{Cunningham} {et~al.}(2019){Cunningham}, {Cenko}, {Burns},
  {Goldstein}, {Lien}, {Kocevski}, {Briggs}, {Connaughton}, {Miller},
  {Racusin}, \& {Stanbro}}]{2019ApJ...879...40C}
{Cunningham}, V., {Cenko}, S.~B., {Burns}, E., {et~al.} 2019, \apj, 879, 40,
  \dodoi{10.3847/1538-4357/ab2235}

\bibitem[{{Cutri} {et~al.}(2003){Cutri}, {Skrutskie}, {van Dyk}, {Beichman},
  {Carpenter}, {Chester}, {Cambresy}, {Evans}, {Fowler}, {Gizis}, {Howard},
  {Huchra}, {Jarrett}, {Kopan}, {Kirkpatrick}, {Light}, {Marsh}, {McCallon},
  {Schneider}, {Stiening}, {Sykes}, {Weinberg}, {Wheaton}, {Wheelock}, \&
  {Zacarias}}]{2003tmc..book.....C}
{Cutri}, R.~M., {Skrutskie}, M.~F., {van Dyk}, S., {et~al.} 2003, {2MASS All
  Sky Catalog of point sources.}

\bibitem[{{Dai} {et~al.}(2022){Dai}, {Feng}, {Yang}, {Zhang}, {Li}, {Niu},
  {Wang}, {Xue}, {Zhang}, {Burke-Spolaor}, {Law}, {Lynch}, {Connor},
  {Anna-Thomas}, {Zhang}, {Duan}, {Yao}, {Tsai}, {Zhu}, {Cruces}, {Hobbs},
  {Miao}, {Niu}, {Filipovic}, \& {Zhu}}]{2022arXiv220308151D}
{Dai}, S., {Feng}, Y., {Yang}, Y.~P., {et~al.} 2022, arXiv e-prints,
  arXiv:2203.08151.
\newblock \doarXiv{2203.08151}

\bibitem[{{Dey} {et~al.}(2019){Dey}, {Schlegel}, {Lang}, {Blum}, {Burleigh},
  {Fan}, {Findlay}, {Finkbeiner}, {Herrera}, {Juneau}, {Landriau}, {Levi},
  {McGreer}, {Meisner}, {Myers}, {Moustakas}, {Nugent}, {Patej}, {Schlafly},
  {Walker}, {Valdes}, {Weaver}, {Y{\`e}che}, {Zou}, {Zhou}, {Abareshi},
  {Abbott}, {Abolfathi}, {Aguilera}, {Alam}, {Allen}, {Alvarez}, {Annis},
  {Ansarinejad}, {Aubert}, {Beechert}, {Bell}, {BenZvi}, {Beutler}, {Bielby},
  {Bolton}, {Brice{\~n}o}, {Buckley-Geer}, {Butler}, {Calamida}, {Carlberg},
  {Carter}, {Casas}, {Castander}, {Choi}, {Comparat}, {Cukanovaite}, {Delubac},
  {DeVries}, {Dey}, {Dhungana}, {Dickinson}, {Ding}, {Donaldson}, {Duan},
  {Duckworth}, {Eftekharzadeh}, {Eisenstein}, {Etourneau}, {Fagrelius},
  {Farihi}, {Fitzpatrick}, {Font-Ribera}, {Fulmer}, {G{\"a}nsicke},
  {Gaztanaga}, {George}, {Gerdes}, {Gontcho}, {Gorgoni}, {Green}, {Guy},
  {Harmer}, {Hernandez}, {Honscheid}, {Huang}, {James}, {Jannuzi}, {Jiang},
  {Joyce}, {Karcher}, {Karkar}, {Kehoe}, {Kneib}, {Kueter-Young}, {Lan},
  {Lauer}, {Le Guillou}, {Le Van Suu}, {Lee}, {Lesser}, {Perreault Levasseur},
  {Li}, {Mann}, {Marshall}, {Mart{\'\i}nez-V{\'a}zquez}, {Martini}, {du Mas des
  Bourboux}, {McManus}, {Meier}, {M{\'e}nard}, {Metcalfe},
  {Mu{\~n}oz-Guti{\'e}rrez}, {Najita}, {Napier}, {Narayan}, {Newman}, {Nie},
  {Nord}, {Norman}, {Olsen}, {Paat}, {Palanque-Delabrouille}, {Peng},
  {Poppett}, {Poremba}, {Prakash}, {Rabinowitz}, {Raichoor}, {Rezaie},
  {Robertson}, {Roe}, {Ross}, {Ross}, {Rudnick}, {Safonova}, {Saha},
  {S{\'a}nchez}, {Savary}, {Schweiker}, {Scott}, {Seo}, {Shan}, {Silva},
  {Slepian}, {Soto}, {Sprayberry}, {Staten}, {Stillman}, {Stupak}, {Summers},
  {Sien Tie}, {Tirado}, {Vargas-Maga{\~n}a}, {Vivas}, {Wechsler}, {Williams},
  {Yang}, {Yang}, {Yapici}, {Zaritsky}, {Zenteno}, {Zhang}, {Zhang}, {Zhou}, \&
  {Zhou}}]{2019AJ....157..168D}
{Dey}, A., {Schlegel}, D.~J., {Lang}, D., {et~al.} 2019, \aj, 157, 168,
  \dodoi{10.3847/1538-3881/ab089d}

\bibitem[{{Eastwood} {et~al.}(2018){Eastwood}, {Anderson}, {Monroe},
  {Hallinan}, {Barsdell}, {Bourke}, {Clark}, {Ellingson}, {Dowell}, {Garsden},
  {Greenhill}, {Hartman}, {Kocz}, {Lazio}, {Price}, {Schinzel}, {Taylor},
  {Vedantham}, {Wang}, \& {Woody}}]{2018AJ....156...32E}
{Eastwood}, M.~W., {Anderson}, M.~M., {Monroe}, R.~M., {et~al.} 2018, \aj, 156,
  32, \dodoi{10.3847/1538-3881/aac721}

\bibitem[{{Fang} \& {Metzger}(2017)}]{2017ApJ...849..153F}
{Fang}, K., \& {Metzger}, B.~D. 2017, \apj, 849, 153,
  \dodoi{10.3847/1538-4357/aa8b6a}

\bibitem[{{Farah} {et~al.}(2019){Farah}, {Flynn}, {Bailes}, {Jameson},
  {Bateman}, {Campbell-Wilson}, {Day}, {Deller}, {Green}, {Gupta}, {Hunstead},
  {Lower}, {Os{\l}owski}, {Parthasarathy}, {Price}, {Ravi}, {Shannon},
  {Sutherland}, {Temby}, {Krishnan}, {Caleb}, {Chang}, {Cruces}, {Roy},
  {Morello}, {Onken}, {Stappers}, {Webb}, \& {Wolf}}]{2019MNRAS.488.2989F}
{Farah}, W., {Flynn}, C., {Bailes}, M., {et~al.} 2019, \mnras, 488, 2989,
  \dodoi{10.1093/mnras/stz1748}

\bibitem[{{Fonseca} {et~al.}(2020){Fonseca}, {Andersen}, {Bhardwaj}, {Chawla},
  {Good}, {Josephy}, {Kaspi}, {Masui}, {Mckinven}, {Michilli}, {Pleunis},
  {Shin}, {Tendulkar}, {Bandura}, {Boyle}, {Brar}, {Cassanelli}, {Cubranic},
  {Dobbs}, {Dong}, {Gaensler}, {Hinshaw}, {Landecker}, {Leung}, {Li}, {Lin},
  {Mena-Parra}, {Merryfield}, {Naidu}, {Ng}, {Patel}, {Pen}, {Rafiei-Ravandi},
  {Rahman}, {Ransom}, {Scholz}, {Smith}, {Stairs}, {Vanderlinde}, {Yadav}, \&
  {Zwaniga}}]{2020ApJ...891L...6F}
{Fonseca}, E., {Andersen}, B.~C., {Bhardwaj}, M., {et~al.} 2020, \apjl, 891,
  L6, \dodoi{10.3847/2041-8213/ab7208}

\bibitem[{{Good} {et~al.}(2021){Good}, {Andersen}, {Chawla}, {Crowter}, {Dong},
  {Fonseca}, {Meyers}, {Ng}, {Pleunis}, {Ransom}, {Stairs}, {Tan}, {Bhardwaj},
  {Boyle}, {Dobbs}, {Gaensler}, {Kaspi}, {Masui}, {Naidu}, {Rafiei-Ravandi},
  {Scholz}, {Smith}, \& {Tendulkar}}]{2021ApJ...922...43G}
{Good}, D.~C., {Andersen}, B.~C., {Chawla}, P., {et~al.} 2021, \apj, 922, 43,
  \dodoi{10.3847/1538-4357/ac1da6}

\bibitem[{{Guzm{\'a}n} {et~al.}(2011){Guzm{\'a}n}, {May}, {Alvarez}, \&
  {Maeda}}]{2011A&A...525A.138G}
{Guzm{\'a}n}, A.~E., {May}, J., {Alvarez}, H., \& {Maeda}, K. 2011, \aap, 525,
  A138, \dodoi{10.1051/0004-6361/200913628}

\bibitem[{{Hashimoto} {et~al.}(2019){Hashimoto}, {Goto}, {Wang}, {Kim}, {Wu},
  \& {Ho}}]{2019MNRAS.488.1908H}
{Hashimoto}, T., {Goto}, T., {Wang}, T.-W., {et~al.} 2019, \mnras, 488, 1908,
  \dodoi{10.1093/mnras/stz1715}

\bibitem[{{Hashimoto} {et~al.}(2020){Hashimoto}, {Goto}, {On}, {Lu}, {Santos},
  {Ho}, {Kim}, {Wang}, \& {Hsiao}}]{2020MNRAS.498.3927H}
{Hashimoto}, T., {Goto}, T., {On}, A. Y.~L., {et~al.} 2020, \mnras, 498, 3927,
  \dodoi{10.1093/mnras/staa2490}

\bibitem[{{Hashimoto} {et~al.}(2021){Hashimoto}, {Goto}, {Santos}, {Ho},
  {Hsiao}, {Wong}, {On}, {Kim}, {Lu}, \& {Kilerci-Eser}}]{2021PhRvD.104l4026H}
{Hashimoto}, T., {Goto}, T., {Santos}, D. J.~D., {et~al.} 2021, \prd, 104,
  124026, \dodoi{10.1103/PhysRevD.104.124026}

\bibitem[{{Haslam} {et~al.}(1981){Haslam}, {Klein}, {Salter}, {Stoffel},
  {Wilson}, {Cleary}, {Cooke}, \& {Thomasson}}]{1981A&A...100..209H}
{Haslam}, C.~G.~T., {Klein}, U., {Salter}, C.~J., {et~al.} 1981, \aap, 100, 209

\bibitem[{{Haslam} {et~al.}(1982){Haslam}, {Salter}, {Stoffel}, \&
  {Wilson}}]{1982A&AS...47....1H}
{Haslam}, C.~G.~T., {Salter}, C.~J., {Stoffel}, H., \& {Wilson}, W.~E. 1982,
  \aaps, 47, 1

\bibitem[{{Heald} {et~al.}(2015){Heald}, {Pizzo}, {Orr{\'u}}, {Breton},
  {Carbone}, {Ferrari}, {Hardcastle}, {Jurusik}, {Macario}, {Mulcahy},
  {Rafferty}, {Asgekar}, {Brentjens}, {Fallows}, {Frieswijk}, {Toribio},
  {Adebahr}, {Arts}, {Bell}, {Bonafede}, {Bray}, {Broderick}, {Cantwell},
  {Carroll}, {Cendes}, {Clarke}, {Croston}, {Daiboo}, {de Gasperin}, {Gregson},
  {Harwood}, {Hassall}, {Heesen}, {Horneffer}, {van der Horst}, {Iacobelli},
  {Jeli{\'c}}, {Jones}, {Kant}, {Kokotanekov}, {Martin}, {McKean}, {Morabito},
  {Nikiel-Wroczy{\'n}ski}, {Offringa}, {Pandey}, {Pandey-Pommier}, {Pietka},
  {Pratley}, {Riseley}, {Rowlinson}, {Sabater}, {Scaife}, {Scheers},
  {Sendlinger}, {Shulevski}, {Sipior}, {Sobey}, {Stewart}, {Stroe}, {Swinbank},
  {Tasse}, {Tr{\"u}stedt}, {Varenius}, {van Velzen}, {Vilchez}, {van Weeren},
  {Wijnholds}, {Williams}, {de Bruyn}, {Nijboer}, {Wise}, {Alexov}, {Anderson},
  {Avruch}, {Beck}, {Bell}, {van Bemmel}, {Bentum}, {Bernardi}, {Best},
  {Breitling}, {Brouw}, {Br{\"u}ggen}, {Butcher}, {Ciardi}, {Conway}, {de
  Geus}, {de Jong}, {de Vos}, {Deller}, {Dettmar}, {Duscha}, {Eisl{\"o}ffel},
  {Engels}, {Falcke}, {Fender}, {Garrett}, {Grie{\ss}meier}, {Gunst},
  {Hamaker}, {Hessels}, {Hoeft}, {H{\"o}randel}, {Holties}, {Intema},
  {Jackson}, {J{\"u}tte}, {Karastergiou}, {Klijn}, {Kondratiev}, {Koopmans},
  {Kuniyoshi}, {Kuper}, {Law}, {van Leeuwen}, {Loose}, {Maat}, {Markoff},
  {McFadden}, {McKay-Bukowski}, {Mevius}, {Miller-Jones}, {Morganti}, {Munk},
  {Nelles}, {Noordam}, {Norden}, {Paas}, {Polatidis}, {Reich}, {Renting},
  {R{\"o}ttgering}, {Schoenmakers}, {Schwarz}, {Sluman}, {Smirnov}, {Stappers},
  {Steinmetz}, {Tagger}, {Tang}, {ter Veen}, {Thoudam}, {Vermeulen}, {Vocks},
  {Vogt}, {Wijers}, {Wucknitz}, {Yatawatta}, \& {Zarka}}]{2015A&A...582A.123H}
{Heald}, G.~H., {Pizzo}, R.~F., {Orr{\'u}}, E., {et~al.} 2015, \aap, 582, A123,
  \dodoi{10.1051/0004-6361/201425210}

\bibitem[{{Heintz} {et~al.}(2020){Heintz}, {Prochaska}, {Simha}, {Platts},
  {Fong}, {Tejos}, {Ryder}, {Aggerwal}, {Bhandari}, {Day}, {Deller},
  {Kilpatrick}, {Law}, {Macquart}, {Mannings}, {Marnoch}, {Sadler}, \&
  {Shannon}}]{2020ApJ...903..152H}
{Heintz}, K.~E., {Prochaska}, J.~X., {Simha}, S., {et~al.} 2020, \apj, 903,
  152, \dodoi{10.3847/1538-4357/abb6fb}

\bibitem[{{Herrmann}(2021)}]{2021ATel14556....1H}
{Herrmann}, W. 2021, The Astronomer's Telegram, 14556, 1

\bibitem[{{Hewish} {et~al.}(1964){Hewish}, {Scott}, \&
  {Wills}}]{1964Natur.203.1214H}
{Hewish}, A., {Scott}, P.~F., \& {Wills}, D. 1964, \nat, 203, 1214,
  \dodoi{10.1038/2031214a0}

\bibitem[{{Ichiki}(2014)}]{2014PTEP.2014fB109I}
{Ichiki}, K. 2014, Progress of Theoretical and Experimental Physics, 2014,
  06B109, \dodoi{10.1093/ptep/ptu065}

\bibitem[{{Intema} {et~al.}(2017){Intema}, {Jagannathan}, {Mooley}, \&
  {Frail}}]{2017A&A...598A..78I}
{Intema}, H.~T., {Jagannathan}, P., {Mooley}, K.~P., \& {Frail}, D.~A. 2017,
  \aap, 598, A78, \dodoi{10.1051/0004-6361/201628536}

\bibitem[{{Jiang} {et~al.}(2019){Jiang}, {Yue}, {Gan}, {Yao}, {Li}, {Pan},
  {Sun}, {Yu}, {Liu}, {Tang}, {Qian}, {Lu}, {Yan}, {Peng}, {Zhang}, {Wang},
  {Li}, \& {Li}}]{2019SCPMA..6259502J}
{Jiang}, P., {Yue}, Y., {Gan}, H., {et~al.} 2019, Science China Physics,
  Mechanics, and Astronomy, 62, 959502, \dodoi{10.1007/s11433-018-9376-1}

\bibitem[{{Josephy} {et~al.}(2021){Josephy}, {Chawla}, {Curtin}, {Kaspi},
  {Bhardwaj}, {Boyle}, {Brar}, {Cassanelli}, {Fonseca}, {Gaensler}, {Leung},
  {Lin}, {Masui}, {Mckinven}, {Mena-Parra}, {Michilli}, {Ng}, {Pleunis},
  {Rafiei-Ravandi}, {Rahman}, {Sanghavi}, {Scholz}, {Shin}, {Smith}, {Stairs},
  {Tendulkar}, \& {Zwaniga}}]{2021ApJ...923....2J}
{Josephy}, A., {Chawla}, P., {Curtin}, A.~P., {et~al.} 2021, \apj, 923, 2,
  \dodoi{10.3847/1538-4357/ac33ad}

\bibitem[{{Karachentsev} {et~al.}(2013){Karachentsev}, {Makarov}, \&
  {Kaisina}}]{2013AJ....145..101K}
{Karachentsev}, I.~D., {Makarov}, D.~I., \& {Kaisina}, E.~I. 2013, \aj, 145,
  101, \dodoi{10.1088/0004-6256/145/4/101}

\bibitem[{{Kerr} {et~al.}(2018){Kerr}, {Coles}, {Ward}, {Johnston}, {Tuntsov},
  \& {Shannon}}]{2018MNRAS.474.4637K}
{Kerr}, M., {Coles}, W.~A., {Ward}, C.~A., {et~al.} 2018, \mnras, 474, 4637,
  \dodoi{10.1093/mnras/stx3101}

\bibitem[{{Kimura} {et~al.}(2018){Kimura}, {Murase}, {Bartos}, {Ioka}, {Heng},
  \& {M{\'e}sz{\'a}ros}}]{2018PhRvD..98d3020K}
{Kimura}, S.~S., {Murase}, K., {Bartos}, I., {et~al.} 2018, \prd, 98, 043020,
  \dodoi{10.1103/PhysRevD.98.043020}

\bibitem[{{Kirsten} {et~al.}(2021){Kirsten}, {Snelders}, {Jenkins}, {Nimmo},
  {van den Eijnden}, {Hessels}, {Gawro{\'n}ski}, \&
  {Yang}}]{2021NatAs...5..414K}
{Kirsten}, F., {Snelders}, M.~P., {Jenkins}, M., {et~al.} 2021, Nature
  Astronomy, 5, 414, \dodoi{10.1038/s41550-020-01246-3}

\bibitem[{{Kirsten} {et~al.}(2022){Kirsten}, {Marcote}, {Nimmo}, {Hessels},
  {Bhardwaj}, {Tendulkar}, {Keimpema}, {Yang}, {Snelders}, {Scholz},
  {Pearlman}, {Law}, {Peters}, {Giroletti}, {Paragi}, {Bassa}, {Hewitt},
  {Bach}, {Bezrukovs}, {Burgay}, {Buttaccio}, {Conway}, {Corongiu}, {Feiler},
  {Forss{\'e}n}, {Gawro{\'n}ski}, {Karuppusamy}, {Kharinov}, {Lindqvist},
  {Maccaferri}, {Melnikov}, {Ould-Boukattine}, {Possenti}, {Surcis}, {Wang},
  {Yuan}, {Aggarwal}, {Anna-Thomas}, {Bower}, {Blaauw}, {Burke-Spolaor},
  {Cassanelli}, {Clarke}, {Fonseca}, {Gaensler}, {Gopinath}, {Kaspi}, {Kassim},
  {Lazio}, {Leung}, {Li}, {Lin}, {Masui}, {Mckinven}, {Michilli}, {Mikhailov},
  {Ng}, {Orbidans}, {Pen}, {Petroff}, {Rahman}, {Ransom}, {Shin}, {Smith},
  {Stairs}, \& {Vlemmings}}]{2022Natur.602..585K}
{Kirsten}, F., {Marcote}, B., {Nimmo}, K., {et~al.} 2022, \nat, 602, 585,
  \dodoi{10.1038/s41586-021-04354-w}

\bibitem[{{Kocz} {et~al.}(2010){Kocz}, {Briggs}, \&
  {Reynolds}}]{2010AJ....140.2086K}
{Kocz}, J., {Briggs}, F.~H., \& {Reynolds}, J. 2010, \aj, 140, 2086,
  \dodoi{10.1088/0004-6256/140/6/2086}

\bibitem[{{Kriele} {et~al.}(2022){Kriele}, {Wayth}, {Bentum}, {Juswardy}, \&
  {Trott}}]{2022PASA...39...17K}
{Kriele}, M.~A., {Wayth}, R.~B., {Bentum}, M.~J., {Juswardy}, B., \& {Trott},
  C.~M. 2022, \pasa, 39, e017, \dodoi{10.1017/pasa.2022.2}

\bibitem[{{Law} {et~al.}(2018){Law}, {Bower}, {Burke-Spolaor}, {Butler},
  {Demorest}, {Halle}, {Khudikyan}, {Lazio}, {Pokorny}, {Robnett}, \&
  {Rupen}}]{2018ApJS..236....8L}
{Law}, C.~J., {Bower}, G.~C., {Burke-Spolaor}, S., {et~al.} 2018, \apjs, 236,
  8, \dodoi{10.3847/1538-4365/aab77b}

\bibitem[{{Leung} {et~al.}(2021){Leung}, {Mena-Parra}, {Masui}, {Bandura},
  {Bhardwaj}, {Boyle}, {Brar}, {Bruneault}, {Cassanelli}, {Cubranic},
  {Kaczmarek}, {Kaspi}, {Landecker}, {Michilli}, {Milutinovic}, {Patel},
  {Pleunis}, {Rahman}, {Renard}, {Sanghavi}, {Stairs}, {Scholz}, {Vanderlinde},
  \& {Chime/Frb Collaboration}}]{2021AJ....161...81L}
{Leung}, C., {Mena-Parra}, J., {Masui}, K., {et~al.} 2021, \aj, 161, 81,
  \dodoi{10.3847/1538-3881/abd174}

\bibitem[{{Leung} {et~al.}(2022){Leung}, {Kader}, {Masui}, {Dobbs}, {Michilli},
  {Mena-Parra}, {Mckinven}, {Ng}, {Bandura}, {Bhardwaj}, {Brar}, {Cassanelli},
  {Chawla}, {Dong}, {Good}, {Kaspi}, {Lanman}, {Lin}, {Meyers}, {Pearlman},
  {Pen}, {Petroff}, {Pleunis}, {Rafiei-Ravandi}, {Rahman}, {Sanghavi},
  {Scholz}, {Shin}, {Siegel}, {Smith}, {Stairs}, {Tendulkar}, \&
  {Vanderlinde}}]{2022PhRvD.106d3017L}
{Leung}, C., {Kader}, Z., {Masui}, K.~W., {et~al.} 2022, \prd, 106, 043017,
  \dodoi{10.1103/PhysRevD.106.043017}

\bibitem[{{Li} {et~al.}(2021){Li}, {Lin}, {Xiong}, {Ge}, {Li}, {Li}, {Lu},
  {Zhang}, {Tuo}, {Nang}, {Zhang}, {Xiao}, {Chen}, {Song}, {Xu}, {Liu}, {Jia},
  {Cao}, {Qu}, {Zhang}, {Gu}, {Liao}, {Zhao}, {Tan}, {Nie}, {Zhao}, {Zheng},
  {Zheng}, {Luo}, {Cai}, {Li}, {Xue}, {Bu}, {Chang}, {Chen}, {Chen}, {Chen},
  {Chen}, {Chen}, {Cui}, {Cui}, {Deng}, {Dong}, {Du}, {Fu}, {Gao}, {Gao},
  {Gao}, {Gu}, {Guan}, {Guo}, {Han}, {Huang}, {Huo}, {Jiang}, {Jiang}, {Jin},
  {Jin}, {Kong}, {Li}, {Li}, {Li}, {Li}, {Li}, {Li}, {Li}, {Liang}, {Liu},
  {Liu}, {Liu}, {Liu}, {Liu}, {Lu}, {Lu}, {Luo}, {Ma}, {Meng}, {Ou}, {Sai},
  {Shang}, {Song}, {Sun}, {Tao}, {Wang}, {Wang}, {Wang}, {Wang}, {Wang}, {Wen},
  {Wu}, {Wu}, {Wu}, {Xiao}, {Xu}, {Yang}, {Yang}, {Yang}, {Yang}, {Yi}, {Yin},
  {You}, {Zhang}, {Zhang}, {Zhang}, {Zhang}, {Zhang}, {Zhang}, {Zhang},
  {Zhang}, {Zhang}, {Zhang}, {Zhang}, {Zhang}, {Zhang}, {Zhang}, {Zhang},
  {Zhang}, {Zhou}, {Zhou}, {Zhu}, {Zhu}, \& {Zhuang}}]{2021NatAs...5..378L}
{Li}, C.~K., {Lin}, L., {Xiong}, S.~L., {et~al.} 2021, Nature Astronomy, 5,
  378, \dodoi{10.1038/s41550-021-01302-6}

\bibitem[{{Li} {et~al.}(2014){Li}, {Zhou}, {He}, {Fan}, \&
  {Wei}}]{2014ApJ...797...33L}
{Li}, X., {Zhou}, B., {He}, H.-N., {Fan}, Y.-Z., \& {Wei}, D.-M. 2014, \apj,
  797, 33, \dodoi{10.1088/0004-637X/797/1/33}

\bibitem[{{Li} \& {Zhang}(2020)}]{2020ApJ...899L...6L}
{Li}, Y., \& {Zhang}, B. 2020, \apjl, 899, L6, \dodoi{10.3847/2041-8213/aba907}

\bibitem[{{Liu} {et~al.}(2019){Liu}, {Li}, {Gao}, \&
  {Zhu}}]{2019PhRvD..99l3517L}
{Liu}, B., {Li}, Z., {Gao}, H., \& {Zhu}, Z.-H. 2019, \prd, 99, 123517,
  \dodoi{10.1103/PhysRevD.99.123517}

\bibitem[{{Lorimer} {et~al.}(2007){Lorimer}, {Bailes}, {McLaughlin},
  {Narkevic}, \& {Crawford}}]{2007Sci...318..777L}
{Lorimer}, D.~R., {Bailes}, M., {McLaughlin}, M.~A., {Narkevic}, D.~J., \&
  {Crawford}, F. 2007, Science, 318, 777, \dodoi{10.1126/science.1147532}

\bibitem[{{LSST Science Collaboration} {et~al.}(2009){LSST Science
  Collaboration}, {Abell}, {Allison}, {Anderson}, {Andrew}, {Angel}, {Armus},
  {Arnett}, {Asztalos}, {Axelrod}, {Bailey}, {Ballantyne}, {Bankert},
  {Barkhouse}, {Barr}, {Barrientos}, {Barth}, {Bartlett}, {Becker}, {Becla},
  {Beers}, {Bernstein}, {Biswas}, {Blanton}, {Bloom}, {Bochanski}, {Boeshaar},
  {Borne}, {Bradac}, {Brandt}, {Bridge}, {Brown}, {Brunner}, {Bullock},
  {Burgasser}, {Burge}, {Burke}, {Cargile}, {Chandrasekharan}, {Chartas},
  {Chesley}, {Chu}, {Cinabro}, {Claire}, {Claver}, {Clowe}, {Connolly}, {Cook},
  {Cooke}, {Cooray}, {Covey}, {Culliton}, {de Jong}, {de Vries}, {Debattista},
  {Delgado}, {Dell'Antonio}, {Dhital}, {Di Stefano}, {Dickinson}, {Dilday},
  {Djorgovski}, {Dobler}, {Donalek}, {Dubois-Felsmann}, {Durech},
  {Eliasdottir}, {Eracleous}, {Eyer}, {Falco}, {Fan}, {Fassnacht}, {Ferguson},
  {Fernandez}, {Fields}, {Finkbeiner}, {Figueroa}, {Fox}, {Francke}, {Frank},
  {Frieman}, {Fromenteau}, {Furqan}, {Galaz}, {Gal-Yam}, {Garnavich},
  {Gawiser}, {Geary}, {Gee}, {Gibson}, {Gilmore}, {Grace}, {Green}, {Gressler},
  {Grillmair}, {Habib}, {Haggerty}, {Hamuy}, {Harris}, {Hawley}, {Heavens},
  {Hebb}, {Henry}, {Hileman}, {Hilton}, {Hoadley}, {Holberg}, {Holman},
  {Howell}, {Infante}, {Ivezic}, {Jacoby}, {Jain}, {R}, {Jedicke}, {Jee},
  {Garrett Jernigan}, {Jha}, {Johnston}, {Jones}, {Juric}, {Kaasalainen},
  {Styliani}, {Kafka}, {Kahn}, {Kaib}, {Kalirai}, {Kantor}, {Kasliwal},
  {Keeton}, {Kessler}, {Knezevic}, {Kowalski}, {Krabbendam}, {Krughoff},
  {Kulkarni}, {Kuhlman}, {Lacy}, {Lepine}, {Liang}, {Lien}, {Lira}, {Long},
  {Lorenz}, {Lotz}, {Lupton}, {Lutz}, {Macri}, {Mahabal}, {Mandelbaum},
  {Marshall}, {May}, {McGehee}, {Meadows}, {Meert}, {Milani}, {Miller},
  {Miller}, {Mills}, {Minniti}, {Monet}, {Mukadam}, {Nakar}, {Neill}, {Newman},
  {Nikolaev}, {Nordby}, {O'Connor}, {Oguri}, {Oliver}, {Olivier}, {Olsen},
  {Olsen}, {Olszewski}, {Oluseyi}, {Padilla}, {Parker}, {Pepper}, {Peterson},
  {Petry}, {Pinto}, {Pizagno}, {Popescu}, {Prsa}, {Radcka}, {Raddick},
  {Rasmussen}, {Rau}, {Rho}, {Rhoads}, {Richards}, {Ridgway}, {Robertson},
  {Roskar}, {Saha}, {Sarajedini}, {Scannapieco}, {Schalk}, {Schindler},
  {Schmidt}, {Schmidt}, {Schneider}, {Schumacher}, {Scranton}, {Sebag},
  {Seppala}, {Shemmer}, {Simon}, {Sivertz}, {Smith}, {Allyn Smith}, {Smith},
  {Spitz}, {Stanford}, {Stassun}, {Strader}, {Strauss}, {Stubbs}, {Sweeney},
  {Szalay}, {Szkody}, {Takada}, {Thorman}, {Trilling}, {Trimble}, {Tyson}, {Van
  Berg}, {Vanden Berk}, {VanderPlas}, {Verde}, {Vrsnak}, {Walkowicz},
  {Wandelt}, {Wang}, {Wang}, {Warner}, {Wechsler}, {West}, {Wiecha},
  {Williams}, {Willman}, {Wittman}, {Wolff}, {Wood-Vasey}, {Wozniak}, {Young},
  {Zentner}, \& {Zhan}}]{2009arXiv0912.0201L}
{LSST Science Collaboration}, {Abell}, P.~A., {Allison}, J., {et~al.} 2009,
  arXiv e-prints, arXiv:0912.0201.
\newblock \doarXiv{0912.0201}

\bibitem[{{Luo} {et~al.}(2018){Luo}, {Lee}, {Lorimer}, \&
  {Zhang}}]{2018MNRAS.481.2320L}
{Luo}, R., {Lee}, K., {Lorimer}, D.~R., \& {Zhang}, B. 2018, \mnras, 481, 2320,
  \dodoi{10.1093/mnras/sty2364}

\bibitem[{{Macquart} {et~al.}(2010){Macquart}, {Bailes}, {Bhat}, {Bower},
  {Bunton}, {Chatterjee}, {Colegate}, {Cordes}, {D'Addario}, {Deller},
  {Dodson}, {Fender}, {Haines}, {Halll}, {Harris}, {Hotan}, {Johnston},
  {Jones}, {Keith}, {Koay}, {Lazio}, {Majid}, {Murphy}, {Navarro}, {Phillips},
  {Quinn}, {Preston}, {Stansby}, {Stairs}, {Stappers}, {Staveley-Smith},
  {Tingay}, {Thompson}, {van Straten}, {Wagstaff}, {Warren}, {Wayth}, {Wen}, \&
  {CRAFT Collaboration}}]{2010PASA...27..272M}
{Macquart}, J.-P., {Bailes}, M., {Bhat}, N.~D.~R., {et~al.} 2010, \pasa, 27,
  272, \dodoi{10.1071/AS09082}

\bibitem[{{Macquart} {et~al.}(2020){Macquart}, {Prochaska}, {McQuinn},
  {Bannister}, {Bhandari}, {Day}, {Deller}, {Ekers}, {James}, {Marnoch},
  {Os{\l}owski}, {Phillips}, {Ryder}, {Scott}, {Shannon}, \&
  {Tejos}}]{2020Natur.581..391M}
{Macquart}, J.~P., {Prochaska}, J.~X., {McQuinn}, M., {et~al.} 2020, \nat, 581,
  391, \dodoi{10.1038/s41586-020-2300-2}

\bibitem[{{MAGIC Collaboration} {et~al.}(2018){MAGIC Collaboration}, {Acciari},
  {Ansoldi}, {Antonelli}, {Arbet Engels}, {Arcaro}, {Baack}, {Babi{\'c}}, {},
  {Banerjee}, {Bangale}, {Barres de Almeida}, {Barrio}, {Becerra Gonz{\'a}lez},
  {Bednarek}, {Bernardini}, {Berti}, {Besenrieder}, {Bhattacharyya},
  {Bigongiari}, {Biland}, {Blanch}, {Bonnoli}, {Carosi}, {Ceribella},
  {Chatterjee}, {Colak}, {Colin}, {Colombo}, {Contreras}, {Cortina}, {Covino},
  {Cumani}, {D'Elia}, {da Vela}, {Dazzi}, {de Angelis}, {de Lotto}, {Delfino},
  {Delgado}, {di Pierro}, {Dom{\'\i}nguez}, {Dominis Prester}, {Dorner},
  {Doro}, {Einecke}, {Elsaesser}, {Fallah Ramazani}, {Fattorini},
  {Fern{\'a}ndez-Barral}, {Ferrara}, {Fidalgo}, {Foffano}, {Fonseca}, {Font},
  {Fruck}, {Gallozzi}, {Garc{\'\i}a L{\'o}pez}, {Garczarczyk}, {Gaug},
  {Giammaria}, {Godinovi{\'c}}, {}, {Guberman}, {Hadasch}, {Hahn}, {Hassan},
  {Herrera}, {Hoang}, {Hrupec}, {Inoue}, {Ishio}, {Iwamura}, {Kubo}, {Kushida},
  {Kuve{\v{z}}di{\'c}}, {}, {Lamastra}, {Lelas}, {Leone}, {Lindfors},
  {Lombardi}, {Longo}, {L{\'o}pez}, {L{\'o}pez-Oramas}, {Maggio}, {Majumdar},
  {Makariev}, {Maneva}, {Manganaro}, {Mannheim}, {Maraschi}, {Mariotti},
  {Mart{\'\i}nez}, {Masuda}, {Mazin}, {Minev}, {Miranda}, {Mirzoyan}, {Molina},
  {Moralejo}, {Moreno}, {Moretti}, {Neustroev}, {Niedzwiecki}, {Nievas
  Rosillo}, {Nigro}, {Nilsson}, {Ninci}, {Nishijima}, {Noda}, {Nogu{\'e}s},
  {Paiano}, {Palacio}, {Paneque}, {Paoletti}, {Paredes}, {Pedaletti},
  {Pe{\~n}il}, {Peresano}, {Persic}, {Prada Moroni}, {Prandini}, {Puljak},
  {Garcia}, {Rhode}, {Rib{\'o}}, {Rico}, {Righi}, {Rugliancich}, {Saha},
  {Saito}, {Satalecka}, {Schweizer}, {Sitarek}, {{\v{S}}nidari{\'c}}, {},
  {Sobczynska}, {Somero}, {Stamerra}, {Strzys}, {Suri{\'c}}, {}, {Tavecchio},
  {Temnikov}, {Terzi{\'c}}, {}, {Teshima}, {Torres-Alb{\`a}}, {Tsujimoto},
  {Vanzo}, {Vazquez Acosta}, {Vovk}, {Ward}, {Will}, {Zari{\'c}}, {Marcote},
  {Spitler}, {Hessels}, {Kashiyama}, {Murase}, {Bosch-Ramon}, {Michilli}, \&
  {Seymour}}]{2018MNRAS.481.2479M}
{MAGIC Collaboration}, {Acciari}, V.~A., {Ansoldi}, S., {et~al.} 2018, \mnras,
  481, 2479, \dodoi{10.1093/mnras/sty2422}

\bibitem[{{Majid} {et~al.}(2021){Majid}, {Pearlman}, {Prince}, {Wharton},
  {Naudet}, {Bansal}, {Connor}, {Bhardwaj}, \&
  {Tendulkar}}]{2021ApJ...919L...6M}
{Majid}, W.~A., {Pearlman}, A.~B., {Prince}, T.~A., {et~al.} 2021, \apjl, 919,
  L6, \dodoi{10.3847/2041-8213/ac1921}

\bibitem[{{Marcote} {et~al.}(2017){Marcote}, {Paragi}, {Hessels}, {Keimpema},
  {van Langevelde}, {Huang}, {Bassa}, {Bogdanov}, {Bower}, {Burke-Spolaor},
  {Butler}, {Campbell}, {Chatterjee}, {Cordes}, {Demorest}, {Garrett}, {Ghosh},
  {Kaspi}, {Law}, {Lazio}, {McLaughlin}, {Ransom}, {Salter}, {Scholz},
  {Seymour}, {Siemion}, {Spitler}, {Tendulkar}, \&
  {Wharton}}]{2017ApJ...834L...8M}
{Marcote}, B., {Paragi}, Z., {Hessels}, J.~W.~T., {et~al.} 2017, \apjl, 834,
  L8, \dodoi{10.3847/2041-8213/834/2/L8}

\bibitem[{{Marcote} {et~al.}(2020){Marcote}, {Nimmo}, {Hessels}, {Tendulkar},
  {Bassa}, {Paragi}, {Keimpema}, {Bhardwaj}, {Karuppusamy}, {Kaspi}, {Law},
  {Michilli}, {Aggarwal}, {Andersen}, {Archibald}, {Bandura}, {Bower}, {Boyle},
  {Brar}, {Burke-Spolaor}, {Butler}, {Cassanelli}, {Chawla}, {Demorest},
  {Dobbs}, {Fonseca}, {Giri}, {Good}, {Gourdji}, {Josephy}, {Kirichenko},
  {Kirsten}, {Landecker}, {Lang}, {Lazio}, {Li}, {Lin}, {Linford}, {Masui},
  {Mena-Parra}, {Naidu}, {Ng}, {Patel}, {Pen}, {Pleunis}, {Rafiei-Ravandi},
  {Rahman}, {Renard}, {Scholz}, {Siegel}, {Smith}, {Stairs}, {Vanderlinde}, \&
  {Zwaniga}}]{2020Natur.577..190M}
{Marcote}, B., {Nimmo}, K., {Hessels}, J.~W.~T., {et~al.} 2020, \nat, 577, 190,
  \dodoi{10.1038/s41586-019-1866-z}

\bibitem[{{Margalit} \& {Loeb}(2016)}]{2016MNRAS.460L..25M}
{Margalit}, B., \& {Loeb}, A. 2016, \mnras, 460, L25,
  \dodoi{10.1093/mnrasl/slw068}

\bibitem[{{Masui} {et~al.}(2015){Masui}, {Lin}, {Sievers}, {Anderson}, {Chang},
  {Chen}, {Ganguly}, {Jarvis}, {Kuo}, {Li}, {Liao}, {McLaughlin}, {Pen},
  {Peterson}, {Roman}, {Timbie}, {Voytek}, \& {Yadav}}]{2015Natur.528..523M}
{Masui}, K., {Lin}, H.-H., {Sievers}, J., {et~al.} 2015, \nat, 528, 523,
  \dodoi{10.1038/nature15769}

\bibitem[{{McQuinn}(2014)}]{2014ApJ...780L..33M}
{McQuinn}, M. 2014, \apjl, 780, L33, \dodoi{10.1088/2041-8205/780/2/L33}

\bibitem[{{Mena-Parra} {et~al.}(2022){Mena-Parra}, {Leung}, {Cary}, {Masui},
  {Kaczmarek}, {Amiri}, {Bandura}, {Boyle}, {Cassanelli}, {Cliche}, {Dobbs},
  {Kaspi}, {Landecker}, {Lanman}, {Sievers}, {Sievers}, \& {Chime/Frb
  Collaboration}}]{2022AJ....163...48M}
{Mena-Parra}, J., {Leung}, C., {Cary}, S., {et~al.} 2022, \aj, 163, 48,
  \dodoi{10.3847/1538-3881/ac397a}

\bibitem[{{Metzger} {et~al.}(2020){Metzger}, {Fang}, \&
  {Margalit}}]{2020ApJ...902L..22M}
{Metzger}, B.~D., {Fang}, K., \& {Margalit}, B. 2020, \apjl, 902, L22,
  \dodoi{10.3847/2041-8213/abbb88}

\bibitem[{{Michilli} {et~al.}(2018){Michilli}, {Seymour}, {Hessels}, {Spitler},
  {Gajjar}, {Archibald}, {Bower}, {Chatterjee}, {Cordes}, {Gourdji}, {Heald},
  {Kaspi}, {Law}, {Sobey}, {Adams}, {Bassa}, {Bogdanov}, {Brinkman},
  {Demorest}, {Fernandez}, {Hellbourg}, {Lazio}, {Lynch}, {Maddox}, {Marcote},
  {McLaughlin}, {Paragi}, {Ransom}, {Scholz}, {Siemion}, {Tendulkar}, {van
  Rooy}, {Wharton}, \& {Whitlow}}]{2018Natur.553..182M}
{Michilli}, D., {Seymour}, A., {Hessels}, J.~W.~T., {et~al.} 2018, \nat, 553,
  182, \dodoi{10.1038/nature25149}

\bibitem[{{Mu{\~n}oz} {et~al.}(2016){Mu{\~n}oz}, {Kovetz}, {Dai}, \&
  {Kamionkowski}}]{2016PhRvL.117i1301M}
{Mu{\~n}oz}, J.~B., {Kovetz}, E.~D., {Dai}, L., \& {Kamionkowski}, M. 2016,
  \prl, 117, 091301, \dodoi{10.1103/PhysRevLett.117.091301}

\bibitem[{{Mu{\~n}oz} \& {Loeb}(2018)}]{2018PhRvD..98j3518M}
{Mu{\~n}oz}, J.~B., \& {Loeb}, A. 2018, \prd, 98, 103518,
  \dodoi{10.1103/PhysRevD.98.103518}

\bibitem[{{Newburgh} {et~al.}(2016){Newburgh}, {Bandura}, {Bucher}, {Chang},
  {Chiang}, {Cliche}, {Dav{\'e}}, {Dobbs}, {Clarkson}, {Ganga}, {Gogo},
  {Gumba}, {Gupta}, {Hilton}, {Johnstone}, {Karastergiou}, {Kunz}, {Lokhorst},
  {Maartens}, {Macpherson}, {Mdlalose}, {Moodley}, {Ngwenya}, {Parra},
  {Peterson}, {Recnik}, {Saliwanchik}, {Santos}, {Sievers}, {Smirnov},
  {Stronkhorst}, {Taylor}, {Vanderlinde}, {Van Vuuren}, {Weltman}, \&
  {Witzemann}}]{2016SPIE.9906E..5XN}
{Newburgh}, L.~B., {Bandura}, K., {Bucher}, M.~A., {et~al.} 2016, in Society of
  Photo-Optical Instrumentation Engineers (SPIE) Conference Series, Vol. 9906,
  Ground-based and Airborne Telescopes VI, ed. H.~J. {Hall}, R.~{Gilmozzi}, \&
  H.~K. {Marshall}, 99065X, \dodoi{10.1117/12.2234286}

\bibitem[{{Ng} {et~al.}(2017){Ng}, {Vanderlinde}, {Paradise}, {Klages},
  {Masui}, {Smith}, {Bandura}, {Boyle}, {Dobbs}, {Kaspi}, {Renard}, {Shaw},
  {Stairs}, \& {Tretyakov}}]{2017ursi.confE...4N}
{Ng}, C., {Vanderlinde}, K., {Paradise}, A., {et~al.} 2017, in XXXII
  International Union of Radio Science General Assembly \& Scientific Symposium
  (URSI GASS) 2017, 4, \dodoi{10.23919/URSIGASS.2017.8105318}

\bibitem[{{Niu} {et~al.}(2021){Niu}, {Li}, {Luo}, {Wang}, {Yao}, {Zhang},
  {Zhu}, {Wang}, {Ye}, {Zhang}, {Niu}, {Tang}, {Duan}, {Krco}, {Dai}, {Feng},
  {Miao}, {Pan}, {Qian}, {Xue}, {Yuan}, {Yue}, {Zhang}, \&
  {Zhang}}]{2021ApJ...909L...8N}
{Niu}, C.-H., {Li}, D., {Luo}, R., {et~al.} 2021, \apjl, 909, L8,
  \dodoi{10.3847/2041-8213/abe7f0}

\bibitem[{{Niu} {et~al.}(2022){Niu}, {Aggarwal}, {Li}, {Zhang}, {Chatterjee},
  {Tsai}, {Yu}, {Law}, {Burke-Spolaor}, {Cordes}, {Zhang}, {Ocker}, {Yao},
  {Wan}, {Feng}, {Niino}, {Bochenek}, {Cruces}, {Connor}, {Jiang}, {Dai},
  {Luo}, {Li}, {Miao}, {Niu}, {Anna-Thomas}, {Sydnor}, {Stern}, {Wang}, {Yuan},
  {Yue}, {Zhou}, {Yan}, {Zhu}, \& {Zhang}}]{2022Natur.606..873N}
{Niu}, C.~H., {Aggarwal}, K., {Li}, D., {et~al.} 2022, \nat, 606, 873,
  \dodoi{10.1038/s41586-022-04755-5}

\bibitem[{{Ocker} {et~al.}(2021){Ocker}, {Cordes}, \&
  {Chatterjee}}]{2021ApJ...911..102O}
{Ocker}, S.~K., {Cordes}, J.~M., \& {Chatterjee}, S. 2021, \apj, 911, 102,
  \dodoi{10.3847/1538-4357/abeb6e}

\bibitem[{{Oppermann} {et~al.}(2018){Oppermann}, {Yu}, \&
  {Pen}}]{2018MNRAS.475.5109O}
{Oppermann}, N., {Yu}, H.-R., \& {Pen}, U.-L. 2018, \mnras, 475, 5109,
  \dodoi{10.1093/mnras/sty004}

\bibitem[{{Pen}(2018)}]{2018NatAs...2..842P}
{Pen}, U.-L. 2018, Nature Astronomy, 2, 842, \dodoi{10.1038/s41550-018-0620-z}

\bibitem[{{Petroff} {et~al.}(2019){Petroff}, {Hessels}, \&
  {Lorimer}}]{2019A&ARv..27....4P}
{Petroff}, E., {Hessels}, J.~W.~T., \& {Lorimer}, D.~R. 2019, \aapr, 27, 4,
  \dodoi{10.1007/s00159-019-0116-6}

\bibitem[{{Petroff} {et~al.}(2014){Petroff}, {van Straten}, {Johnston},
  {Bailes}, {Barr}, {Bates}, {Bhat}, {Burgay}, {Burke-Spolaor}, {Champion},
  {Coster}, {Flynn}, {Keane}, {Keith}, {Kramer}, {Levin}, {Ng}, {Possenti},
  {Stappers}, {Tiburzi}, \& {Thornton}}]{2014ApJ...789L..26P}
{Petroff}, E., {van Straten}, W., {Johnston}, S., {et~al.} 2014, \apjl, 789,
  L26, \dodoi{10.1088/2041-8205/789/2/L26}

\bibitem[{{Petroff} {et~al.}(2016){Petroff}, {Barr}, {Jameson}, {Keane},
  {Bailes}, {Kramer}, {Morello}, {Tabbara}, \& {van
  Straten}}]{2016PASA...33...45P}
{Petroff}, E., {Barr}, E.~D., {Jameson}, A., {et~al.} 2016, \pasa, 33, e045,
  \dodoi{10.1017/pasa.2016.35}

\bibitem[{{Petroff} {et~al.}(2017){Petroff}, {Houben}, {Bannister},
  {Burke-Spolaor}, {Cordes}, {Falcke}, {van Haren}, {Karastergiou}, {Kramer},
  {Law}, {van Leeuwen}, {Lorimer}, {Martinez-Rubi}, {Rachen}, {Spitler}, \&
  {Weltman}}]{2017arXiv171008155P}
{Petroff}, E., {Houben}, L., {Bannister}, K., {et~al.} 2017, arXiv e-prints,
  arXiv:1710.08155.
\newblock \doarXiv{1710.08155}

\bibitem[{{Piro} {et~al.}(2021){Piro}, {Bruni}, {Troja}, {O'Connor}, {Panessa},
  {Ricci}, {Zhang}, {Burgay}, {Dichiara}, {Lee}, {Lotti}, {Niu}, {Pilia},
  {Possenti}, {Trudu}, {Xu}, {Zhu}, {Kutyrev}, \&
  {Veilleux}}]{2021A&A...656L..15P}
{Piro}, L., {Bruni}, G., {Troja}, E., {et~al.} 2021, \aap, 656, L15,
  \dodoi{10.1051/0004-6361/202141903}

\bibitem[{{Platts} {et~al.}(2019){Platts}, {Weltman}, {Walters}, {Tendulkar},
  {Gordin}, \& {Kandhai}}]{2019PhR...821....1P}
{Platts}, E., {Weltman}, A., {Walters}, A., {et~al.} 2019, \physrep, 821, 1,
  \dodoi{10.1016/j.physrep.2019.06.003}

\bibitem[{{Popov} \& {Postnov}(2013)}]{2013arXiv1307.4924P}
{Popov}, S.~B., \& {Postnov}, K.~A. 2013, arXiv e-prints, arXiv:1307.4924.
\newblock \doarXiv{1307.4924}

\bibitem[{{Popov} \& {Pshirkov}(2016)}]{2016MNRAS.462L..16P}
{Popov}, S.~B., \& {Pshirkov}, M.~S. 2016, \mnras, 462, L16,
  \dodoi{10.1093/mnrasl/slw118}

\bibitem[{{Prochaska} \& {Zheng}(2019)}]{2019MNRAS.485..648P}
{Prochaska}, J.~X., \& {Zheng}, Y. 2019, \mnras, 485, 648,
  \dodoi{10.1093/mnras/stz261}

\bibitem[{{Rafiei-Ravandi} {et~al.}(2021){Rafiei-Ravandi}, {Smith}, {Li},
  {Masui}, {Josephy}, {Dobbs}, {Lang}, {Bhardwaj}, {Patel}, {Bandura},
  {Berger}, {Boyle}, {Brar}, {Breitman}, {Cassanelli}, {Chawla}, {Adam Dong},
  {Fonseca}, {Gaensler}, {Giri}, {Good}, {Halpern}, {Kaczmarek}, {Kaspi},
  {Leung}, {Lin}, {Mena-Parra}, {Meyers}, {Michilli}, {M{\"u}nchmeyer}, {Ng},
  {Petroff}, {Pleunis}, {Rahman}, {Sanghavi}, {Scholz}, {Shin}, {Stairs},
  {Tendulkar}, {Vanderlinde}, \& {Zwaniga}}]{2021ApJ...922...42R}
{Rafiei-Ravandi}, M., {Smith}, K.~M., {Li}, D., {et~al.} 2021, \apj, 922, 42,
  \dodoi{10.3847/1538-4357/ac1dab}

\bibitem[{{Rajwade} {et~al.}(2021){Rajwade}, {Stappers}, {Williams}, {Barr},
  {Christiaan Bezuidenhout}, {Caleb}, {Driessen}, {Jankowski}, {Malenta},
  {Morello}, {Sanidas}, \& {Surnis}}]{2021arXiv210308410R}
{Rajwade}, K., {Stappers}, B., {Williams}, C., {et~al.} 2021, arXiv e-prints,
  arXiv:2103.08410.
\newblock \doarXiv{2103.08410}

\bibitem[{{Rajwade} {et~al.}(2020){Rajwade}, {Mickaliger}, {Stappers},
  {Morello}, {Agarwal}, {Bassa}, {Breton}, {Caleb}, {Karastergiou}, {Keane}, \&
  {Lorimer}}]{2020MNRAS.495.3551R}
{Rajwade}, K.~M., {Mickaliger}, M.~B., {Stappers}, B.~W., {et~al.} 2020,
  \mnras, 495, 3551, \dodoi{10.1093/mnras/staa1237}

\bibitem[{{Ravi}(2019{\natexlab{a}})}]{2019NatAs...3..928R}
{Ravi}, V. 2019{\natexlab{a}}, Nature Astronomy, 3, 928,
  \dodoi{10.1038/s41550-019-0831-y}

\bibitem[{{Ravi}(2019{\natexlab{b}})}]{2019ApJ...872...88R}
---. 2019{\natexlab{b}}, \apj, 872, 88, \dodoi{10.3847/1538-4357/aafb30}

\bibitem[{{Ravi} {et~al.}(2019){Ravi}, {Catha}, {D'Addario}, {Djorgovski},
  {Hallinan}, {Hobbs}, {Kocz}, {Kulkarni}, {Shi}, {Vedantham}, {Weinreb}, \&
  {Woody}}]{2019Natur.572..352R}
{Ravi}, V., {Catha}, M., {D'Addario}, L., {et~al.} 2019, \nat, 572, 352,
  \dodoi{10.1038/s41586-019-1389-7}

\bibitem[{{Schr{\"o}der}(2017)}]{2017PrPNP..93....1S}
{Schr{\"o}der}, F.~G. 2017, Progress in Particle and Nuclear Physics, 93, 1,
  \dodoi{10.1016/j.ppnp.2016.12.002}

\bibitem[{{Shin} {et~al.}(2022){Shin}, {Masui}, {Bhardwaj}, {Cassanelli},
  {Chawla}, {Dobbs}, {Dong}, {Fonseca}, {Gaensler}, {Herrera-Mart{\'\i}n},
  {Kaczmarek}, {Kaspi}, {Leung}, {Merryfield}, {M{\"u}nchmeyer}, {Pearlman},
  {Rafiei-Ravandi}, {Smith}, \& {Tendulkar}}]{2022arXiv220714316S}
{Shin}, K., {Masui}, K.~W., {Bhardwaj}, M., {et~al.} 2022, arXiv e-prints,
  arXiv:2207.14316.
\newblock \doarXiv{2207.14316}

\bibitem[{{Slosar} {et~al.}(2019){Slosar}, {Ahmed}, {Alonso}, {Amin}, {Arena},
  {Bandura}, {Battaglia}, {Blazek}, {Bull}, {Castorina}, {Chang}, {Connor},
  {Dav{\'e}}, {Dvorkin}, {van Engelen}, {Ferraro}, {Flauger}, {Foreman},
  {Frisch}, {Green}, {Holder}, {Jacobs}, {Johnson}, {Dillon}, {Karagiannis},
  {Kaurov}, {Knox}, {Liu}, {Loverde}, {Ma}, {Masui}, {McClintock}, {Moodley},
  {Munchmeyer}, {Newburgh}, {Ng}, {Nomerotski}, {O'Connor}, {Obuljen},
  {Padmanabhan}, {Parkinson}, {Prochaska}, {Rajendran}, {Rapetti},
  {Saliwanchik}, {Schaan}, {Sehgal}, {Shaw}, {Sheehy}, {Sheldon}, {Shirley},
  {Silverstein}, {Slatyer}, {Slosar}, {Stankus}, {Stebbins}, {Timbie},
  {Tucker}, {Tyndall}, {Villaescusa Navarro}, {Wallisch}, \&
  {White}}]{2019BAAS...51g..53S}
{Slosar}, A., {Ahmed}, Z., {Alonso}, D., {et~al.} 2019, in Bulletin of the
  American Astronomical Society, Vol.~51, 53.
\newblock \doarXiv{1907.12559}

\bibitem[{{Spinelli} {et~al.}(2021){Spinelli}, {Bernardi}, {Garsden},
  {Greenhill}, {Fialkov}, {Dowell}, \& {Price}}]{2021MNRAS.505.1575S}
{Spinelli}, M., {Bernardi}, G., {Garsden}, H., {et~al.} 2021, \mnras, 505,
  1575, \dodoi{10.1093/mnras/stab1363}

\bibitem[{{Spitler} {et~al.}(2016){Spitler}, {Scholz}, {Hessels}, {Bogdanov},
  {Brazier}, {Camilo}, {Chatterjee}, {Cordes}, {Crawford}, {Deneva}, {Ferdman},
  {Freire}, {Kaspi}, {Lazarus}, {Lynch}, {Madsen}, {McLaughlin}, {Patel},
  {Ransom}, {Seymour}, {Stairs}, {Stappers}, {van Leeuwen}, \&
  {Zhu}}]{2016Natur.531..202S}
{Spitler}, L.~G., {Scholz}, P., {Hessels}, J.~W.~T., {et~al.} 2016, \nat, 531,
  202, \dodoi{10.1038/nature17168}

\bibitem[{{Tavani} {et~al.}(2021){Tavani}, {Casentini}, {Ursi}, {Verrecchia},
  {Addis}, {Antonelli}, {Argan}, {Barbiellini}, {Baroncelli}, {Bernardi},
  {Bianchi}, {Bulgarelli}, {Caraveo}, {Cardillo}, {Cattaneo}, {Chen}, {Costa},
  {Del Monte}, {Di Cocco}, {Di Persio}, {Donnarumma}, {Evangelista}, {Feroci},
  {Ferrari}, {Fioretti}, {Fuschino}, {Galli}, {Gianotti}, {Giuliani},
  {Labanti}, {Lazzarotto}, {Lipari}, {Longo}, {Lucarelli}, {Magro},
  {Marisaldi}, {Mereghetti}, {Morelli}, {Morselli}, {Naldi}, {Pacciani},
  {Parmiggiani}, {Paoletti}, {Pellizzoni}, {Perri}, {Perotti}, {Piano},
  {Picozza}, {Pilia}, {Pittori}, {Puccetti}, {Pupillo}, {Rapisarda},
  {Rappoldi}, {Rubini}, {Setti}, {Soffitta}, {Trifoglio}, {Trois},
  {Vercellone}, {Vittorini}, {Giommi}, \& {D'Amico}}]{2021NatAs...5..401T}
{Tavani}, M., {Casentini}, C., {Ursi}, A., {et~al.} 2021, Nature Astronomy, 5,
  401, \dodoi{10.1038/s41550-020-01276-x}

\bibitem[{{Taylor}(1974)}]{1974A&AS...15..367T}
{Taylor}, J.~H. 1974, \aaps, 15, 367

\bibitem[{{Tegmark} \& {Zaldarriaga}(2009)}]{2009PhRvD..79h3530T}
{Tegmark}, M., \& {Zaldarriaga}, M. 2009, \prd, 79, 083530,
  \dodoi{10.1103/PhysRevD.79.083530}

\bibitem[{{Tendulkar} {et~al.}(2017){Tendulkar}, {Bassa}, {Cordes}, {Bower},
  {Law}, {Chatterjee}, {Adams}, {Bogdanov}, {Burke-Spolaor}, {Butler},
  {Demorest}, {Hessels}, {Kaspi}, {Lazio}, {Maddox}, {Marcote}, {McLaughlin},
  {Paragi}, {Ransom}, {Scholz}, {Seymour}, {Spitler}, {van Langevelde}, \&
  {Wharton}}]{2017ApJ...834L...7T}
{Tendulkar}, S.~P., {Bassa}, C.~G., {Cordes}, J.~M., {et~al.} 2017, \apjl, 834,
  L7, \dodoi{10.3847/2041-8213/834/2/L7}

\bibitem[{{The LIGO Scientific Collaboration} {et~al.}(2022){The LIGO
  Scientific Collaboration}, {the Virgo Collaboration}, {the KAGRA
  Collaboration}, {the CHIME/FRB Collaboration}, {:}, {Abbott}, {Abbott},
  {Acernese}, {Ackley}, {Adams}, {Adhikari}, {Adhikari}, {Adya}, {Affeldt},
  {Agarwal}, {Agathos}, {Agatsuma}, {Aggarwal}, {Aguiar}, {Aiello}, {Ain},
  {Ajith}, {Akutsu}, {Albanesi}, {Allocca}, {Altin}, {Amato}, {Anand}, {Anand},
  {Ananyeva}, {Anderson}, {Anderson}, {Ando}, {Andrade}, {Andres},
  {Andri{\'c}}, {Angelova}, {Ansoldi}, {Antelis}, {Antier}, {Appert}, {Arai},
  {Arai}, {Arai}, {Araki}, {Araya}, {Araya}, {Areeda}, {Ar{\`e}ne}, {Aritomi},
  {Arnaud}, {Aronson}, {Arun}, {Asada}, {Asali}, {Ashton}, {Aso}, {Assiduo},
  {Aston}, {Astone}, {Aubin}, {Austin}, {Babak}, {Badaracco}, {Bader},
  {Badger}, {Bae}, {Bae}, {Baer}, {Bagnasco}, {Bai}, {Baiotti}, {Baird},
  {Bajpai}, {Ball}, {Ballardin}, {Ballmer}, {Balsamo}, {Baltus}, {Banagiri},
  {Bankar}, {Barayoga}, {Barbieri}, {Barish}, {Barker}, {Barneo}, {Barone},
  {Barr}, {Barsotti}, {Barsuglia}, {Barta}, {Bartlett}, {Barton}, {Bartos},
  {Bassiri}, {Basti}, {Bawaj}, {Bayley}, {Baylor}, {Bazzan}, {B{\'e}csy},
  {Bedakihale}, {Bejger}, {Belahcene}, {Benedetto}, {Beniwal}, {Bennett},
  {Bentley}, {BenYaala}, {Bergamin}, {Berger}, {Bernuzzi}, {Berry},
  {Bersanetti}, {Bertolini}, {Betzwieser}, {Beveridge}, {Bhandare}, {Bhardwaj},
  {Bhattacharjee}, {Bhaumik}, {Bilenko}, {Billingsley}, {Bini}, {Birney},
  {Birnholtz}, {Biscans}, {Bischi}, {Biscoveanu}, {Bisht}, {Biswas}, {Bitossi},
  {Bizouard}, {Blackburn}, {Blair}, {Blair}, {Blair}, {Bobba}, {Bode}, {Boer},
  {Bogaert}, {Boldrini}, {Bonavena}, {Bondu}, {Bonilla}, {Bonnand}, {Booker},
  {Boom}, {Bork}, {Boschi}, {Bose}, {Bose}, {Bossilkov}, {Boudart},
  {Bouffanais}, {Bozzi}, {Bradaschia}, {Brady}, {Bramley}, {Branch},
  {Branchesi}, {Brau}, {Breschi}, {Briant}, {Briggs}, {Brillet}, {Brinkmann},
  {Brockill}, {Brooks}, {Brooks}, {Brown}, {Brunett}, {Bruno}, {Bruntz},
  {Bryant}, {Buchanan}, {Bulik}, {Bulten}, {Buonanno}, {Buscicchio},
  {Buskulic}, {Buy}, {Byer}, {Cadonati}, {Cagnoli}, {Cahillane}, {Calder{\'o}n
  Bustillo}, {Callaghan}, {Callister}, {Calloni}, {Cameron}, {Camp}, {Canepa},
  {Canevarolo}, {Cannavacciuolo}, {Cannon}, {Cao}, {Cao}, {Capocasa}, {Capote},
  {Carapella}, {Carbognani}, {Carlin}, {Carney}, {Carpinelli}, {Carrillo},
  {Carullo}, {Carver}, {Casanueva Diaz}, {Casentini}, {Castaldi}, {Caudill},
  {Cavagli{\`a}}, {Cavalier}, {Cavalieri}, {Ceasar}, {Cella},
  {Cerd{\'a}-Dur{\'a}n}, {Cesarini}, {Chaibi}, {Chakravarti}, {Chalathadka
  Subrahmanya}, {Champion}, {Chan}, {Chan}, {Chan}, {Chan}, {Chan}, {Chandra},
  {Chanial}, {Chao}, {Charlton}, {Chase}, {Chassande-Mottin}, {Chatterjee},
  {Chatterjee}, {Chatterjee}, {Chaturvedi}, {Chaty}, {Chen}, {Chen}, {Chen},
  {Chen}, {Chen}, {Chen}, {Chen}, {Chen}, {Cheng}, {Cheong}, {Cheung}, {Chia},
  {Chiadini}, {Chiang}, {Chiarini}, {Chierici}, {Chincarini}, {Chiofalo},
  {Chiummo}, {Cho}, {Cho}, {Choudhary}, {Choudhary}, {Christensen}, {Chu},
  {Chu}, {Chu}, {Chua}, {Chung}, {Ciani}, {Ciecielag}, {Cie{\'s}lar},
  {Cifaldi}, {Ciobanu}, {Ciolfi}, {Cipriano}, {Cirone}, {Clara}, {Clark},
  {Clark}, {Clarke}, {Clearwater}, {Clesse}, {Cleva}, {Coccia}, {Codazzo},
  {Cohadon}, {Cohen}, {Cohen}, {Colleoni}, {Collette}, {Colombo}, {Colpi},
  {Compton}, {Constancio}, {Conti}, {Cooper}, {Corban}, {Corbitt},
  {Cordero-Carri{\'o}n}, {Corezzi}, {Corley}, {Cornish}, {Corre}, {Corsi},
  {Cortese}, {Costa}, {Cotesta}, {Coughlin}, {Coulon}, {Countryman}, {Cousins},
  {Couvares}, {Coward}, {Cowart}, {Coyne}, {Coyne}, {Creighton}, {Creighton},
  {Criswell}, {Croquette}, {Crowder}, {Cudell}, {Cullen}, {Cumming},
  {Cummings}, {Cunningham}, {Cuoco}, {Cury{\l}o}, {Dabadie}, {Dal Canton},
  {Dall'Osso}, {D{\'a}lya}, {Dana}, {DaneshgaranBajastani}, {D'Angelo},
  {Danilishin}, {D'Antonio}, {Danzmann}, {Darsow-Fromm}, {Dasgupta}, {Datrier},
  {Datta}, {Dattilo}, {Dave}, {Davier}, {Davies}, {Davis}, {Davis}, {Daw},
  {Dean}, {DeBra}, {Deenadayalan}, {Degallaix}, {De Laurentis},
  {Del{\'e}glise}, {Del Favero}, {De Lillo}, {De Lillo}, {Del Pozzo},
  {DeMarchi}, {De Matteis}, {D'Emilio}, {Demos}, {Dent}, {Depasse}, {De
  Pietri}, {De Rosa}, {De Rossi}, {DeSalvo}, {De Simone}, {Dhurandhar},
  {D{\'\i}az}, {Diaz-Ortiz}, {Didio}, {Dietrich}, {Di Fiore}, {Di Fronzo}, {Di
  Giorgio}, {Di Giovanni}, {Di Giovanni}, {Di Girolamo}, {Di Lieto}, {Ding},
  {Di Pace}, {Di Palma}, {Di Renzo}, {Divakarla}, {Dmitriev}, {Doctor},
  {D'Onofrio}, {Donovan}, {Dooley}, {Doravari}, {Dorrington}, {Drago},
  {Driggers}, {Drori}, {Ducoin}, {Dupej}, {Durante}, {D'Urso}, {Duverne},
  {Dwyer}, {Eassa}, {Easter}, {Ebersold}, {Eckhardt}, {Eddolls}, {Edelman},
  {Edo}, {Edy}, {Effler}, {Eguchi}, {Eichholz}, {Eikenberry}, {Eisenmann},
  {Eisenstein}, {Ejlli}, {Engelby}, {Enomoto}, {Errico}, {Essick},
  {Estell{\'e}s}, {Estevez}, {Etienne}, {Etzel}, {Evans}, {Evans}, {Ewing},
  {Fafone}, {Fair}, {Fairhurst}, {Farah}, {Farinon}, {Farr}, {Farr}, {Farrow},
  {Fauchon-Jones}, {Favaro}, {Favata}, {Fays}, {Fazio}, {Feicht}, {Fejer},
  {Fenyvesi}, {Ferguson}, {Fernandez-Galiana}, {Ferrante}, {Ferreira},
  {Fidecaro}, {Figura}, {Fiori}, {Fishbach}, {Fisher}, {Fittipaldi}, {Fiumara},
  {Flaminio}, {Floden}, {Fong}, {Font}, {Fornal}, {Forsyth}, {Franke},
  {Frasca}, {Frasconi}, {Frederick}, {Freed}, {Frei}, {Freise}, {Frey},
  {Fritschel}, {Frolov}, {Fronz{\'e}}, {Fujii}, {Fujikawa}, {Fukunaga},
  {Fukushima}, {Fulda}, {Fyffe}, {Gabbard}, {Gadre}, {Gair}, {Gais},
  {Galaudage}, {Gamba}, {Ganapathy}, {Ganguly}, {Gao}, {Gaonkar}, {Garaventa},
  {Garc{\'\i}a-N{\'u}{\~n}ez}, {Garc{\'\i}a-Quir{\'o}s}, {Garufi}, {Gateley},
  {Gaudio}, {Gayathri}, {Ge}, {Gemme}, {Gennai}, {George}, {Gerberding},
  {Gergely}, {Gewecke}, {Ghonge}, {Ghosh}, {Ghosh}, {Ghosh}, {Ghosh},
  {Giacomazzo}, {Giacoppo}, {Giaime}, {Giardina}, {Gibson}, {Gier}, {Giesler},
  {Giri}, {Gissi}, {Glanzer}, {Gleckl}, {Godwin}, {Goetz}, {Goetz}, {Gohlke},
  {Goncharov}, {Gonz{\'a}lez}, {Gopakumar}, {Gosselin}, {Gouaty}, {Gould},
  {Grace}, {Grado}, {Granata}, {Granata}, {Grant}, {Gras}, {Grassia}, {Gray},
  {Gray}, {Greco}, {Green}, {Green}, {Gretarsson}, {Gretarsson}, {Griffith},
  {Griffiths}, {Griggs}, {Grignani}, {Grimaldi}, {Grimm}, {Grote}, {Grunewald},
  {Gruning}, {Guerra}, {Guidi}, {Guimaraes}, {Guix{\'e}}, {Gulati}, {Guo},
  {Guo}, {Gupta}, {Gupta}, {Gupta}, {Gustafson}, {Gustafson}, {Guzman}, {Ha},
  {Haegel}, {Hagiwara}, {Haino}, {Halim}, {Hall}, {Hamilton}, {Hammond}, {Han},
  {Haney}, {Hanks}, {Hanna}, {Hannam}, {Hannuksela}, {Hansen}, {Hansen},
  {Hanson}, {Harder}, {Hardwick}, {Haris}, {Harms}, {Harry}, {Harry},
  {Hartwig}, {Hasegawa}, {Haskell}, {Hasskew}, {Haster}, {Hattori}, {Haughian},
  {Hayakawa}, {Hayama}, {Hayes}, {Healy}, {Heidmann}, {Heidt}, {Heintze},
  {Heinze}, {Heinzel}, {Heitmann}, {Hellman}, {Hello}, {Helmling-Cornell},
  {Hemming}, {Hendry}, {Heng}, {Hennes}, {Hennig}, {Hennig}, {Hernandez},
  {Hernandez Vivanco}, {Heurs}, {Hild}, {Hill}, {Himemoto}, {Hines},
  {Hiranuma}, {Hirata}, {Hirose}, {Hochheim}, {Hofman}, {Hohmann}, {Holcomb},
  {Holland}, {Hollows}, {Holmes}, {Holt}, {Holz}, {Hong}, {Hopkins}, {Hough},
  {Hourihane}, {Howell}, {Hoy}, {Hoyland}, {Hreibi}, {Hsieh}, {Hsu}, {Huang},
  {Huang}, {Huang}, {Huang}, {Huang}, {Huang}, {H{\"u}bner}, {Huddart},
  {Hughey}, {Hui}, {Hui}, {Husa}, {Huttner}, {Huxford}, {Huynh-Dinh}, {Ide},
  {Idzkowski}, {Iess}, {Ikenoue}, {Imam}, {Inayoshi}, {Ingram}, {Inoue},
  {Ioka}, {Isi}, {Isleif}, {Ito}, {Itoh}, {Iyer}, {Izumi}, {JaberianHamedan},
  {Jacqmin}, {Jadhav}, {Jadhav}, {James}, {Jan}, {Jani}, {Janquart},
  {Janssens}, {Janthalur}, {Jaranowski}, {Jariwala}, {Jaume}, {Jenkins},
  {Jenner}, {Jeon}, {Jeunon}, {Jia}, {Jin}, {Johns}, {Jones}, {Jones}, {Jones},
  {Jones}, {Jones}, {Jonker}, {Ju}, {Jung}, {Jung}, {Junker}, {Juste},
  {Kaihotsu}, {Kajita}, {Kakizaki}, {Kalaghatgi}, {Kalogera}, {Kamai},
  {Kamiizumi}, {Kanda}, {Kandhasamy}, {Kang}, {Kanner}, {Kao}, {Kapadia},
  {Kapasi}, {Karat}, {Karathanasis}, {Karki}, {Kashyap}, {Kasprzack},
  {Kastaun}, {Katsanevas}, {Katsavounidis}, {Katzman}, {Kaur}, {Kawabe},
  {Kawaguchi}, {Kawai}, {Kawasaki}, {K{\'e}f{\'e}lian}, {Keitel}, {Key},
  {Khadka}, {Khalili}, {Khan}, {Khazanov}, {Khetan}, {Khursheed}, {Kijbunchoo},
  {Kim}, {Kim}, {Kim}, {Kim}, {Kim}, {Kim}, {Kimball}, {Kimura},
  {Kinley-Hanlon}, {Kirchhoff}, {Kissel}, {Kita}, {Kitazawa}, {Kleybolte},
  {Klimenko}, {Knee}, {Knowles}, {Knyazev}, {Koch}, {Koekoek}, {Kojima},
  {Kokeyama}, {Koley}, {Kolitsidou}, {Kolstein}, {Komori}, {Kondrashov},
  {Kong}, {Kontos}, {Koper}, {Korobko}, {Kotake}, {Kovalam}, {Kozak},
  {Kozakai}, {Kozu}, {Kringel}, {Krishnendu}, {Kr{\'o}lak}, {Kuehn}, {Kuei},
  {Kuijer}, {Kumar}, {Kumar}, {Kumar}, {Kumar}, {Kume}, {Kuns}, {Kuo}, {Kuo},
  {Kuromiya}, {Kuroyanagi}, {Kusayanagi}, {Kuwahara}, {Kwak}, {Lagabbe},
  {Laghi}, {Lalande}, {Lam}, {Lamberts}, {Landry}, {Lane}, {Lang}, {Lange},
  {Lantz}, {La Rosa}, {Lartaux-Vollard}, {Lasky}, {Laxen}, {Lazzarini},
  {Lazzaro}, {Leaci}, {Leavey}, {Lecoeuche}, {Lee}, {Lee}, {Lee}, {Lee}, {Lee},
  {Lee}, {Lehmann}, {Lema{\^\i}tre}, {Leonardi}, {Leroy}, {Letendre},
  {Levesque}, {Levin}, {Leviton}, {Leyde}, {Li}, {Li}, {Li}, {Li}, {Li}, {Li},
  {Lin}, {Lin}, {Lin}, {Lin}, {Lin}, {Linde}, {Linker}, {Linley}, {Littenberg},
  {Liu}, {Liu}, {Liu}, {Liu}, {Llamas}, {Llorens-Monteagudo}, {Lo}, {Lockwood},
  {London}, {Longo}, {Lopez}, {Lopez Portilla}, {Lorenzini}, {Loriette},
  {Lormand}, {Losurdo}, {Lott}, {Lough}, {Lousto}, {Lovelace}, {Lucaccioni},
  {L{\"u}ck}, {Lumaca}, {Lundgren}, {Luo}, {Lynam}, {Macas}, {MacInnis},
  {Macleod}, {MacMillan}, {Macquet}, {Maga{\~n}a Hernandez}, {Magazz{\`u}},
  {Magee}, {Maggiore}, {Magnozzi}, {Mahesh}, {Majorana}, {Makarem},
  {Maksimovic}, {Maliakal}, {Malik}, {Man}, {Mandic}, {Mangano}, {Mango},
  {Mansell}, {Manske}, {Mantovani}, {Mapelli}, {Marchesoni}, {Marchio},
  {Marion}, {Mark}, {M{\'a}rka}, {M{\'a}rka}, {Markakis}, {Markosyan},
  {Markowitz}, {Maros}, {Marquina}, {Marsat}, {Martelli}, {Martin}, {Martin},
  {Martinez}, {Martinez}, {Martinez}, {Martinovic}, {Martynov}, {Marx},
  {Masalehdan}, {Mason}, {Massera}, {Masserot}, {Massinger}, {Masso-Reid},
  {Mastrogiovanni}, {Matas}, {Mateu-Lucena}, {Matichard}, {Matiushechkina},
  {Mavalvala}, {McCann}, {McCarthy}, {McClelland}, {McClincy}, {McCormick},
  {McCuller}, {McGhee}, {McGuire}, {McIsaac}, {McIver}, {McRae}, {McWilliams},
  {Meacher}, {Mehmet}, {Mehta}, {Meijer}, {Melatos}, {Melchor}, {Mendell},
  {Menendez-Vazquez}, {Menoni}, {Mercer}, {Mereni}, {Merfeld}, {Merilh},
  {Merritt}, {Merzougui}, {Meshkov}, {Messenger}, {Messick}, {Meyers},
  {Meylahn}, {Mhaske}, {Miani}, {Miao}, {Michaloliakos}, {Michel}, {Michimura},
  {Middleton}, {Milano}, {Miller}, {Miller}, {Miller}, {Millhouse}, {Mills},
  {Milotti}, {Minazzoli}, {Minenkov}, {Mio}, {Mir}, {Miravet-Ten{\'e}s},
  {Mishra}, {Mishra}, {Mistry}, {Mitra}, {Mitrofanov}, {Mitselmakher},
  {Mittleman}, {Miyakawa}, {Miyamoto}, {Miyazaki}, {Miyo}, {Miyoki}, {Mo},
  {Moguel}, {Mogushi}, {Mohapatra}, {Mohite}, {Molina}, {Molina-Ruiz},
  {Mondin}, {Montani}, {Moore}, {Moraru}, {Morawski}, {More}, {Moreno},
  {Moreno}, {Mori}, {Morisaki}, {Moriwaki}, {Mours}, {Mow-Lowry}, {Mozzon},
  {Muciaccia}, {Mukherjee}, {Mukherjee}, {Mukherjee}, {Mukherjee}, {Mukherjee},
  {Mukund}, {Mullavey}, {Munch}, {Mu{\~n}iz}, {Murray}, {Musenich}, {Muusse},
  {Nadji}, {Nagano}, {Nagano}, {Nagar}, {Nakamura}, {Nakano}, {Nakano},
  {Nakashima}, {Nakayama}, {Napolano}, {Nardecchia}, {Narikawa}, {Naticchioni},
  {Nayak}, {Nayak}, {Negishi}, {Neil}, {Neilson}, {Nelemans}, {Nelson}, {Nery},
  {Neubauer}, {Neunzert}, {Ng}, {Ng}, {Nguyen}, {Nguyen}, {Nguyen}, {Nguyen
  Quynh}, {Ni}, {Nichols}, {Nishizawa}, {Nissanke}, {Nitoglia}, {Nocera},
  {Norman}, {North}, {Nozaki}, {Nuttall}, {Oberling}, {O'Brien}, {Obuchi},
  {O'Dell}, {Oelker}, {Ogaki}, {Oganesyan}, {Oh}, {Oh}, {Oh}, {Ohashi},
  {Ohishi}, {Ohkawa}, {Ohme}, {Ohta}, {Okada}, {Okutani}, {Okutomi},
  {Olivetto}, {Oohara}, {Ooi}, {Oram}, {O'Reilly}, {Ormiston}, {Ormsby},
  {Ortega}, {O'Shaughnessy}, {O'Shea}, {Oshino}, {Ossokine}, {Osthelder},
  {Otabe}, {Ottaway}, {Overmier}, {Pace}, {Pagano}, {Page}, {Pagliaroli},
  {Pai}, {Pai}, {Palamos}, {Palashov}, {Palomba}, {Pan}, {Pan}, {Panda},
  {Pang}, {Pang}, {Pankow}, {Pannarale}, {Pant}, {Panther}, {Paoletti},
  {Paoli}, {Paolone}, {Parisi}, {Park}, {Park}, {Parker}, {Pascucci},
  {Pasqualetti}, {Passaquieti}, {Passuello}, {Patel}, {Pathak}, {Patricelli},
  {Patron}, {Patrone}, {Paul}, {Payne}, {Pedraza}, {Pegoraro}, {Pele},
  {Pe{\~n}a Arellano}, {Penn}, {Perego}, {Pereira}, {Pereira}, {Perez},
  {P{\'e}rigois}, {Perkins}, {Perreca}, {Perri{\`e}s}, {Petermann},
  {Petterson}, {Pfeiffer}, {Pham}, {Phukon}, {Piccinni}, {Pichot},
  {Piendibene}, {Piergiovanni}, {Pierini}, {Pierro}, {Pillant}, {Pillas},
  {Pilo}, {Pinard}, {Pinto}, {Pinto}, {Piotrzkowski}, {Pirello}, {Pitkin},
  {Placidi}, {Planas}, {Plastino}, {Pluchar}, {Poggiani}, {Polini}, {Pong},
  {Ponrathnam}, {Popolizio}, {Porter}, {Poulton}, {Powell}, {Pracchia},
  {Pradier}, {Prajapati}, {Prasai}, {Prasanna}, {Pratten}, {Principe}, {Prodi},
  {Prokhorov}, {Prosposito}, {Prudenzi}, {Puecher}, {Punturo}, {Puosi},
  {Puppo}, {P{\"u}rrer}, {Qi}, {Quetschke}, {Quitzow-James}, {Raab},
  {Raaijmakers}, {Radkins}, {Radulesco}, {Raffai}, {Rail}, {Raja}, {Rajan},
  {Ramirez}, {Ramirez}, {Ramos-Buades}, {Rana}, {Rapagnani}, {Rapol}, {Ray},
  {Raymond}, {Raza}, {Razzano}, {Read}, {Rees}, {Regimbau}, {Rei}, {Reid},
  {Reid}, {Reitze}, {Relton}, {Renzini}, {Rettegno}, {Rezac}, {Ricci},
  {Richards}, {Richardson}, {Richardson}, {Riemenschneider}, {Riles},
  {Rinaldi}, {Rink}, {Rizzo}, {Robertson}, {Robie}, {Robinet}, {Rocchi},
  {Rodriguez}, {Rolland}, {Rollins}, {Romanelli}, {Romano}, {Romel},
  {Romero-Rodr{\'\i}guez}, {Romero-Shaw}, {Romie}, {Ronchini}, {Rosa}, {Rose},
  {Rosi{\'n}ska}, {Ross}, {Rowan}, {Rowlinson}, {Roy}, {Roy}, {Roy}, {Rozza},
  {Ruggi}, {Ryan}, {Sachdev}, {Sadecki}, {Sadiq}, {Sago}, {Saito}, {Saito},
  {Sakai}, {Sakai}, {Sakellariadou}, {Sakuno}, {Salafia}, {Salconi}, {Saleem},
  {Salemi}, {Samajdar}, {Sanchez}, {Sanchez}, {Sanchez}, {Sanchis-Gual},
  {Sanders}, {Sanuy}, {Saravanan}, {Sarin}, {Sassolas}, {Satari}, {Sato},
  {Sato}, {Sauter}, {Savage}, {Sawada}, {Sawant}, {Sawant}, {Sayah},
  {Schaetzl}, {Scheel}, {Scheuer}, {Schiworski}, {Schmidt}, {Schmidt},
  {Schnabel}, {Schneewind}, {Schofield}, {Sch{\"o}nbeck}, {Schulte}, {Schutz},
  {Schwartz}, {Scott}, {Scott}, {Seglar-Arroyo}, {Sekiguchi}, {Sekiguchi},
  {Sellers}, {Sengupta}, {Sentenac}, {Seo}, {Sequino}, {Sergeev}, {Setyawati},
  {Shaffer}, {Shahriar}, {Shams}, {Shao}, {Sharma}, {Sharma}, {Shawhan},
  {Shcheblanov}, {Shibagaki}, {Shikauchi}, {Shimizu}, {Shimoda}, {Shimode},
  {Shinkai}, {Shishido}, {Shoda}, {Shoemaker}, {Shoemaker}, {ShyamSundar},
  {Sieniawska}, {Sigg}, {Singer}, {Singh}, {Singh}, {Singha}, {Sintes},
  {Sipala}, {Skliris}, {Slagmolen}, {Slaven-Blair}, {Smetana}, {Smith},
  {Smith}, {Soldateschi}, {Somala}, {Somiya}, {Son}, {Soni}, {Soni}, {Sordini},
  {Sorrentino}, {Sorrentino}, {Sotani}, {Soulard}, {Souradeep}, {Sowell},
  {Spagnuolo}, {Spencer}, {Spera}, {Srinivasan}, {Srivastava}, {Srivastava},
  {Staats}, {Stachie}, {Steer}, {Steinlechner}, {Steinlechner}, {Stops},
  {Stover}, {Strain}, {Strang}, {Stratta}, {Strunk}, {Sturani}, {Stuver},
  {Sudhagar}, {Sudhir}, {Sugimoto}, {Suh}, {Summerscales}, {Sun}, {Sun},
  {Sunil}, {Sur}, {Suresh}, {Sutton}, {Suzuki}, {Suzuki}, {Swinkels},
  {Szczepa{\'n}czyk}, {Szewczyk}, {Tacca}, {Tagoshi}, {Tait}, {Takahashi},
  {Takahashi}, {Takamori}, {Takano}, {Takeda}, {Takeda}, {Talbot}, {Talbot},
  {Tanaka}, {Tanaka}, {Tanaka}, {Tanaka}, {Tanaka}, {Tanasijczuk}, {Tanioka},
  {Tanner}, {Tao}, {Tao}, {Tapia San Mart{\'\i}n}, {Taranto}, {Tasson},
  {Telada}, {Tenorio}, {Terhune}, {Terkowski}, {Thirugnanasambandam}, {Thomas},
  {Thomas}, {Thompson}, {Thondapu}, {Thorne}, {Thrane}, {Tiwari}, {Tiwari},
  {Tiwari}, {Toivonen}, {Toland}, {Tolley}, {Tomaru}, {Tomigami}, {Tomura},
  {Tonelli}, {Torres-Forn{\'e}}, {Torrie}, {Tosta e Melo}, {T{\"o}yr{\"a}},
  {Trapananti}, {Travasso}, {Traylor}, {Trevor}, {Tringali}, {Tripathee},
  {Troiano}, {Trovato}, {Trozzo}, {Trudeau}, {Tsai}, {Tsai}, {Tsang}, {Tsang},
  {Tsao}, {Tse}, {Tso}, {Tsubono}, {Tsuchida}, {Tsukada}, {Tsuna}, {Tsutsui},
  {Tsuzuki}, {Turbang}, {Turconi}, {Tuyenbayev}, {Ubhi}, {Uchikata},
  {Uchiyama}, {Udall}, {Ueda}, {Uehara}, {Ueno}, {Ueshima}, {Unnikrishnan},
  {Uraguchi}, {Urban}, {Ushiba}, {Utina}, {Vahlbruch}, {Vajente}, {Vajpeyi},
  {Valdes}, {Valentini}, {Valsan}, {van Bakel}, {van Beuzekom}, {van den
  Brand}, {Van Den Broeck}, {Vander-Hyde}, {van der Schaaf}, {van Heijningen},
  {Vanosky}, {van Putten}, {van Remortel}, {Vardaro}, {Vargas}, {Varma},
  {Vas{\'u}th}, {Vecchio}, {Vedovato}, {Veitch}, {Veitch}, {Venneberg},
  {Venugopalan}, {Verkindt}, {Verma}, {Verma}, {Veske}, {Vetrano},
  {Vicer{\'e}}, {Vidyant}, {Viets}, {Vijaykumar}, {Villa-Ortega}, {Vinet},
  {Virtuoso}, {Vitale}, {Vo}, {Vocca}, {von Reis}, {von Wrangel}, {Vorvick},
  {Vyatchanin}, {Wade}, {Wade}, {Wagner}, {Walet}, {Walker}, {Wallace},
  {Wallace}, {Walsh}, {Wang}, {Wang}, {Wang}, {Ward}, {Warner}, {Was},
  {Washimi}, {Washington}, {Watada}, {Watchi}, {Weaver}, {Webster}, {Weinert},
  {Weinstein}, {Weiss}, {Weller}, {Wellmann}, {Wen}, {We{\ss}els}, {Wette},
  {Whelan}, {White}, {Whiting}, {Whittle}, {Wilken}, {Williams}, {Williams},
  {Williamson}, {Willis}, {Willke}, {Wilson}, {Winkler}, {Wipf}, {Wlodarczyk},
  {Woan}, {Woehler}, {Wofford}, {Wong}, {Wu}, {Wu}, {Wu}, {Wu}, {Wysocki},
  {Xiao}, {Xu}, {Yamada}, {Yamamoto}, {Yamamoto}, {Yamamoto}, {Yamamoto},
  {Yamashita}, {Yamazaki}, {Yang}, {Yang}, {Yang}, {Yang}, {Yang}, {Yap},
  {Yeeles}, {Yelikar}, {Ying}, {Yokogawa}, {Yokoyama}, {Yokozawa}, {Yoo},
  {Yoshioka}, {Yu}, {Yu}, {Yuzurihara}, {Zadro{\.z}ny}, {Zanolin}, {Zeidler},
  {Zelenova}, {Zendri}, {Zevin}, {Zhan}, {Zhang}, {Zhang}, {Zhang}, {Zhang},
  {Zhang}, {Zhao}, {Zhao}, {Zhao}, {Zhao}, {Zhou}, {Zhou}, {Zhu}, {Zhu},
  {Zimmerman}, {Zucker}, {Zweizig}, {Bhardwaj}, {Boyle}, {Cassanelli}, {Dong},
  {Fonseca}, {Kaspi}, {Leung}, {Masui}, {Meyers}, {Michilli}, {Ng}, {Pearlman},
  {Petroff}, {Pleunis}, {Rafiei-Ravandi}, {Rahman}, {Ransom}, {Scholz}, {Shin},
  {Smith}, {Stairs}, {Tendulkar}, \& {Zwaniga}}]{2022arXiv220312038T}
{The LIGO Scientific Collaboration}, {the Virgo Collaboration}, {the KAGRA
  Collaboration}, {et~al.} 2022, arXiv e-prints, arXiv:2203.12038.
\newblock \doarXiv{2203.12038}

\bibitem[{{Thulasiram} \& {Lin}(2021)}]{2021MNRAS.508.1947T}
{Thulasiram}, P., \& {Lin}, H.-H. 2021, \mnras, 508, 1947,
  \dodoi{10.1093/mnras/stab2692}

\bibitem[{{Tominaga} {et~al.}(2018){Tominaga}, {Niino}, {Totani}, {Yasuda},
  {Furusawa}, {Tanaka}, {Bhandari}, {Dodson}, {Keane}, {Morokuma}, {Petroff},
  \& {Possenti}}]{2018PASJ...70..103T}
{Tominaga}, N., {Niino}, Y., {Totani}, T., {et~al.} 2018, \pasj, 70, 103,
  \dodoi{10.1093/pasj/psy101}

\bibitem[{{Torchinsky} {et~al.}(2016){Torchinsky}, {Broderick}, {Gunst},
  {Faulkner}, \& {van Cappellen}}]{2016arXiv161000683T}
{Torchinsky}, S.~A., {Broderick}, J.~W., {Gunst}, A., {Faulkner}, A.~J., \&
  {van Cappellen}, W. 2016, arXiv e-prints, arXiv:1610.00683.
\newblock \doarXiv{1610.00683}

\bibitem[{{Totani}(2013)}]{2013PASJ...65L..12T}
{Totani}, T. 2013, \pasj, 65, L12, \dodoi{10.1093/pasj/65.5.L12}

\bibitem[{{Vanderlinde} {et~al.}(2019){Vanderlinde}, {Liu}, {Gaensler}, {Bond},
  {Hinshaw}, {Ng}, {Chiang}, {Stairs}, {Brown}, {Sievers}, {Mena}, {Smith},
  {Bandura}, {Masui}, {Spekkens}, {Belostotski}, {Dobbs}, {Turok}, {Boyle},
  {Rupen}, {Landecker}, {Pen}, \& {Kaspi}}]{2019clrp.2020...28V}
{Vanderlinde}, K., {Liu}, A., {Gaensler}, B., {et~al.} 2019, in Canadian Long
  Range Plan for Astronomy and Astrophysics White Papers, Vol. 2020, 28,
  \dodoi{10.5281/zenodo.3765414}

\bibitem[{{Wayth} {et~al.}(2015){Wayth}, {Lenc}, {Bell}, {Callingham},
  {Dwarakanath}, {Franzen}, {For}, {Gaensler}, {Hancock}, {Hindson},
  {Hurley-Walker}, {Jackson}, {Johnston-Hollitt}, {Kapi{\'n}ska}, {McKinley},
  {Morgan}, {Offringa}, {Procopio}, {Staveley-Smith}, {Wu}, {Zheng}, {Trott},
  {Bernardi}, {Bowman}, {Briggs}, {Cappallo}, {Corey}, {Deshpande}, {Emrich},
  {Goeke}, {Greenhill}, {Hazelton}, {Kaplan}, {Kasper}, {Kratzenberg},
  {Lonsdale}, {Lynch}, {McWhirter}, {Mitchell}, {Morales}, {Morgan}, {Oberoi},
  {Ord}, {Prabu}, {Rogers}, {Roshi}, {Shankar}, {Srivani}, {Subrahmanyan},
  {Tingay}, {Waterson}, {Webster}, {Whitney}, {Williams}, \&
  {Williams}}]{2015PASA...32...25W}
{Wayth}, R.~B., {Lenc}, E., {Bell}, M.~E., {et~al.} 2015, \pasa, 32, e025,
  \dodoi{10.1017/pasa.2015.26}

\bibitem[{{Wei} {et~al.}(2015){Wei}, {Gao}, {Wu}, \&
  {M{\'e}sz{\'a}ros}}]{2015PhRvL.115z1101W}
{Wei}, J.-J., {Gao}, H., {Wu}, X.-F., \& {M{\'e}sz{\'a}ros}, P. 2015, \prl,
  115, 261101, \dodoi{10.1103/PhysRevLett.115.261101}

\bibitem[{{Wei} {et~al.}(2018){Wei}, {Wu}, \& {Gao}}]{2018ApJ...860L...7W}
{Wei}, J.-J., {Wu}, X.-F., \& {Gao}, H. 2018, \apjl, 860, L7,
  \dodoi{10.3847/2041-8213/aac8e2}

\bibitem[{{Yalinewich} \& {Pen}(2022)}]{2022MNRAS.515.5682Y}
{Yalinewich}, A., \& {Pen}, U.-L. 2022, \mnras, 515, 5682,
  \dodoi{10.1093/mnras/stac2087}

\bibitem[{{Yamasaki} {et~al.}(2018){Yamasaki}, {Totani}, \&
  {Kiuchi}}]{2018PASJ...70...39Y}
{Yamasaki}, S., {Totani}, T., \& {Kiuchi}, K. 2018, \pasj, 70, 39,
  \dodoi{10.1093/pasj/psy029}

\bibitem[{{Yang} {et~al.}(2019){Yang}, {Zhang}, \& {Wei}}]{2019ApJ...878...89Y}
{Yang}, Y.-P., {Zhang}, B., \& {Wei}, J.-Y. 2019, \apj, 878, 89,
  \dodoi{10.3847/1538-4357/ab1fe2}

\bibitem[{{Zackay} \& {Ofek}(2017)}]{2017ApJ...835...11Z}
{Zackay}, B., \& {Ofek}, E.~O. 2017, \apj, 835, 11,
  \dodoi{10.3847/1538-4357/835/1/11}

\end{thebibliography}

\bibliographystyle{aasjournal}

\end{document}